\documentclass[paper]{JHEP}
\usepackage{epsfig}
\usepackage{amssymb}
\def\beq{\begin{equation}}
\def\eeq{\end{equation}}

\def\beqn{\begin{eqnarray}}
\def\eeqn{\end{eqnarray}}
\def \ep{\epsilon}
\def \as{\alpha_{\rm s}}
\def\VEV#1{\left\langle #1\right\rangle}
\def\abs#1{\left|#1\right|}
\def\HW{{\small HERWIG}}

\newcommand\OO{\overline O}

\newcommand\sss{\scriptscriptstyle\rm}
\newcommand\xsec{\frac{d\sigma}{dx}}
\newcommand\xsecy{\frac{d\sigma}{dy}}
\newcommand\xsecz{\frac{d\sigma}{dz}}
\newcommand\xsecO{\frac{d\sigma}{dO}}
\newcommand\xsecOO{\frac{d\sigma}{d\OO}}
\newcommand\xsecONLO{\left(\xsecO\right)_{\sss NLO}}
\newcommand\xsecOONLO{\left(\xsecOO\right)_{\sss NLO}}
\newcommand\xsecOH{\left(\xsecO\right)_{\sss H}}
\newcommand\xsecORSz{\left(\xsecO\right)_{\sss RS0}}
\newcommand\xsecORS{\left(\xsecO\right)_{\sss RS}}
\newcommand\xdsecO{d\sigma/dO}
\newcommand\xdsecONLO{(d\sigma/dO)_{\sss NLO}}
\newcommand\xdsecOH{(d\sigma/dO)_{\sss H}}
\newcommand\xdsecORSz{(d\sigma/dO)_{\sss RS0}}
\newcommand\xdsecORS{(d\sigma/dO)_{\sss RS}}
\newcommand\xsborn{\left(\xsec\right)_{\sss B}}
\newcommand\xsvirt{\left(\xsec\right)_{\sss V}}
\newcommand\xsreal{\left(\xsec\right)_{\sss R}}
\newcommand\xm{x_{\sss M}}
\newcommand\IMC{I_{\sss MC}}

\newcommand\wEV{w_{\sss EV}}
\newcommand\wCNT{w_{\sss CT}}
\newcommand\MCatNLO{{\rm MC}@{\rm NLO}}
\newcommand\dyNLO{\left(\xsecy\right)_{\sss NLO}}
\newcommand\dyMC{\left(\xsecy\right)_{\sss MC}}
\newcommand\dzNLO{\left(\xsecz\right)_{\sss NLO}}
\newcommand\dzMC{\left(\xsecz\right)_{\sss MC}}

\newcommand\clH{{\mathbb H}}
\newcommand\clS{{\mathbb S}}
\newcommand\Prob{{\mathbb P}}
\newcommand\xsecOOclH{\frac{d\sigma_{\clH}}{d\OO}}
\newcommand\xsecOOclS{\frac{d\sigma_{\clS}}{d\OO}}
\newcommand\dszero{\frac{d\sigma_{\clS}}{dy}}
\newcommand\dsone{\frac{d\sigma_{\clH}}{dy}}
\newcommand\dszerond{d\sigma_{\clS}/dy}
\newcommand\dsonend{d\sigma_{\clH}/dy}
\newcommand\orda{{\cal O}(a)}
\newcommand\ordat{{\cal O}(a^2)}
\newcommand\stot{\sigma_{tot}}
\newcommand\stepf{\Theta}
\newcommand\mborn{{\cal M}^{(b)}}
\newcommand\mvreg{{\cal M}^{(v,reg)}}
\newcommand\mreal{{\cal M}^{(r)}}
\newcommand\ev{{|_{\rm ev}}}
\newcommand\cnt{|_{\rm ct}}
\newcommand\xMC{|_{\sss {\rm MC}}}

\newcommand\evB{{\Big|_{\rm ev}}}
\newcommand\cntB{\Big|_{\rm ct}}

\newcommand\xMCB{\Big|_{\sss {\rm MC}}}
\newcommand\evBB{{\Bigg|_{\rm ev}}}
\newcommand\cntBB{\Bigg|_{\rm ct}}
\newcommand\xMCBB{{\Bigg|_{\sss {\rm MC}}}}

\newcommand\Ktwo{{\bf 2}}
\newcommand\Kqtwo{{\bf \tilde{2}}}
\newcommand\Kthree{{\bf 3}}
\newcommand\KK{{\bf k}}
\newcommand\Itwo{{\cal I}_2}
\newcommand\Iqtwo{{\cal I}_{\tilde {2}}}

\newcommand\mwwt{M_{\sss WW}^2}
\newcommand\mwt{m_{\sss W}^2}
\newcommand\kww{k_{\sss WW}}
\newcommand\kt{k_{\sss {\rm T}}}
\newcommand\pt{p_{\sss {\rm T}}}
\newcommand\Et{E_{\sss {\rm T}}}
\newcommand\ptwp{p_{\sss {\rm T}}^{\sss W^+}}
\newcommand\Etwp{M_{\sss {\rm T}}^{\sss W^+}}
\newcommand\Etwm{M_{\sss {\rm T}}^{\sss W^-}}
\newcommand\ywp{y^{\sss W^+}}
\newcommand\yww{y_{\sss WW}}
\newcommand\ptww{p_{\sss {\rm T}}^{\sss (WW)}}
\newcommand\dphiww{\Delta\phi^{\sss (WW)}}

\newcommand\thw{\theta_{\sss W}}
\newcommand\CF{C_{\rm {\sss F}}}

\newcommand\omxrhot{\left(\frac{1}{1-x}\right)_{\tilde{\rho}}}
\newcommand\lomxrhot{\left(\frac{\log(1-x)}{1-x}\right)_{\tilde{\rho}}}
\newcommand\omyom{\left(\frac{1}{1-y}\right)_{\omega}}
\newcommand\opyom{\left(\frac{1}{1+y}\right)_{\omega}}
\newcommand\opmyom{\left(\frac{1}{1\pm y}\right)_{\omega}}
\newcommand\pfun{{\cal P}}
\newcommand\Gfun{{\cal G}}
\newcommand\omxP{\left(\frac{1}{1-x}\right)_{\pfun}}
\newcommand\lomxP{\left(\frac{\log(1-x)}{1-x}\right)_{\pfun}}
\newcommand\FomxP{\left(\frac{F(x)}{1-x}\right)_{\pfun}}
\newcommand\omyP{\left(\frac{1}{1-y}\right)_{\pfun}}
\newcommand\opyP{\left(\frac{1}{1+y}\right)_{\pfun}}
\newcommand\opmyP{\left(\frac{1}{1\pm y}\right)_{\pfun}}
\newcommand\Fsoft{{\cal F}_s}
\newcommand\Fcoll{{\cal F}_c}
\newcommand\by{\bar{y}}
\newcommand\bSigma{\overline{\Sigma}}
\newcommand\Dfun{{\cal D}}
\newcommand\muz{\mu_0}
\newcommand\muf{\mu_{\sss F}}
\newcommand\mur{\mu_{\sss R}}
\newcommand\muMC{\mu_{\sss MC}}
\newcommand\muNLO{\mu_{\sss NLO}}
\newcommand\muMCatNLO{\mu_{\sss \MCatNLO}}
\newcommand\xoMC{x_1^{({\sss MC})}}
\newcommand\xtMC{x_2^{({\sss MC})}}
\newcommand\xos{x_1^{(s)}}
\newcommand\xts{x_2^{(s)}}
\newcommand\xocp{x_1^{(c+)}}
\newcommand\xtcp{x_2^{(c+)}}
\newcommand\xocm{x_1^{(c-)}}
\newcommand\xtcm{x_2^{(c-)}}
\newcommand\Hone{(H_1)}
\newcommand\Htwo{(H_2)}
\newcommand\flav{{\bf f}}
\preprint{
 Cavendish--HEP--02/01\hfill\\
 LAPTH--905/02\hfill\\
 GEF--th--2/2002\hfill\\
 hep-ph/0204244}
\title{\boldmath Matching NLO QCD Computations and
Parton Shower Simulations%
\footnote{Work supported in part by the UK Particle Physics and
Astronomy Research Council and by the EU Fourth Framework Programme
`Training and Mobility of Researchers', Network `Quantum Chromodynamics
and the Deep Structure of Elementary Particles',
contract FMRX-CT98-0194 (DG 12 - MIHT).}}
\author{Stefano Frixione%
  \thanks{On leave of absence from INFN, Sez. di Genova, Italy}\\
  Laboratoire d'Annecy-le-Vieux de Physique de Particules\\
  Chemin de Bellevue, BP 110,
  74941 Annecy-le-Vieux CEDEX, France\\
  E-mail: \email{Stefano.Frixione@cern.ch}}
\author{Bryan R.\ Webber\\
  Cavendish Laboratory, University of Cambridge,\\
  Madingley Road, Cambridge CB3 0HE, U.K.\\
  E-mail: \email{webber@hep.phy.cam.ac.uk}}
\abstract{We propose a method for matching the next-to-leading order
  (NLO) calculation of a given QCD process with a parton shower Monte
  Carlo (MC) simulation. The method has the following features: fully
  exclusive events are generated, with hadronization according to the
  MC model; total exclusive rates are accurate to NLO; NLO results for
  distributions are recovered upon expansion in $\as$; hard emissions
  are treated as in NLO computations while soft/collinear emissions
  are handled by the MC simulation, with the same logarithmic accuracy
  as the MC; and matching between the hard- and soft/collinear-emission 
  regions is smooth. A fraction of events with negative weight is
  generated, but unweighting remains possible with reasonable
  efficiency.  The method is clarified using a simple toy model, and
  illustrated by application to the hadroproduction of W$^+$W$^-$
  pairs.  }
\begin{document}
\section{Introduction}
The reliable prediction of cross sections and final-state distributions
for QCD processes is important not only as a test of QCD but also for
the design of collider experiments and new particle searches.  All
systematic approaches to this problem are based on perturbation theory,
usually truncated at next-to-leading order (NLO), or in a few cases
next-next-to-leading order (NNLO).  Such calculations yield the
best available results for sufficiently inclusive observables, in
kinematic regions where higher-order corrections are not enhanced.
However, in many cases a more exclusive description of final states
and/or a wider kinematic coverage are needed.

For the description of exclusive hadronic final states, perturbative
calculations have to be combined with a model for the conversion of
partonic final states into hadrons (hadronization).  Existing
hadronization models are in remarkably good agreement with a wide
range of data, after tuning of model parameters
\cite{Marchesini:1992ch,Sjostrand:2000wi}. However, these models operate
on partonic states with high multiplicity and low relative transverse
momenta, which are obtained from a parton shower
\cite{Sjostrand:1986ys,Marchesini:1988cf}
or dipole cascade~\cite{Lonnblad:1992tz}
approximation to QCD dynamics and not from fixed-order calculations.

To extend the coverage of perturbative calculations into regions where
higher-order contributions are enhanced, one needs to identify and resum the
enhanced terms to all orders. The resummed expression can then be combined
with the fixed-order result, after subtracting resummed terms of the
same order to avoid double-counting.  This has been done explicitly for
many important observables and processes, but a general procedure has not
yet been developed. In principle, the parton shower provides such a
procedure, if the problem of double-counting can be overcome. Combining
NLO and parton-shower computations also
offers the possibility of interfacing to hadronization models, thus
overcoming the main deficiencies of existing calculations.

Methods for combining fixed-order and parton-shower or dipole-cascade
computations in QCD have been discussed in
refs.~\cite{Seymour:1995df}-\cite{Lonnblad:2001iq}.
Many of those papers concern the rather different problem of combining
leading-order (tree level) matrix elements for hard multi-jet production
with parton showers or dipole cascades for jet fragmentation.
The only fully-implemented schemes for matching NLO
computations and parton showers that have been used to produce
phenomenological results~\cite{Baer:1991ca}-\cite{Dobbs:2001gb}
have been based on the so-called phase-space slicing
method for regularising the infrared divergences encountered at NLO.
The slicing method is simple but has the disadvantage of introducing a
slicing parameter, which should in principle be sent to zero at the end of
the calculation but in practice cannot be chosen too small. This can
introduce uncontrolled errors and/or large negative event weights.
The slicing method has therefore tended to be replaced where
possible in NLO calculations by subtraction methods, in which no
approximations need to be made and event weights can be
controlled by improving the choice of subtraction terms.

The subtraction method has been considered in 
refs.~\cite{Collins:2000qd,Chen:2001ci,Collins:2001fm,Chen:2001nf}. The main
concern of those papers was the correct matching of NLO matrix element
calculations to the scale and scheme for next-to-leading logarithmic (NLL)
parton showers.  Up to now, no satisfactory algorithms for NLL showering
have been developed.  When such algorithms are available, the methods
proposed there will make it possible for event generators
to generate the NLO evolution of parton distribution functions (PDFs)
and hard process cross sections, without the explicit input of NLO PDFs.
Existing parton shower generators, however, correspond only to LL evolution, 
and therefore the matching criteria of 
refs.~\cite{Collins:2000qd,Chen:2001ci,Collins:2001fm,Chen:2001nf}
operate beyond the level of precision that can be achieved at present.

In the present paper, we present a method for matching NLO matrix elements 
and existing parton shower generators based on the subtraction method 
for NLO calculations. 
First, in sect.~\ref{sec:gen}, we outline the general objectives and
limitations of our approach. The method is then illustrated in
sect.~\ref{sec:toy} in the context of a simplified toy model. 
The case of QCD is dealt with in sects.~\ref{sec:QCDMC} 
and~\ref{sec:other}. In sect.~\ref{sec:QCDMC} results are
presented for the process of W$^+$W$^-$ production in hadronic
collisions. In sect.~\ref{sec:other} we discuss briefly the
application of the method to other, more complex, processes.
Concluding comments are given in sect.~\ref{sec:conc}.
All the technical details are collected in the appendices, where we
also comment on the matching between NLO computations and parton shower 
approaches in the framework of the slicing method.

\section{General framework}\label{sec:gen}
\subsection{Objectives\label{sec:obj}}
It is undeniable that parton shower Monte Carlos (MC's) provide more general 
and flexible tools compared to NLO computations. By running an MC we
get events, which we believe to be a faithful description of the real
events detected by real experiments; this is one of the reasons why
the MC's are so popular with experimentalists. On the other hand,
NLO computations emphasize the role of (infrared-safe) observables; the 
presence of negative contributions, and the necessity of considering
inclusive quantities in order to get rid of infinities, make it very
problematic to even talk about events. Still, NLO computations have
their virtues; they can handle hard emissions, and they can estimate
total rates with better accuracy than MC's.

In this paper, we aim to construct an MC that works like a regular
MC, but on top of that knows how to treat hard emissions, and can
compute rates to NLO accuracy. For brevity, we call this tool
an $\MCatNLO$. Clearly, there is a lot of freedom in constructing
an $\MCatNLO$. We shall follow a minimal approach. Namely, we want
to keep the original MC as unmodified as possible; all major changes
have to be carried out on the NLO codes. Amongst other things, this
implies that our $\MCatNLO$ will not improve the logarithmic accuracy
of the original MC, in those regions of the phase space where resummation
is needed. As far as the NLO calculations are concerned, we shall not
perform any approximations, neither at the level of matrix elements,
nor at the level of phase space; thus, we shall adopt the subtraction
method in order to cancel the infrared divergences that arise in the
intermediate steps of the computation.

To be more definite, we require $\MCatNLO$ to have the following 
characteristics:
\begin{itemize}
\item The output is a set of events, which are fully exclusive.
\item Total rates are accurate to NLO.
\item NLO results for all observables are recovered upon expansion 
 of $\MCatNLO$ results in $\as$. 
\item Hard emissions are treated as in NLO computations.
\item Soft/collinear emissions are treated as in MC.
\item The matching between hard- and soft-emission regions is smooth.
\item MC hadronization models are adopted.
\end{itemize}
In sect.~\ref{sec:QCDMC}, we shall construct an $\MCatNLO$ fulfilling 
these conditions. However, we find that the technical 
complications of QCD tend to hide the basic ideas upon which
our approach is based. Thus, we prefer to start from
an oversimplified (and unrealistic) case, where the structure
of the $\MCatNLO$ is apparent; this is done in sect.~\ref{sec:toy}.
The reason for doing this is that we shall see that 
the approach that leads to the $\MCatNLO$ in a simple case
can actually be adopted in the QCD case as well, which will
only be technically more involved.
\newpage
\subsection{Negative weights}
We do not require the $\MCatNLO$ to be positive definite; events
may have negative weights. However, 
it should be pointed out that these negative-weight events are
of a completely different nature from the negative 
contributions that appear in NLO computations. In particular, the 
distributions of the positive- and negative-weight events are
{\em separately} finite in our $\MCatNLO$ (the same is not true
in an NLO computation). This implies the possibility of unweighted
event generation,\footnote{In this paper, we adopt a slightly different
notation from that commonly used by experimentalists. Namely,
we call unweighted events any set of events whose weights are all identical, 
{\em up to a sign}. In other words, the weights of our unweighted events are
of the form $\pm w$, where $w$ is a constant.}
as is customary in ordinary MC's. The situation is
actually completely analogous to that occurring in polarized collisions,
where physical observables are obtained as the difference of two
positive-definite and finite quantities. Furthermore, we shall see that 
in our case the number of negative-weight events is reasonably small, and
thus the number of $\MCatNLO$ events necessary to get a smooth distribution
is comparable to that in ordinary MC's.

\subsection{Double counting\label{sec:double}}
When the MC generates events with real parton emission, it generates 
kinematical configurations that are also taken into account by the
NLO computation: the possibility of having the same kinematical
configuration from the MC {\em and} from the NLO may lead to
double counting. We shall adopt the following quantitative
definition of double counting:

\noindent{\em
An $\MCatNLO$ is affected by double counting if 
its prediction for any observable, at the first order beyond the
Born approximation\footnote{By the Born approximation we mean
the lowest order at which the hard process contributes to any
observable.} in the expansion in the coupling constant,
is not equal to the NLO prediction.}

According to this definition, double counting may correspond to
either an excess or a deficit in the prediction, at any point in
phase space. This includes contributions from real emission and
virtual corrections.
Generally speaking, the MC fills the phase space efficiently only
in the soft and collinear regions. In these regions, the leading 
behaviour of the kinematical distribution of one-particle 
emission is given by the Born cross section, times a kernel
that describes the soft or collinear emission. This holds
both for the MC and for the NLO. Thus, in the soft and collinear 
regions of the one-particle-emission phase space, the MC and NLO 
results coincide when only the leading terms are considered.
Elsewhere, the MC is not reliable and the NLO provides the best
estimate.  The problem is to merge these two descriptions.

\section{Toy model studies}\label{sec:toy}
In this section we study a toy model that allows a simple discussion of 
the key features of the NLO computation, of the MC approach, and of the
matching between the two. We assume that a system can radiate massless
particles (which we call photons), whose energy we denote by $x$,
with $0\le x \le x_s\le 1$, $x_s$ being the energy of the system
before the radiation. After the radiation, the energy of the system is
$x_s^\prime=x_s-x$. The system can undergo several further emissions; 
on the other hand, one photon cannot split further.

In a perturbative computation, the Born term corresponds to no
emissions. The first non-trivial order in perturbation theory
gets contribution from those diagrams with one and only one
emission, with either virtual or real photons. We write the 
corresponding contributions to the cross section as follows:
\beqn
\xsborn&=&B\delta(x),
\label{born}
\\
\xsvirt&=&a\left(\frac{B}{2\ep}+V\right)\delta(x),
\label{virt}
\\
\xsreal&=&a\frac{R(x)}{x},
\label{real}
\eeqn
for the Born, virtual, and real contributions respectively; $a$ is
the coupling constant, possibly times a colour factor; $B$ and $V$ are
constant with respect to $x$, and
\beq
\lim_{x\to 0}R(x)=B.
\label{limreal}
\eeq
The constant $B$ appears in eqs.~(\ref{virt}) and~(\ref{limreal}) 
since we expect the residue of the leading singularity of the virtual
and real contributions to be given by the Born term, times a suitable
kernel. We take this kernel equal to 1, since this simplifies the
computations, and it is not restrictive. Finally, $\ep$ is the 
parameter entering dimensional regularization in $4-2\ep$ dimensions.

The task of predicting an infrared-safe observable $O$ to NLO accuracy 
amounts to computing the following quantity
\beq
\VEV{O}=\lim_{\ep\to 0}\int_0^1 dx\,x^{-2\ep} O(x)\left[\xsborn + \xsvirt 
+\xsreal\right],
\label{nlopred}
\eeq
where $O(x)$ is the observable as a function of $x$, possibly times a 
set of $\stepf$ functions defining a histogram bin. By ``infrared-safe''
here we simply mean that the integral in eq.~(\ref{nlopred}) exists. 
The main technical problem in eq.~(\ref{nlopred}) is due to the
presence of the regularising parameter $\ep$; in order to have an
efficient numerical procedure, it is mandatory to extract the pole in
$\ep$ from the real contribution, in this way cancelling analytically
the pole explicitly present in the virtual contribution.

\subsection{NLO: slicing versus subtraction}
Within the context of this toy model, the usual methods of cancelling
the pole in $\ep$ are as follows.
In the {\em slicing method}, a small parameter $\delta$ is introduced
into the real contribution in the following way:
\beq
\VEV{O}_{\sss R}=\int_0^\delta dx\, x^{-2\ep} O(x)\xsreal + 
\int_\delta^1 dx\, x^{-2\ep} O(x)\xsreal .
\label{Rslicing}
\eeq
In the first term on the r.h.s.\ of this equation we expand $O(x)$ and $R(x)$
in Taylor series around 0, and keep only the first term; the smaller
$\delta$, the better the approximation. On the other hand, the second
term in eq.~(\ref{Rslicing}) does not contain any singularity, and
we can just set $\ep=0$. We obtain
\beqn
\VEV{O}_{\sss R}&=&aBO(0)\int_0^\delta dx\, \frac{x^{-2\ep}}{x} +
\int_\delta^1 dx\, O(x)\xsreal + {\cal O}(\delta)
\\
&=&a\left(-\frac{1}{2\ep}+\log\delta\right)BO(0)
+a\int_\delta^1 dx\, \frac{O(x)R(x)}{x} + {\cal O}(\delta,\ep).
\eeqn
Using this result in eq.~(\ref{nlopred}), we get the NLO prediction
for $O$ as given in the slicing method:
\beq
\VEV{O}_{\sss slice}=BO(0)+a\left[\left(B\log\delta +V\right)O(0) + 
\int_\delta^1 dx\, \frac{O(x)R(x)}{x}\right]
+ {\cal O}(\delta).
\label{nloslicing}
\eeq
The terms ${\cal O}(\delta)$ cannot be computed, and are neglected; 
in general, one must check that these neglected terms are numerically 
small, by plotting $\VEV{O}_{\sss slice}$ versus $\delta$ in a suitable range 
in $\delta$.

In the {\em subtraction method}, no approximation is performed.
One writes the real contribution as follows:
\beq
\VEV{O}_{\sss R}=aBO(0)\int_0^1 dx\, \frac{x^{-2\ep}}{x} 
+a\int_0^1 dx\, \frac{O(x)R(x)-BO(0)}{x^{1+2\ep}}\;.
\label{subtraction}
\eeq
The second term on the r.h.s.\ does not 
contain singularities, and we can set $\ep=0$:
\beq
\VEV{O}_{\sss R}=-a\frac{B}{2\ep}O(0)
+a\int_0^1 dx\, \frac{O(x)R(x)-BO(0)}{x}\;.
\eeq
Therefore, the NLO prediction as given in the subtraction method is:
\beq
\VEV{O}_{\sss sub}=BO(0)+a\left[V O(0) + 
\int_0^1 dx\, \frac{O(x)R(x)-BO(0)}{x}\right].
\label{nlosubt}
\eeq
We rewrite this in a slightly different form, which will be more
suited to discussion of the matching with the MC:
\beq
\VEV{O}_{\sss sub}=\int_0^1 dx \left[O(x)\frac{aR(x)}{x}
+O(0)\left(B+aV-\frac{aB}{x}\right)\right].
\label{nlosubtint}
\eeq

\subsection{Toy Monte Carlo\label{sec:toyMC}}
In a treatment based on Monte Carlo methods, the system can undergo
an arbitrary number of emissions (branchings), with probability controlled 
by the {\em Sudakov form factor}, defined for our toy model as follows:
\beq
\Delta(x_1,x_2)=\exp\left[-a\int_{x_1}^{x_2}dz\frac{Q(z)}{z}\right],
\label{Deltadef}
\eeq
where $Q(z)$ is a monotonic function with the following general properties:
\beq
0\le Q(z)\le 1,\;\;\;\;
\lim_{z\to 0}Q(z)=1,\;\;\;\;
\lim_{z\to 1}Q(z)=0.
\label{fconditions}
\eeq
Note that the MC is also well defined without the last
condition in eq.~(\ref{fconditions}).
We discuss later the specific functional form assumed for $Q$. 
If $x_s$ is the energy of the system before the first branching occurs,
then $\Delta(x,x_s)$ is the probability that no photon be emitted with 
energy $z$ such that $x\le z\le x_s$. At fixed $x$, this probability 
tends to 1 when $a\to 0$; that is, no emission is possible in the 
zero-coupling limit. On the other hand, when $a\to\infty$, we have 
$\Delta(x,x_s)\to 0$, which means that it is impossible that the 
system does not emit. The toy MC code we have written implements
ordered emissions in the photon energies\footnote{This may violate
energy-momentum conservation, as is usual in a leading logarithm
implementation.} and proceeds through the following steps:

\begin{itemize}

\item[0)] Define $x_0$ as the lower bound on the photon energy after 
 a branching; thus $x_0$ plays the same role as the cutoffs used to
 define the Sudakov form factor in QCD MC's.

\item[1)] Define $\xm$ to be the maximum energy available to the photon;
 before the first branching, we have $\xm=x_s$, where $x_s$ is the energy 
 of the system.

\item[2)] Pick a random number $R_1\in [0,1]$, and solve for $z$ the equation
\beq
\exp\left[-a\int_{z}^{\xm}\frac{dz^\prime}{z^\prime}\right]
\equiv\left(\frac{z}{\xm}\right)^a = 
R_1\;\;\;\;\Longrightarrow\;\;\;\;z=\xm R_1^{1/a}.
\label{photonenergy}
\eeq

\item[3)]Pick a random number $R_2\in [0,1]$. If $Q(z)<R_2$, let $\xm=z$, 
reject the branching, and return to step 2).

\item[4)] If $z<x_0$, reject the branching and exit. If $z>x_0$,
 a photon with energy $z$ has been emitted. We let $\xm=z$, and 
 iterate the procedure going back to 2).

\end{itemize}
This procedure ensures that the photon emissions have the distribution
prescribed by the Sudakov form factor in eq.~(\ref{Deltadef}).
As can be seen from that equation, the probability of
having soft emission (i.e., with $z<x_0$), and thus of terminating 
the shower, is $\Delta(x_0,x_s)$, which tends to one when $a\to 0$ 
(weak coupling), and when $x_s\to x_0$ (no energy left for 
non-soft branching). In this toy model, the hardest emission
is always the first one.

\subsection{Matching NLO and MC\label{sec:Matching}}
In order to introduce some useful notation, we start from the LO perturbative
result, which does not pose any problems, since it is in fact what is
implemented in ordinary MC's. At the LO, the system has not yet lost
any energy to perturbative photon emission. Thus, the maximum energy
available to the photon in the first branching is 1. For each event,
we set $\xm=1$, and proceed as described in sect.~\ref{sec:toyMC}. The 
total exclusive rate will have to be equal to $B$, according to 
eq.~(\ref{born});
thus, if we generate $N$ events, each event has a weight equal to $B/N$.

We can formally read this procedure from eq.~(\ref{nlosubtint}), by setting
$a=0$ and performing the formal substitution
\beq
O(x)\;\longrightarrow\;\IMC(O,\xm(x))\,,
\label{IMCdef}
\eeq
where $\IMC(O,\xm)$ stands for interface-to-MC, and indicates symbolically
the distribution in the observable $O$ as obtained by running the MC
starting from a given $\xm$, and giving each event the weight $1/N$; 
its precise definition is given in
app.~\ref{app:IMCdef}. For $x=0$ we need $\xm=1$; we do not need
to specify $\xm(x)$ for other values of $x$ at present. Interfacing
LO with MC, we thus obtain
\beq
\left(\xsecO\right)_{\sss MC@LO}=B\IMC(O,1).
\label{IMCborn}
\eeq
This equation reminds us of the total rate ($B$), and the maximum energy
available for photons at the first branching ($\xm=1$).

\subsubsection{Naive subtraction\label{sec:nsub}}
The first attempt at extending eq.~(\ref{IMCborn}) to NLO amounts
simply to substituting eq.~(\ref{IMCdef}) into eq.~(\ref{nlosubtint}). 
We get
\beq
\left(\xsecO\right)_{\sss naive}
=\int_0^1 dx \Bigg[\IMC(O,\xm(x))\frac{aR(x)}{x}
+\IMC(O,1)\left(B+aV-\frac{aB}{x}\right)\Bigg].
\label{IMCnlonaive}
\eeq
This equation suggests the following procedure:
\begin{itemize}
\item Pick at random $0\le x\le 1$.
\item Generate an MC event with $\xm(x)$ as maximum energy available to the
 photon in the first branching; attach to this event the weight $\wEV=aR(x)/x$.
\item Generate another MC event (a ``counter-event'') with $\xm=1$; 
 attach to this event the weight $\wCNT=B+aV-aB/x$.
\item Repeat the first three steps $N$ times, and normalize with $1/N$.
\end{itemize}
Unfortunately, this procedure is bound to fail, since the weights
$\wEV$ and $\wCNT$ diverge as $x\to 0$. A small cutoff $\rho$ can
then be introduced; as $x\to\rho$, we would have $\wEV\simeq aB/\rho$
and $\wCNT\simeq -aB/\rho$. Both these weights are very large in
absolute value; this means that the eventual unweighting procedure,
performed by the MC, will be highly inefficient. 
Furthermore, as we shall see in sect.~\ref{sec:obs},
eq.~(\ref{IMCnlonaive}) has problems of double counting, defined
according to the criterion in sect.~\ref{sec:double}.

\subsubsection{Modified subtraction\label{sec:msub}}
To avoid the problems outlined above, we define the {\em modified
subtraction method} by
\beqn
\left(\xsecO\right)_{\sss msub}
&=&\int_0^1 dx \Bigg[\IMC(O,\xm(x))\frac{a[R(x)-BQ(x)]}{x}
\nonumber \\*
&&+\IMC(O,1)\left(B+aV+\frac{aB[Q(x)-1]}{x}\right)\Bigg]\;.
\label{IMCfive}
\eeqn
This amounts to saying that, in eq.~(\ref{IMCnlonaive}),
we subtract and add the quantity
\beq
\IMC(O,\xm)\frac{aBQ(x)}{x},
\label{addsubt}
\eeq
using $\xm=\xm(x)$ in the first and $\xm=1$ in the second term introduced
in this way. We point out that $Q$ is the same function that appears in the
definition of the Sudakov form factor, eq.~(\ref{Deltadef}). By virtue of
the second condition in eq.~(\ref{fconditions}), the weights $\wEV$ and 
$\wCNT$ attached to events and counter-events are now separately convergent 
as $x\to 0$.

The two terms involving $Q(x)$, that we inserted in eq.~(\ref{IMCfive}),
are not identical; therefore, the procedure which leads from 
eq.~(\ref{IMCnlonaive}) to eq.~(\ref{IMCfive}) is not a subtraction 
in the usual sense of an NLO computation. Still, these two terms do not
contribute to the observable $O$ at $\orda$, because they are compensated 
by analogous terms due to the parton shower $B\IMC(O,1)$; this will be 
shown with explicit examples in sect.~\ref{sec:obs}, and in general in 
app.~\ref{app:expansion}. We point out that the function
$aQ(x)/x$ that appears in eq.~(\ref{addsubt}) does {\em not} depend
upon the MC cutoff $x_0$. It follows that the dependence upon $x_0$
of the $\MCatNLO$, eq.~(\ref{IMCfive}), is identical to that of the
ordinary MC, being entirely due to $\IMC$, and is therefore power-like.

It should be clear from the description of the procedure given above
that the dependence upon $O$ in eqs.~(\ref{IMCnlonaive}) and
(\ref{IMCfive}) is purely formal, serving only to allow 
us to write such equations in a compact form.  We stress that,
in the actual implementation of {\em all} the
$\MCatNLO$'s described in this paper, the computation of observables
proceeds exactly as in standard MC's: namely, events are generated,
showered, and hadronized without reference to any specific observable.

Equation~(\ref{IMCfive}) is our master equation for the toy model.
We shall show in what follows that the $\MCatNLO$ defined in this
way meets the requirements given in sect.~\ref{sec:obj}.
In sect.~\ref{sec:QCDMC}, we shall use eq.~(\ref{IMCfive}) to 
construct a QCD $\MCatNLO$ by analogy with the toy model.

\subsection{Observables\label{sec:obs}}
\subsubsection{Exclusive observable}
As an example of an ``exclusive'' observable in the toy model,
we shall consider the quantity
\beq
y=\max(x_1,\ldots,x_n),
\label{ydef}
\eeq
where $n$ denotes the number of emissions in one given event (in the 
case of no emission, the event is not used to fill the histograms).
This observable is the energy of the hardest photon of the event.
It is particularly convenient because in the MC approach it coincides
with the energy of the first photon emitted in the shower;
when $\MCatNLO$ is considered, $y$
is the larger of the energy of the photon emitted perturbatively,
and that of the first photon in the shower (if any). This fact allows
us to obtain analytical results for MC, NLO and $\MCatNLO$.

The NLO prediction for this observable is (here and in the following,
we consider only $y>x_0$ in order to simplify the analytical results)
\beq
\dyNLO = a\frac{R(y)}{y}\,,
\label{NLOdy}
\eeq
while the MC result, normalized to the Born cross section, is
\beq
\dyMC = aB\frac{Q(y)}{y}\Delta(y,1)\,,
\label{MCdy}
\eeq
which is just the differential cross section at $x=y$, computed
using eq.~(\ref{Deltadef}), times the probability that there is no 
harder emission at $x>y$.

In order to study the $\MCatNLO$ results, we have to consider
the quantities $\IMC(y,1)$ and $\IMC(y,\xm(x))$ that will be substituted
into eq.~(\ref{IMCnlonaive}) or (\ref{IMCfive}). In the case of
$\IMC(y,1)$, the hardest photon comes from the MC and we have
\beq
\IMC(y,1)= a\frac{Q(y)}{y}\Delta(y,1)\,.
\label{eqIMCy1}
\eeq
In the case of $\IMC(y,\xm(x))$, the hardest photon may come
from the NLO contribution or the MC, and so we find
\beqn
\IMC(y,\xm(x)) &=& \Delta\Big(\!\min\{x,\xm(x)\},\xm(x)\Big)\delta(y-x)
\nonumber\\*
&&+a\frac{Q(y)}{y}\Delta(y,\xm(x))\stepf(y-x)\stepf(\xm(x)-y)\;.
\label{eqIMCyxm}
\eeqn
As a shorthand notation, in the following we shall denote the
contributions to $d\sigma/dy$ from $\IMC(y,1)$ and $\IMC(y,\xm(x))$
by $\dszerond$ and $\dsonend$ respectively, since they correspond to
a $\clS$tandard MC evolution (i.e., with no prior NLO real emission), 
and to a $\clH$ard MC evolution (whose initial condition includes
one NLO real emission). This notation will be extensively used
from sect.~\ref{sec:generalization} on.
In the case of the naive subtraction method, eq.~(\ref{IMCnlonaive}),
these contributions are separately divergent, but their sum is not.
We find
\beq
\left(\dszero\right)_{\sss naive}
= a\frac{Q(y)}{y}\Delta(y,1)\left[B+aV
-aB\int_0^1\frac{dx}{x}\right]\;.
\label{mdszeronaive}
\eeq
The results for $\dsonend$ depend somewhat on the form of the function
$\xm(x)$ introduced in eq.~(\ref{IMCdef}). We define $x_e$ to be the
solution of $x=\xm(x)$, i.e., $x_e=\xm(x_e)$, and we assume this solution
to be unique. We further assume $\xm(x)$ to be a monotonically decreasing
function; thus, $x<\xm(x)$ for $x<x_e$. Then:

\noindent
$\diamond$ for $y<x_e:$ 
\beq
\left(\dsone\right)_{\sss naive}=a\frac{R(y)}{y}\Delta(y,\xm(y))
+a^2\frac{Q(y)}{y}\int_0^y \frac{dx}{x}R(x)\Delta(y,\xm(x))\;;
\label{mnaivecaseone}
\eeq

\noindent
$\diamond$ for $y>x_e:$ 
\beq
\left(\dsone\right)_{\sss naive}=a\frac{R(y)}{y}
+a^2\frac{Q(y)}{y}\int_0^{\xm^{-1}(y)}
\frac{dx}{x}R(x)\Delta(y,\xm(x))\;.
\label{mnaivecasetwo}
\eeq
Expanding in $a$, we find
\beq
\left(\xsecy\right)_{\sss naive}
\equiv \left(\dszero\right)_{\sss naive}+\left(\dsone\right)_{\sss naive}
=\frac a y [R(y)+BQ(y)]+\ordat\;.
\label{xsecynaiveexp}
\eeq
Therefore the naive subtraction method suffers from double counting,
according to the definition in sect.~\ref{sec:double}, especially
at small $y$, where $Q(y)\to 1$.

For the modified subtraction method, eq.~(\ref{IMCfive}),
we find instead that both contributions are finite:
\beq
\left(\dszero\right)_{\sss msub}
= a\frac{Q(y)}{y}\Delta(y,1)\left[B+aV
+aB\int_0^1\,dx\frac{Q(x)-1}{x}\right]\;,
\label{mdszeroappfive}
\eeq
while

\noindent
$\diamond$ for $y<x_e:$ 
\beqn
\left(\dsone\right)_{\sss msub}&=&\frac{a}{y}[R(y)-BQ(y)]\,\Delta(y,\xm(y))
\nonumber \\*
&+&a^2\frac{Q(y)}{y}\int_0^y \frac{dx}{x}[R(x)-BQ(x)]\Delta(y,\xm(x))\;;
\label{mfivecaseone}
\eeqn

\noindent
$\diamond$ for $y>x_e:$ 
\beq
\left(\dsone\right)_{\sss msub}=\frac a y [R(y)-BQ(y)]
+a^2\frac{Q(y)}{y}\int_0^{\xm^{-1}(y)}
\frac{dx}{x}[R(x)-BQ(x)]\Delta(y,\xm(x))\;.
\label{mfivecasetwo}
\eeq
Expanding in $a$, we now get
\beq
\left(\xsecy\right)_{\sss msub}=\frac{aR(y)}{y}+\ordat\;,
\label{xsecyfiveexp}
\eeq
in agreement with the NLO prediction for all values of $y$. 
Furthermore, for $y\to 0$ we have the expected
resummation of leading logarithms,
\beq
\left(\xsecy\right)_{\sss msub}\longrightarrow
\left(\frac{aB}{y}+\ordat\right)\Delta(y,1)
+{\rm PST}\;\;\;\;\;\;\;\;
{\rm for}~y\to 0\,,
\label{smallyxsec}
\eeq
where PST stands for power-suppressed terms, i.e., terms
that are not singular for $y\to 0$. We stress that the behaviour 
of \mbox{$d\sigma/dy$} in eq.~(\ref{smallyxsec}) is
dictated by eq.~(\ref{mdszeroappfive}), i.e., the contribution of
$\clH$ events is power suppressed for $y\to 0$, since $(R(y)-BQ(y))/y=
{\cal O}(1)$ in this limit. We also notice that the functional form in
eq.~(\ref{smallyxsec}) is the same as that in eq.~(\ref{MCdy}). Thus,
the $\MCatNLO$ resums large logarithms of $y$ in the same way as does
the MC. A general derivation of this result for a wide class of
observables is given in app.~\ref{app:logacc}.

\subsubsection{Inclusive observable}
As an ``inclusive'' observable, we shall consider the fully inclusive
distribution
of the photon energies; thus, each photon emitted, either at the NLO
level or by the MC shower, will contribute to the physical observable,
which we shall denote by $z$. The NLO and MC results (for $z>x_0$) are:
\beqn
\dzNLO&=&a\frac{R(z)}{z},
\label{NLOdz}
\\
\dzMC&=& aB\frac{Q(z)}{z}.
\label{MCdz}
\eeqn
Eq.~(\ref{NLOdz}) coincides with eq.~(\ref{NLOdy}), which is a trivial
consequence of the fact that only one emission occurs at NLO. On the
other hand, the MC result in eq.~(\ref{MCdz}) differs from the exclusive
case, eq.~(\ref{MCdy}), since all emissions now contribute, not only
the first.  As a consequence, eq.~(\ref{MCdz})
has a logarithmically-divergent integral at $z\to 0$, 
whereas this does not happen in eq.~(\ref{MCdy}).

In order to obtain the predictions of the $\MCatNLO$, we need to find
the analogues of eqs.~(\ref{eqIMCy1}) and (\ref{eqIMCyxm}), in
the case of the variable $z$. We obtain
\beqn
\IMC(z,1) &=& a\frac{Q(z)}{z},
\label{eqIMCzsoft}
\\
\IMC(z,\xm(x)) &=& \delta(z-x)+
a\frac{Q(z)}{z}\stepf(\xm(x)-z).
\label{eqIMCz}
\eeqn
The term with the $\delta$-function in eq.~(\ref{eqIMCz}) accounts for
real emission at the NLO level, which always contributes to the
inclusive spectrum when it occurs; the other term is the contribution
of the MC shower.

To compute the $z$ distribution for the naive subtraction method,
we use eq.~(\ref{eqIMCzsoft}) and eq.~(\ref{eqIMCz}) in 
eq.~(\ref{IMCnlonaive}), to obtain
\beq
\left(\xsecz\right)_{\sss naive} = \frac{a}{z}\Bigg[R(z)
+Q(z)\Bigg(\stot
-a\int_{\xm^{-1}(z)}^1 \frac{dx}x R(x)\Bigg)\Bigg]\,,
\label{dsigmadznaive}
\eeq
where $\stot$ is the NLO total cross section
\beq
\stot = B + aV +a\int_0^1\frac{dx}x [R(x)-B]\;.
\eeq
As before, we may expand in $a$ to find
\beq
\left(\xsecz\right)_{\sss naive}=\frac a z [R(z)+BQ(z)]+\ordat\;,
\label{xsecznaiveexp}
\eeq
which shows again that the naive subtraction method suffers from double
counting.

For the modified subtraction method we use eq.~(\ref{eqIMCzsoft})
and eq.~(\ref{eqIMCz}) in eq.~(\ref{IMCfive}) to find
\beqn
\left(\xsecz\right)_{\sss msub}&=&\frac{a}{z}\Bigg[R(z)-BQ(z)
\nonumber \\*&&\phantom{\frac{a}{z}}
+Q(z)\Bigg(\stot
-a\int_{\xm^{-1}(z)}^1 dx\frac{R(x)-BQ(x)}{x}\Bigg)\Bigg].
\label{dsigmadzfive}
\eeqn
Expanding in $a$ we now find
\beq
\left(\xsecz\right)_{\sss msub}=\frac{aR(z)}{z}+\ordat ;
\label{xseczfiveexp}
\eeq
thus, as in the case of the observable $y$, the $\MCatNLO$ prediction
at $\orda$ coincides with the NLO result for all $z$.
The small-$z$ limit of eq.~(\ref{dsigmadzfive}) is
\beq
\lim_{z\to 0}z\left(\xsecz\right)_{\sss msub}=a\stot\;.
\label{smallzxsec}
\eeq
From eq.~(\ref{MCdz}) we obtain instead
\beq
\lim_{z\to 0}z\dzMC=aB\;.
\label{smallzborn}
\eeq
Therefore, at $z\to 0$ the $\MCatNLO$ result is equal to that of the
standard MC, times $\stot/B$, which is the K-factor for the total rate.

\subsection{Results}
\begin{figure}[t]
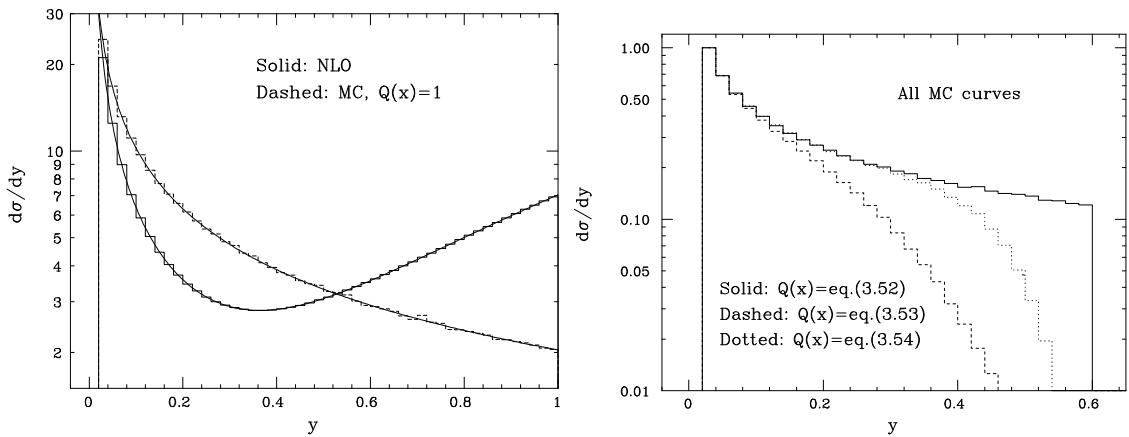

  \begin{center}
    \epsfig{figure=nlovsmc.eps,width=0.49\textwidth}
    \epsfig{figure=mcwithdead.eps,width=0.49\textwidth}
\caption{\label{fig:NLOvsMC} 
Left panel: Toy model NLO (solid histogram) and MC results (dashed histogram),
overlaid on analytical results of eqs.~(\ref{NLOdy}) and~(\ref{MCdy}),
with $Q(x)\equiv 1$.
Right panel: MC results, in the case of the dead zone with $x_{dead}=0.6$.
$Q$ is chosen as in eq.~(\ref{Qone}) (solid), eq.~(\ref{Qtwo}) (dashed), 
and eq.~(\ref{Qthree}) (dotted); the first bin is normalized to unity.
}
  \end{center}
\end{figure}

In this section we present toy model results for NLO, MC, 
and $\MCatNLO$. The following parameters have been used in
eqs.~(\ref{born}), (\ref{virt}) and (\ref{real}):
\beq
a=0.3\;,\;\;\;\;B=2\;,\;\;\;\;V=1\;.
\eeq
For the real emission distribution in eq.~(\ref{real}) we have taken
\beq
R(x) = B+x (1+x/2+20 x^2)\;,
\label{Rdef}
\eeq
so that the total cross section is
\beq
\stot=B+a\left(V+\frac{95}{12}\right)\;.
\label{stotres}
\eeq
Note that this implies a large K-factor, $\stot/B=2.34$.  Finally,
for the MC cutoff parameter we used
\beq
x_0=0.02\;,
\eeq
and for the maximum energy function in eq.~(\ref{IMCdef}) we chose
the form $\xm(x)=1-x$.
In the left panel of fig.~\ref{fig:NLOvsMC} we show the NLO result obtained 
with eq.~(\ref{nlosubtint}) for the observable $y$ defined in eq.~(\ref{ydef})
(solid histogram), compared to the result obtained by running
the MC (dashed histogram), with $Q(x)\equiv 1$; 
the integral of the latter result has been
normalized to the NLO rate (i.e., each event is given a weight
equal to $\stot/N$). As a cross check for the numerical
implementation against analytical results, we also plot as solid
lines the functions 
given in eqs.~(\ref{NLOdy}) and~(\ref{MCdy}) (for the latter, $B$ has 
been substituted with $\stot$, in order to be consistent with the 
normalization of the dashed histogram), which in fact appear 
to be in perfect agreement with the corresponding histograms.
The shapes of the NLO and MC results are completely different. 
In fact, the function $R(x)$ has been chosen to introduce a clear
difference between the NLO and MC results, for a better understanding
of the effect of the matching between NLO and MC.

\begin{figure}[t]
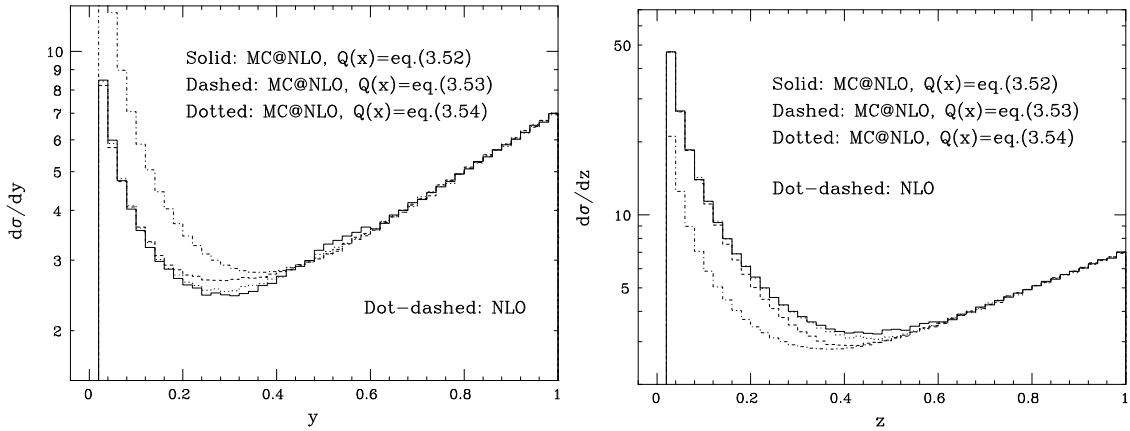

  \begin{center}
    \epsfig{figure=appr_5.eps,width=0.49\textwidth}
    \epsfig{figure=mcatnloz.eps,width=0.49\textwidth}
\caption{\label{fig:apprfive} 
Toy model $\MCatNLO$ results; a dead zone is present ($x_{dead}=0.6$)
and $Q$ is as in eq.~(\ref{Qone}) (solid), eq.~(\ref{Qtwo}) (dashed), 
or eq.~(\ref{Qthree}) (dotted). The results are for the exclusive
observable $y$ (left panel), and for the fully inclusive 
observable $z$ (right panel). The normalization is absolute, and the 
dot-dashed histogram is the pure NLO result.
}
  \end{center}
\end{figure}

To study the matching of NLO and MC, we consider
an MC with a {\em dead zone}. In our toy model, this corresponds to an MC
that cannot emit photons with energy $x>x_{dead}$ (in the numerical
simulations we set $x_{dead}=0.6$). We can easily get a toy MC
with a dead zone by suitably choosing the function $Q$ in
eq.~(\ref{Deltadef}). We first introduce a smoothing function of the form
\beq
G(x)=\frac{c^2(1-x)^{2\beta}}{x^{2\alpha}+c^2(1-x)^{2\beta}},
\label{gfun}
\eeq
where $\alpha$, $\beta$ and $c$ are free parameters,
and then consider three different implementations of the dead zone:
\beqn
&&1.\phantom{aa}
Q(x)=\stepf(x_{dead}-x);
\label{Qone}
\\
&&2.\phantom{aa}
Q(x)=\stepf(x_{dead}-x)G(x/x_{dead}),\;\;\;\;{\rm with}\;\;\;\;
\alpha=1, \beta=1, c=1;
\label{Qtwo}
\\
&&3.\phantom{aa}
Q(x)=\stepf(x_{dead}-x)G(x/x_{dead}),\;\;\;\;{\rm with}\;\;\;\;
\alpha=2, \beta=1, c=8.
\label{Qthree}
\eeqn
The results for the pure MC evolution are presented in the
right panel of fig.~\ref{fig:NLOvsMC}; the dependence upon 
$Q$ is actually fairly big, since we made extreme choices in
eqs.~(\ref{Qone})--(\ref{Qthree}). It is thus striking that the
corresponding results obtained with the $\MCatNLO$, shown in the left
panel of fig.~\ref{fig:apprfive}, do not display big differences
from each other.
All three $\MCatNLO$ curves are quite close to the pure NLO result
(and coincide with it for $y>x_{dead}$), and differ sizably from it 
only in the region of small $x$, where the effect of multiple emission
is expected to be important.\footnote{Since $y$ is an exclusive variable,
and the total rate is that of the NLO result, the differences between
the $\MCatNLO$ and NLO curves at $y>x_0$ are in fact compensated by
the region $y<x_0$ in the total rate.}
In this region, however, the three $\MCatNLO$ curves
are close to each other, notwithstanding the differences shown
in the right panel of fig.~\ref{fig:NLOvsMC}. We are here in a region
where the leading logarithmic behaviour of the MC is highly dominant,
and thus the impact of $Q$ (which is subleading in logarithmic accuracy)
is moderate.  We see that the curve corresponding
to eq.~(\ref{Qone}) displays 
a less regular behaviour than the others at $y\simeq x_{dead}$; 
however, the effect is very moderate, being at worst of $\ordat$. 

The results for the inclusive $z$ distribution
are presented in the right panel of fig.~\ref{fig:apprfive}. 
Again, we can see that the three $\MCatNLO$ curves are remarkably close
to each other, in spite of the differences in the implementations
of the dead zone. The ratio of the cross section in the leftmost bin
of the solid histogram over the NLO result for the same quantity is 2.23.
This value is fairly close to the K-factor $\stot/B=2.34$, that we
should find according to eqs.~(\ref{smallzxsec}) and~(\ref{smallzborn});
the small difference is due to fact that this bin is not quite at the 
limit $z\to 0$.

We can thus conclude that $\MCatNLO$ reduces the numerical dependence
upon subleading logarithms, which is seen when running the MC alone;
this is because
those regions where the subleading terms are becoming numerically
important get sizable contributions from the hard NLO emissions in
the $\MCatNLO$, which are not present in the MC, and which eventually
become dominant. On the other hand, in the soft-dominated regions, the 
shapes of the $\MCatNLO$ curves just reproduce those of the MC curves, 
as we expect.

\section{\boldmath The QCD case: W$^+$W$^-$ production\label{sec:QCDMC}}
We now face the problem of constructing an $\MCatNLO$ in QCD. In order
to simplify as much as possible all the technical details related to
the colour structure of the hard process, we consider here the
production of W$^+$W$^-$ pairs in hadronic collisions. This process also
has a fairly simple structure of soft and collinear singularities.
However, we argue that the method we propose is actually valid for any
kind of hard production process.

We shall use \HW\ \cite{Marchesini:1992ch} as our reference
parton shower MC. Other choices are of course possible, and our
approach is not tailored to work with \HW, although some of the
technical details, reported in the appendices, do depend on the
specific implementation of the hard process and parton showering
in that program.

In order to define our $\MCatNLO$ for W pair production, we rely on
the results previously presented for the toy model. First
we emphasize the main points of the toy $\MCatNLO$. Then we make a 
parallel between the contributions to the toy model cross section,
and the contributions to the W$^+$W$^-$ cross section. Using these
results, we proceed to construct the QCD $\MCatNLO$, by analogy 
with the toy $\MCatNLO$.

\subsection{Toy model revisited\label{sec:generalization}}
Let us consider again our master equation for the toy model $\MCatNLO$,
eq.~(\ref{IMCfive}). The $\MCatNLO$ approach consists in running the
ordinary MC, with initial conditions and weights determined by the NLO
calculation. Depending upon the initial conditions, we distinguish
two classes of events, which correspond to $\IMC(O,\xm(x))$ and
$\IMC(O,1)$. We denote the class corresponding to $\IMC(O,\xm(x))$ by
$\clH$, which stands for ``hard'', and reminds us that one emission
already occurred, due to NLO, prior to the MC run. The class corresponding
to $\IMC(O,1)$ is denoted by $\clS$, which stands for ``standard''; this
reminds us that this initial condition is the same as for the
ordinary MC runs. 

It is important to realize that for $\clS$ events the finite part of the
virtual contribution, $aV$, the real counterterm, $-aB/x$, and the term
$aBQ(x)/x$, serve exclusively to fix the normalization of this class
of events, relative to that of the class $\clH$. In other words, it is
irrelevant that the latter two terms explicitly depend upon $x$:
in the $\MCatNLO$, the spectrum of the observable $O$ resulting from
$\clS$ events is {\em entirely} due to MC evolution.  In contrast,
the NLO hard emissions do contribute to the kinematics of $\clH$
events, and thus help to determine $O$.

In order to obtain in QCD a structure analogous to that outlined
above, we therefore have to collect all the terms of the NLO cross
section that do {\em not} contribute to hard emissions, and use them
to compute the normalization of class $\clS$ events; these events
result from MC evolutions whose initial conditions are identical to
those of an ordinary MC. On the other hand, the kinematics of NLO hard
emissions determine the initial conditions for MC evolutions for
$\clH$ events.

Finally, we have already seen in the case of the observables $y$ and $z$
(see also app.~\ref{app:expansion} for a more formal proof) that at
$\orda$ the dependence upon $Q(x)$ disappears. Roughly speaking, this 
happens because the contribution due to the Born term $B$, in 
class $\clS$ events, cancels at $\orda$ with the contribution 
of the terms $\pm aBQ(x)/x$. We observe that, in the toy model,
$aBQ(x)/x$ is the $\orda$ term in the expansion of the MC result
(see eqs.~(\ref{MCdy}) and~(\ref{MCdz})). We therefore guess that
the role of $aBQ(x)/x$ in a QCD $\MCatNLO$ is played by the 
${\cal O}(\as)$ term in the expansion of the result of the
ordinary MC, when only the shower is taken into account (i.e., 
the hadronization step is skipped). We shall eventually prove
(see app.~\ref{app:expansion}) that this is in fact the case.

\subsection{NLO cross section\label{sec:NLOxsec}}
We now deal explicitly with W$^+$W$^-$ production in hadronic
collisions, and we start by writing the NLO cross section. In this
paper, we use the results of ref.~\cite{Frixione:1993yp}, which we
rewrite in a form more suited to our present goal; it should be clear
that any W$^+$W$^-$ cross section (or at least, one based upon the
subtraction method) can be cast in the same form. Owing to the
factorization theorem, we have
\beq
d\sigma(P_1,P_2)=\sum_{ab}\int dx_1 dx_2 f^{(H_1)}_a(x_1)
f^{(H_2)}_b(x_2) d\hat{\sigma}_{ab}(x_1 P_1,x_2 P_2),
\label{factth}
\eeq
for the collision of hadrons $H_1$ and $H_2$ with momenta $P_1$ and
$P_2$ respectively. The sum runs over those parton flavours that give
partonic subprocesses contributing to W pair production at NLO.
We symbolically denote these processes as follows:
\beqn
q\bar{q}\;&\longrightarrow&\;W^+W^-\,,
\label{qqproc}
\\
q\bar{q}\;&\longrightarrow&\;W^+W^-g\,,
\label{qqgproc}
\\
qg\;&\longrightarrow&\;W^+W^-q\,,
\label{qgqproc}
\eeqn
where $q=u,~d,\ldots$ and eq.~(\ref{qgqproc}) includes the corresponding
antiquark processes.  Equation~(\ref{qqproc}) contributes to
Born and virtual emission terms, whereas eqs.~(\ref{qqgproc})
and~(\ref{qgqproc}) contribute to real emission terms.

The information on the hard process in eq.~(\ref{factth}) is contained
in the subtracted partonic cross sections, $d\hat{\sigma}_{ab}$ (the hat
means that counterterms are implicit in the expression).
We write these cross sections as follows (eq.~(3.45) 
of ref.~\cite{Frixione:1993yp}):
\beq
d\hat{\sigma}_{ab}=d\sigma_{ab}^{(b)}+d\sigma_{ab}^{(sv)}
+d\hat{\sigma}_{ab}^{(c+)}+d\hat{\sigma}_{ab}^{(c-)}
+d\hat{\sigma}_{ab}^{(f)},
\label{sigmapart}
\eeq
where:
\begin{itemize}
\item[{\bf a)}] $d\sigma_{ab}^{(b)}$ is the Born contribution. It has
  $2\to 2$ kinematics, and corresponds to eq.~(\ref{born}) of the
  toy model.
\item[{\bf b)}] $d\sigma_{ab}^{(sv)}$ is the contribution of the
  virtual diagrams, and of all those terms, resulting from various
  subtractions, that have $2\to 2$ kinematics and are of ${\cal O}(\as)$. 
  This term corresponds to the finite part of eq.~(\ref{virt}) of the 
  toy model.
\item[{\bf c)}] $d\hat{\sigma}_{ab}^{(c\pm)}$ are the contributions of
  the finite remainders resulting from the subtraction of the
  initial-state collinear divergences ($+$ and $-$ indicate radiation 
  from partons $a$ and $b$ respectively). These terms do not have an
  analogue in the toy model, since in the toy model there are no
  parton densities. They have $2\to 3$ kinematics, with the light
  outgoing parton always collinear to the beam-line; we refer to
  these configurations as quasi-$2\to 2$ kinematics. We stress that
  \mbox{$d\hat{\sigma}_{ab}^{(c\pm)}$} are subtracted quantities, and
  that the counterterms have $2\to 2$ kinematics.
\item[{\bf d)}] $d\hat{\sigma}_{ab}^{(f)}$ is the contribution of the
  real emission diagrams, with its divergences suitably subtracted; it
  has $2\to 3$ kinematics (the counterterms have $2\to 2$ or
  quasi-$2\to 2$ kinematics). The unsubtracted part of this term 
  corresponds to eq.~(\ref{real}) of the toy model.
\end{itemize}
We point out that all the terms on the r.h.s.\ of eq.~(\ref{sigmapart})
are separately finite. Also, they do not contain any large logarithms,
and so no cancellation of large numbers occurs among them.  This
property is a feature of the subtraction method, and does not hold in
the slicing method.

Items {\bf a)}-{\bf d)} above achieve the identification between
terms contributing to the QCD cross section and to the
toy model cross section. We now go into more detail, by writing
the quantities appearing in eq.~(\ref{sigmapart}) more explicitly .

We denote by $d\phi_2(s)$ and $d\phi_3(s)$ the phase spaces of the W$^+$W$^-$
pair and of the W$^+$W$^-$ pair plus a light parton, respectively; they
depend upon the kinematical variables of these particles, and upon
$s=(p_1+p_2)^2$, the c.m.\ energy squared of the partons that
initiate the hard scattering, whose momenta we denote by 
\mbox{$p_1=x_1 P_1$} and \mbox{$p_2=x_2 P_2$}. The Born term is
\beq
d\sigma_{ab}^{(b)}(p_1,p_2)=\mborn_{ab}(p_1,p_2) d\phi_2(s),
\eeq
where $\mborn_{ab}$ is the Born matrix element squared, summed over 
spin and colour degrees of freedom (the average factor for initial-state 
ones is understood), times the flux factor $1/(2s)$. The soft-virtual 
contribution is
\beq
d\sigma_{ab}^{(sv)}(p_1,p_2)=
\left[{\cal Q}_{ab}\,\mborn_{ab}(p_1,p_2) 
+\mvreg_{ab}(p_1,p_2) \right] d\phi_2(s),
\label{sigsv}
\eeq
where ${\cal Q}_{ab}$ is a constant with respect to the kinematics of
the partonic process, and $\mvreg_{ab}$ is the finite part of the virtual
contribution.  As with any other finite part resulting from the
cancellation of singularities, its definition is to a certain extent
arbitrary; however, any redefinition is compensated by suitable
changes in the other terms contributing to the cross section.
The initial-state collinear terms are:
\beq
d\hat{\sigma}_{ab}^{(c\pm)}=d\sigma_{ab}^{(c\pm)}\evB
-d\sigma_{ab}^{(c\pm)}\cntB\,,
\label{sigcpm}
\eeq
where
\beqn
d\sigma_{ab}^{(c+)}(p_1,p_2)\evB&=&\frac{\as}{2\pi}\left[
\log\frac{s(1-x)^2}{\mu^2}P_{ca}^{(0)}(x)-P_{ca}^{(1)}(x)
\right]\mborn_{cb}(xp_1,p_2) d\phi_2(xs)dx,
\nonumber \\*&&
\label{cpev}
\\
d\sigma_{ab}^{(c-)}(p_1,p_2)\evB&=&\frac{\as}{2\pi}\left[
\log\frac{s(1-x)^2}{\mu^2}P_{cb}^{(0)}(x)-P_{cb}^{(1)}(x)
\right]\mborn_{ac}(p_1,xp_2) d\phi_2(xs)dx.
\nonumber \\*&&
\label{cmev}
\eeqn
In eq.~(\ref{sigcpm}), the term $d\sigma_{ab}^{(c\pm)}\cnt$
(the ``ct'' stands for counterterm) is constructed in order to cancel
the divergences present in the collinear ``event'' 
contribution $d\sigma_{ab}^{(c\pm)}\ev$; these divergences are
soft ($x\to 1$), as can be seen explicitly in eqs.~(\ref{cpev})
and~(\ref{cmev}). The counterterm has a form that is fully specified in
a given subtraction scheme (see appendix~\ref{app:WWxsecs}). In
eqs.~(\ref{cpev}) and~(\ref{cmev}),
$\mu$ is the factorization mass scale (which also enters in the
parton densities), and $P_{ab}^{(i)}$ is the $i^{th}$ term in the
$\epsilon$-expansion
of the Altarelli-Parisi kernel in $4-2\epsilon$ dimensions,
taken for $x<1$ (i.e., no $\delta(1-x)$ terms are included).
Finally, we write the real emission term as
\beqn
d\hat{\sigma}_{ab}^{(f)}=d\sigma_{ab}^{(f)}\evB
-d\sigma_{ab}^{(f)}\cntB\,,
\label{sigreal}
\\
d\sigma_{ab}^{(f)}(p_1,p_2)\evB=\mreal_{ab}(p_1,p_2)d\phi_3(s)\,,
\label{rev}
\eeqn
where $\mreal_{ab}$ is the matrix element squared resulting from the real
emission diagrams (summed over all spins and colours, and averaged over
those relevant to incoming partons), 
times the flux factor. This quantity has soft and
initial-state collinear singularities, and thus
$d\sigma_{ab}^{(f)}\ev$ has divergences, which are cancelled by
the counterterms $d\sigma_{ab}^{(f)}\cnt$ in eq.~(\ref{sigreal}).
As in the case of $d\sigma_{ab}^{(c\pm)}\cnt$, these counterterms 
are fully specified only in a given subtraction scheme. However,
one should keep in mind that the physical cross section is always
strictly independent of the choice of the subtraction scheme.

\subsection{Observables at NLO}
As in the case of the toy model, a preliminary step in the construction
of the $\MCatNLO$ is the prediction of an observable at the NLO. We 
start by defining an unintegrated, fully-differential 
cross section
\beq
d\hat{\Sigma}_{ab}(x_1,x_2)=f^{(H_1)}_a(x_1)
f^{(H_2)}_b(x_2)d\hat{\sigma}_{ab}(x_1 P_1,x_2 P_2),
\label{hatSigmadef}
\eeq
and analogous quantities, for all the contributions to the partonic
cross section that appear on the r.h.s.\ of eq.~(\ref{sigmapart}).
Any observable $O$ can depend, at the NLO, upon the kinematical variables
of the W$^+$, of the W$^-$, and of the light outgoing parton. We denote 
by $O(\Kthree)$ the observable computed in non-soft and non-collinear
kinematics; this corresponds to $O(x)$ in the toy model. We also
denote by $O(\Ktwo)$ and $O(\Kqtwo)$ the observable computed in
$2\to 2$ (i.e., soft) and in quasi-$2\to 2$ (i.e., collinear non-soft)
kinematics respectively; these correspond to $O(0)$ in the toy model.
In all cases, we understand the variables of the outgoing particles
to be given in the laboratory frame. We then have
\beqn
\VEV{O}&=&\sum_{ab}\int dx_1 dx_2 d\phi_3\Bigg[
O(\Kthree)\frac{d\Sigma_{ab}^{(f)}}{d\phi_3}\evBB
+O(\Ktwo)\frac{1}{\Itwo}\left(\frac{d\Sigma_{ab}^{(b)}}{d\phi_2}
+\frac{d\Sigma_{ab}^{(sv)}}{d\phi_2}\right)
\nonumber \\*&& \phantom{d\phi_3}
+O(\Kqtwo)\frac{1}{\Iqtwo}\left(
\frac{d\Sigma_{ab}^{(c+)}}{d\phi_2 dx}\evBB
+\frac{d\Sigma_{ab}^{(c-)}}{d\phi_2 dx}\evBB\right)
-O(\Ktwo)\frac{1}{\Iqtwo}
\left(\frac{d\Sigma_{ab}^{(c+)}}{d\phi_2 dx}\cntBB
+\frac{d\Sigma_{ab}^{(c-)}}{d\phi_2 dx}\cntBB\right)
\nonumber \\*&& \phantom{d\phi_3}
-\{O(\Ktwo),O(\Kqtwo)\}\frac{d\Sigma_{ab}^{(f)}}{d\phi_3}\cntBB
\Bigg],
\label{WWnlosubtint}
\eeqn
where in the last term $\{O(\Ktwo),O(\Kqtwo)\}$ indicates that the
counterterms of the real contribution can have both $2\to 2$ and
quasi-$2\to 2$ kinematics. The factors $\Itwo$ 
and $\Iqtwo$ guarantee that the $2\to 2$ and quasi-$2\to 2$ terms
are correctly normalized upon integration over the full three-body
phase space; the following equations must therefore hold
\beqn
\int d\phi_3(s) &=&\Itwo\int d\phi_2(s)\,,
\\
\int d\phi_3(s) &=&\Iqtwo\int d\phi_2(xs)dx\,,
\eeqn
the integration being performed over the whole phase space.
Eq.~(\ref{WWnlosubtint}) should be compared with eq.~(\ref{nlosubtint}).
The factors $1/\Itwo$
and $1/\Iqtwo$ are not present in the toy model, simply because the 
integral over the phase space $dx$ in that case is equal to 1; if this were 
not the case, a factor analogous to $1/\Itwo$ would have to be inserted in
eq.~(\ref{nlosubtint}). The term 
\mbox{$d\Sigma_{ab}^{(f)}/d\phi_3\ev=$}$f_a f_b\mreal_{ab}$
corresponds to $aR(x)/x$ in eq.~(\ref{nlosubtint}). The term
\mbox{$d\Sigma_{ab}^{(b)}/d\phi_2=$}$f_a f_b\mborn_{ab}$ corresponds
to $B$, the Born contribution, while the term
\mbox{$d\Sigma_{ab}^{(sv)}/d\phi_2=$}
\mbox{$f_a f_b ({\cal Q}_{ab}\mborn_{ab}+\mvreg_{ab})$}
corresponds to $V$, the finite part of the virtual contribution.
As we said already, the contributions $d\Sigma_{ab}^{(c\pm)}$ have
no analogues in the toy model, simply because there are no parton
densities there. Finally, \mbox{$d\Sigma_{ab}^{(f)}/d\phi_3\cnt$}
represents the counterterms necessary to cancel the divergences of
the real matrix element; it is thus equivalent to the last term
in eq.~(\ref{nlosubtint}), $aB/x$.

\subsection[Matching NLO and MC: the QCD $\MCatNLO$]{\boldmath Matching
NLO and MC: the QCD $\MCatNLO$}\label{sec:match}
Eq.~(\ref{WWnlosubtint}) is very similar to eq.~(\ref{nlosubtint}).
We could try to follow the same steps that brought us from
eq.~(\ref{nlosubtint}) to eq.~(\ref{IMCfive}):
replace the observables in eq.~(\ref{WWnlosubtint}) with suitable
interface-to-MC's and then modify the equation with a quantity analogous
to that introduced in eq.~(\ref{addsubt}). There is, however, a
technical problem, related to the initial conditions
of class $\clS$ events. In the toy model, these initial conditions are
unambiguously determined by letting \mbox{$x\to 0$} in the corresponding
system-plus-one-photon configurations. This procedure cannot be automatically
generalized to QCD, where matrix elements have soft {\em and collinear}
singularities (corresponding to $\Ktwo$ and $\Kqtwo$ configurations
respectively). This may suggest the necessity of considering two $\clS$
classes, defined as the sets of those events resulting from MC evolutions
that start from $\Ktwo$ and $\Kqtwo$ initial conditions respectively. 
However, the MC cannot handle $\Kqtwo$ initial conditions, since final-state 
light partons are strictly collinear to the beam-line in these cases.

The solution can be found by observing that, at least as far as the
variables of the W$^+$ and W$^-$ are concerned, $\Ktwo$ and $\Kqtwo$
only differ by a longitudinal boost. The variables of the outgoing light
parton are of no concern here, since they cannot contribute to the
kinematics of any observables in the configurations $\Ktwo$ and $\Kqtwo$,
the parton being either soft of collinear to the beam line. The integration 
over the momentum fractions $x_1$ and $x_2$ in eq.~(\ref{WWnlosubtint}) 
implicitly contains such a boost. Roughly speaking, for fixed
$x_1$ and $x_2$ we have $\Ktwo\ne\Kqtwo$, but given $x_1$ and $x_2$, 
there exist $\tilde{x}_1$ and $\tilde{x}_2$ such that
$\Ktwo(x_i)=\Kqtwo(\tilde{x}_i)$. In practice, we adopt the following
procedure.\footnote{To the best of our knowledge, this method was
invented in order to improve the numerical stability of the results
of ref.~\cite{Mangano:jk}.} We treat each term under the integral
sign in eq.~(\ref{WWnlosubtint}) independently (we assume some
regularization procedure has been adopted in the intermediate steps of the
computation), performing a change of integration variables that is
in general different for each term:
\beq
(x_1,x_2)\;\longrightarrow\;(z_1,z_2)\,,\;\;\;\;
x_1=x_1^{(\alpha)}(z_1,z_2,\phi_3)\,,\;\;\;\;
x_2=x_2^{(\alpha)}(z_1,z_2,\phi_3)\,,
\label{xxtozz}
\eeq
where $\phi_3$ denotes here the set of independent three-body phase-space 
variables. The functional forms of these changes of variables will be given 
in app.~\ref{app:proj}. Here we are only interested in the fact that
eq.~(\ref{xxtozz}) must be such that $\Ktwo(x_i)=\Kqtwo(\tilde{x}_i)$.
We define, for each of the functions $\Sigma_{ab}$ that contribute 
to eq.~(\ref{WWnlosubtint}),
\beq
d\bSigma_{ab}^{(\alpha)}(z_1,z_2,\phi_3)=
\frac{\partial(x_1^{(\alpha)},x_2^{(\alpha)})}{\partial(z_1,z_2)}
d\Sigma_{ab}^{(\alpha)}\Big(x_1^{(\alpha)}(z_1,z_2,\phi_3),
x_2^{(\alpha)}(z_1,z_2,\phi_3),\phi_3\Big)\,,
\label{StobS}
\eeq
where $\partial(x_1^{(\alpha)},x_2^{(\alpha)})/\partial(z_1,z_2)$ 
is the Jacobian corresponding
to eq.~(\ref{xxtozz}). Then, as we shall show in app.~\ref{app:proj},
eq.~(\ref{WWnlosubtint}) can be recast in the following form:
\beqn
\VEV{O}&=&\sum_{ab}\int dz_1 dz_2 d\phi_3\Bigg\{
O(\Kthree)\frac{d\bSigma_{ab}^{(f)}}{d\phi_3}\evBB
+O(\Ktwo)\Bigg[
\frac{1}{\Itwo}\Bigg(\frac{d\bSigma_{ab}^{(b)}}{d\phi_2}
+\frac{d\bSigma_{ab}^{(sv)}}{d\phi_2}\Bigg)
\nonumber \\*&& \phantom{d\phi_3}
+\frac{1}{\Iqtwo}\Bigg(
\frac{d\bSigma_{ab}^{(c+)}}{d\phi_2 dx}\evBB
+\frac{d\bSigma_{ab}^{(c-)}}{d\phi_2 dx}\evBB\Bigg)
-\frac{1}{\Iqtwo}
\Bigg(\frac{d\bSigma_{ab}^{(c+)}}{d\phi_2 dx}\cntBB
+\frac{d\bSigma_{ab}^{(c-)}}{d\phi_2 dx}\cntBB\Bigg)
\nonumber \\*&& \phantom{d\phi_3}
-\frac{d\bSigma_{ab}^{(f)}}{d\phi_3}\cntBB
\Bigg]\Bigg\}.
\label{WWevproj}
\eeqn
We stress that the only difference between eq.~(\ref{WWnlosubtint})
and eq.~(\ref{WWevproj}) is that the latter does not contain $O(\Kqtwo)$
any longer, thanks to the definition of the change of variables, which
ensures that $\Ktwo(x_i)=\Kqtwo(\tilde{x}_i)$. We also stress that no 
approximation has been made in eq.~(\ref{WWevproj}); it is just a different 
way of writing the NLO cross section.

It should be clear that the necessity for the change of variables in
eq.~(\ref{xxtozz}) is due to the simultaneous presence in 
eq.~(\ref{WWnlosubtint}) of $O(\Ktwo)$ and $O(\Kqtwo)$, which
is in turn due to the form of the hadronic cross section in
eq.~(\ref{factth}), where the luminosity is factorized. One
could directly start with an NLO cross section that does not contain
$O(\Kqtwo)$; clearly, our formulae hold in this case as well,
since one just needs to set \mbox{$(z_1,z_2)\equiv (x_1,x_2)$}.

It should also be clear that our goal is to obtain an expression
for the NLO cross section that only contains $\Kthree$ and $\Ktwo$
configurations. Although we find that we can easily achieve this
through eq.~(\ref{xxtozz}), we stress that other methods could be
devised, and that our final formula for the $\MCatNLO$ will hold,
regardless of the procedure adopted to eliminate $\Kqtwo$ configurations.

The quantities that contribute to $O(\Kthree)$ and $O(\Ktwo)$ in
eq.~(\ref{WWevproj}) correspond to those that contribute to
$O(x)$ and $O(0)$ in the toy model, respectively, according to the analogies
discussed in sect.~\ref{sec:NLOxsec}. The only exceptions are 
the terms $d\hat{\sigma}_{ab}^{(c\pm)}$, which do not have
analogues in the toy model. However, as shown in app.~\ref{app:proj},
these terms must be treated in the same way as the real matrix element
collinear counterterms. 
This is in fact to be expected: $d\hat{\sigma}_{ab}^{(c\pm)}$
are strictly related to the collinear counterterms, being finite remainders
resulting from the subtraction of the collinear singularities of the
real matrix elements.

Equation~(\ref{WWevproj}) is therefore strictly analogous to
eq.~(\ref{nlosubtint}).  
This allows us to extend to the QCD case the concept of the
event classes $\clH$ and $\clS$. In turn, this implies that we need
two interface-to-MC's, as in the toy model, which we denote by
$\IMC(O,\Kthree)$ and $\IMC(O,\Ktwo)$. These are the analogues of
$\IMC(O,\xm(x))$ and $\IMC(O,1)$ in the toy model. We can thus
turn eq.~(\ref{WWevproj}) into a prescription for a $\MCatNLO$,
by replacing $O(\Kthree)$ by $\IMC(O,\Kthree)$, and $O(\Ktwo)$
by $\IMC(O,\Ktwo)$ . In this way, we get the analogue of
eq.~(\ref{IMCnlonaive}), that is, a ``naive'' $\MCatNLO$ that
suffers from double counting and difficulties in the generation
of unweighted events. However, precisely as in the case of the
toy model, we can modify the naive prescription in order to get
a QCD $\MCatNLO$ that implements a modified subtraction, and
fulfills the requirements given in sect.~\ref{sec:obj}.

To accomplish this task, we need to find the analogue of the quantity
$aBQ(x)/x$ of the toy model. As discussed in
sect.~\ref{sec:generalization}, what we
are looking for is the ${\cal O}(\as)$ term in the expansion of
the result obtained by running the ordinary MC; we denote this 
quantity by
\beq
\frac{d\bSigma_{ab}}{d\phi_3}\xMCBB\,.
\label{MCexp}
\eeq
The specific form of the quantity in eq.~(\ref{MCexp}) is dependent 
upon the implementation; different MC's will induce different forms 
of $d\bSigma_{ab}\xMC$. On the other hand, $d\bSigma_{ab}\xMC$ is 
independent of the procedure through which we eliminated $\Kqtwo$
configurations. The form we use in the present paper, and 
all the technical details concerning its determination, are reported 
in app.~\ref{app:DZ}. Here, we merely need to note
that $d\bSigma_{ab}\xMC$ must have the same leading singular behaviour as
the matrix element contribution $d\bSigma_{ab}^{(f)}$ when approaching
soft or collinear configurations, in order for the former quantity to
be used as a local counterterm to the latter. As we shall discuss
in app.~\ref{app:DZ}, this fact is not trivial, and we shall need
to modify the ${\cal O}(\as)$ result obtained from the MC. For the
time being, we treat $d\bSigma_{ab}\xMC$ in the same way as 
$aBQ(x)/x$ is treated in the toy model (see eq.~(\ref{addsubt})):
we subtract and add the quantity
\beq
\IMC(O,{\bf n})\,\frac{d\bSigma_{ab}}{d\phi_3}\xMCBB
\label{QCDaddsub}
\eeq
under the integral sign, setting ${\bf n}=\Kthree$ in the negative
term and ${\bf n}=\Ktwo$ in the positive one. This gives a modified
subtraction formula analogous to that in eq.~(\ref{IMCfive}):
\beqn
\xsecO&=&\sum_{ab}\int dz_1 dz_2 d\phi_3\Bigg\{
\IMC(O,\Kthree)\Bigg(\frac{d\bSigma_{ab}^{(f)}}{d\phi_3}\evBB
-\frac{d\bSigma_{ab}}{d\phi_3}\xMCBB\Bigg)
\nonumber \\*&& \phantom{a}
+\IMC(O,\Ktwo)\Bigg[
-\frac{d\bSigma_{ab}^{(f)}}{d\phi_3}\cntBB
+\frac{d\bSigma_{ab}}{d\phi_3}\xMCBB
+\frac{1}{\Itwo}\Bigg(\frac{d\bSigma_{ab}^{(b)}}{d\phi_2}
+\frac{d\bSigma_{ab}^{(sv)}}{d\phi_2}\Bigg)
\nonumber \\*&& \phantom{a}
+\frac{1}{\Iqtwo}\Bigg(
\frac{d\bSigma_{ab}^{(c+)}}{d\phi_2 dx}\evBB
+\frac{d\bSigma_{ab}^{(c-)}}{d\phi_2 dx}\evBB\Bigg)
-\frac{1}{\Iqtwo}\Bigg(\frac{d\bSigma_{ab}^{(c+)}}{d\phi_2 dx}\cntBB
+\frac{d\bSigma_{ab}^{(c-)}}{d\phi_2 dx}\cntBB\Bigg)
\Bigg]\Bigg\}.\phantom{aaaa}
\label{rMCatNLO}
\eeqn
Eq.~(\ref{rMCatNLO}) is our proposal for a QCD $\MCatNLO$. In order
to use this prescription, we need to write it in a form more
suited to implementation in a computer code, which will be done
in the next section. We stress that, as in the case of the toy 
model, the dependence upon $O$ in eq.~(\ref{rMCatNLO}) is only
formal: the $\MCatNLO$ generates events without reference 
to any observable. 

\subsection[QCD $\MCatNLO$: practical implementation]{\boldmath QCD
$\MCatNLO$: practical implementation\label{sec:impl}}
The straightforward implementation of eq.~(\ref{rMCatNLO}) results
in a computer code which is not optimally efficient, in particular
in the case of unweighted event generation. We therefore implement
our $\MCatNLO$ through the following procedure.

We start by computing the following integrals:
\beqn
I_{\clH}&=&\sum_{ab}\int dz_1 dz_2 d\phi_3 
\Bigg(\frac{d\bSigma_{ab}^{(f)}}{d\phi_3}\evBB
-\frac{d\bSigma_{ab}}{d\phi_3}\xMCBB\Bigg)
\label{Ithreedef}
\\
I_{\clS}&=&\sum_{ab}\int dz_1 dz_2 d\phi_3 
\Bigg[-\frac{d\bSigma_{ab}^{(f)}}{d\phi_3}\cntBB
+\frac{d\bSigma_{ab}}{d\phi_3}\xMCBB
\nonumber \\*&& \phantom{a}
+\frac{1}{\Itwo}\Bigg(\frac{d\bSigma_{ab}^{(b)}}{d\phi_2}
+\frac{d\bSigma_{ab}^{(sv)}}{d\phi_2}\Bigg)
+\frac{1}{\Iqtwo}\Bigg(
\frac{d\bSigma_{ab}^{(c+)}}{d\phi_2 dx}\evBB
+\frac{d\bSigma_{ab}^{(c-)}}{d\phi_2 dx}\evBB\Bigg)
\nonumber \\*&& \phantom{a}
-\frac{1}{\Iqtwo}\Bigg(\frac{d\bSigma_{ab}^{(c+)}}{d\phi_2 dx}\cntBB
+\frac{d\bSigma_{ab}^{(c-)}}{d\phi_2 dx}\cntBB\Bigg)
\Bigg]\,.
\label{Itwodef}
\eeqn
Note that these two integrals are finite, owing to the properties
of $d\bSigma_{ab}\xMC$. Clearly,
\beq
\stot=I_{\clS}+I_{\clH},
\label{stotII}
\eeq
where $\stot$ is the total rate at the NLO level. We also compute
the integrals $J_{\clH}$ and $J_{\clS}$, defined as in eqs.~(\ref{Ithreedef})
and~(\ref{Itwodef}) respectively, except that the integrands
are taken in absolute value (so that, if the integrands were positive-definite,
we would have $I_i=J_i$). We use a modified version of the adaptive
integration routine {\small BASES}~\cite{Kawabata:1995th}, which allows one to
compute at the same time the integral of a given function, and the 
integral of the absolute value of that function. {\small BASES} is an evolution
of the original {\small VEGAS} package~\cite{Lepage:1977sw}.

We then make use of the routine {\small SPRING}, which is part of the
{\small BASES} package. Roughly speaking, for any positive-definite function
$f(\vec{x})$, {\small SPRING} returns a set of values $\{\vec{x}_i\}_{i=1}^n$
distributed according to the function $f$; the number $n$
must be given as an input. Thus, {\small SPRING} effectively achieves the
generation of $n$ unweighted events (with weights equal to 1).
In our case, the absolute values of the integrands in
eqs.~(\ref{Ithreedef}) and~(\ref{Itwodef}) are given as input functions 
to {\small SPRING}. The numbers of events generated in the two cases are 
(possibly with a roundoff to the next integer)
\beq
N_{\clH}=N_{tot}\frac{J_{\clH}}{J_{\clS}+J_{\clH}}\,,\;\;\;\;\;\;
N_{\clS}=N_{tot}\frac{J_{\clS}}{J_{\clS}+J_{\clH}}\,,
\label{NNdef}
\eeq
where $N_{tot}$ is the total number of $\MCatNLO$ events that we
eventually want to generate. It is easy, within {\small SPRING}, to keep track
of the sign of the integrand function, before its absolute value
is taken. Thus, {\small SPRING} achieves unweighted event generation, with 
the weights equal to +1 or --1. 

We symbolically denote the results of the {\small SPRING} runs as follows:
\beqn
I_{\clH}\;\;&\stackrel{\rm {SPRING}}{\longrightarrow}&\;\;
\{\Kthree_i,\flav_i,w_i^{(\clH)}\}_{i=1}^{N_{\clH}}\,,
\label{clHev}
\\
I_{\clS}\;\;&\stackrel{\rm {SPRING}}{\longrightarrow}&\;\;
\{\Ktwo_i,\flav_i,w_i^{(\clS)}\}_{i=1}^{N_{\clS}}\,;
\label{clSev}
\eeqn
thus, eqs.~(\ref{clHev}) and~(\ref{clSev}) correspond to events belonging
to classes $\clH$ and $\clS$ respectively.
For a given $i$, \mbox{$\{\Kthree_i,\flav_i,w_i^{(\clH)}\}$} gives the 
complete description of a three-body event; $\Kthree_i$ represents the 
kinematics (in the lab frame), \mbox{$\flav_i=\{a_i,b_i,c_i\}$} are
the flavours of the initial- $(a_i,b_i)$ and final-state $(c_i)$
partons, and $w_i^{(\clH)}=\pm 1$ is the weight of the event returned 
by {\small SPRING}. The event \mbox{$\{\Ktwo_i,\flav_i,w_i^{(\clS)}\}$} 
is analogous, except that it has $2\to 2$ kinematics (and thus
\mbox{$\flav_i=\{a_i,b_i\}$}). 

The events in eqs.~(\ref{clHev}) and~(\ref{clSev}) are the initial
conditions for MC evolution. In the case of $\clH$ events, 
eq.~(\ref{clHev}), the kinematics $\Kthree_i$ is determined by 
\mbox{$(z_1,z_2,\phi_3)$}. The choice of the flavours $\flav_i$ is
more involved. One possibility is to fix the pair of 
flavours $(a,b)$ that appears in the sum in eq.~(\ref{Ithreedef}) 
(and thus to fix $\flav$, since $c$ in this case is unambiguously
determined by $a$ and $b$), to compute the integral, and to repeat 
the computation for all possible flavour pairs. This procedure is 
however inconvenient from the point of view of computing time. 
A faster approach is simply to compute three integrals of the type
in eq.~(\ref{Ithreedef}), for the cases $ab = q\bar q$, $qg$ and $\bar q g$
summed over quark flavours.  For each of these integrals, the quark
flavours are then determined by statistical
methods; this can give a wrong assignment of flavours on event-by-event
basis, but guarantees that the ``distribution'' in the flavours is
returned correctly. The reason for not computing a single integral
is that subprocesses with $q\bar q$, $qg$ and $\bar q g$ initial states
often give contributions of different signs to the cross section, which
is a source of errors when statistical methods are used.

In the case of $\clS$ events, eq.~(\ref{clSev}), the kinematic
configuration $\Ktwo_i$ is unambiguously defined as the soft limit of the
corresponding $2\to 3$ kinematics, determined by \mbox{$(z_1,z_2,\phi_3)$}.
For the flavour assignment, we proceed similarly to what was done in
the case of $\clH$ events. However, here the relevant flavours
are those of the partons entering the $2\to 2$ hard process which
factorizes in the case of soft or collinear emissions. Thus, we
use the Born contribution $d\bSigma_{ab}^{(b)}$ to assign the
flavour. Notice that this can result in errors of ${\cal O}(\as)$ 
in flavour assignment. However, the effects induced in this
way are ${\cal O}(\as^2)$, and thus beyond control. On the other hand,
with this prescription one achieves at ${\cal O}(\as)$ the cancellation
of the MC terms $d\bSigma_{ab}\xMC$ introduced in the modified
subtraction, i.e., one avoids double counting. More details on this
point are given in app.~\ref{app:expansion}. It should be noticed
that, since the Born contribution is always positive, there is no
need to compute the integral in eq.~(\ref{Itwodef}) separately
for $q\bar q$, $qg$ and $\bar q g$ initial states.

The above statistical method has the virtue of being easy to
generalize to other processes. We point out that eqs.~(\ref{stotII})
and~(\ref{NNdef}), together with those in the remainder of this section,
are trivially extended to the case in which more than one integral is
computed for $\clH$ and/or $\clS$ events.

The weights in eqs.~(\ref{clHev}) and~(\ref{clSev}) have the 
following properties:
\beq
\frac{J_{\clH}}{N_{\clH}}\sum_{i=1}^{N_{\clH}}w_i^{(\clH)}=I_{\clH}\,,
\;\;\;\;\;\;
\frac{J_{\clS}}{N_{\clS}}\sum_{i=1}^{N_{\clS}}w_i^{(\clS)}=I_{\clS}\,,
\eeq
and thus, using eq.~(\ref{NNdef}),
\beq
\stot=\frac{J_{\clH}+J_{\clS}}{N_{tot}}\left(
\sum_{i=1}^{N_{\clH}}w_i^{(\clH)}+\sum_{i=1}^{N_{\clS}}w_i^{(\clS)}\right).
\eeq
We then define the set of events
\beq
\{\KK_i,\flav_i,w_i\}_{i=1}^{N_{tot}}\,=\,
\{\Kthree_i,\flav_i,w_i^{(\clH)}\}_{i=1}^{N_{\clH}}\bigcup\,
\{\Ktwo_i,\flav_i,w_i^{(\clS)}\}_{i=1}^{N_{\clS}}\,,
\label{NLOevents}
\eeq
where $\KK_i$ denotes either $\Kthree_i$ or $\Ktwo_i$, and
\beq
w_i=w_i^{(\clS,\clH)}\frac{J_{\clH}+J_{\clS}}{N_{tot}}
\;\;\;\;\Rightarrow\;\;\;\;
\sum_{i=1}^{N_{tot}}w_i=\stot\,.
\label{wgttoxsec}
\eeq

We constructed a Fortran code whose output is the set of 
events \mbox{$\{\KK_i,\flav_i,w_i\}_{i=1}^{N_{tot}}$}. 
We point out that the merging of the two event sets on
the r.h.s.\ of eq.~(\ref{NLOevents}) is done randomly;
therefore, the probability of finding a $\clS$ ($\clH$)
event at an arbitrary place in the event list is a constant,
equal to \mbox{$N_{\clS}/N_{tot}$} (\mbox{$N_{\clH}/N_{tot}$}).
This set of events is passed to \HW; there, it
replaces the routine that generates the hard process.
We stress that \HW\ has not been modified; we just added
a routine that reads the parton configurations coming from the
NLO code, and translates them in the standard \HW\ form. Each
configuration generates a shower, and all the other operations
requested by the user. Notice that the unweighting procedure
is entirely dealt with by the NLO code; \HW\ reads the weights 
provided by the NLO code, and does not discard any events
(apart from the few killed in the ordinary shower/hadronization process).
According to eq.~(\ref{wgttoxsec}), in order to get the total cross
section right, exactly $N_{tot}$ events must be processed by 
\HW\ . However, should only $N_{ev}<N_{tot}$ events be processed, the
correct rate will be recovered by multiplying all the physical results
by a factor of \mbox{$N_{tot}/N_{ev}$}, since all weights have the
same absolute value. In other words, any subset of the event
set \mbox{$\{\KK_i,\flav_i,w_i\}_{i=1}^{N_{tot}}$} can act as an
event set itself, provided that the weights are multiplied by a
common factor, in order for eq.~(\ref{wgttoxsec}) to hold.

The presence of negative-weight events is a distinctive feature
of $\MCatNLO$'s. Negative weights 
imply larger statistical errors than in the case of ordinary MC's, 
when the same total number of events is generated. In the $\clH$ 
and $\clS$ samples of eqs.~(\ref{clHev}) and~(\ref{clSev}),
the fraction of negative-weight events is
\beq
f_i=\frac{1}{2}\left(1-\frac{I_i}{J_i}\right),
\label{nfrac}
\eeq
for $i=\clS,\clH$; clearly, the closer $J_i$ is to $I_i$, the smaller is
$f_i$.  We can in fact reduce the number of negative-weight events in our
$\MCatNLO$ by varying $J_{\clS}$.  This quantity depends upon the form of
the counterevents, or, in a given subtraction scheme, upon the free
parameters that define the subtraction\footnote{These are $\tilde{\rho}$ and 
$\omega$ in the standard subtraction, $\zeta$ in the $\zeta$-subtraction:
see app.~\ref{app:SSub} and app.~\ref{app:zetaSub}.}. Notice that
this is not in contradiction with the properties of the subtraction
method, since $J_{\clS}$ is not a physical quantity. Thus, we can adjust these
free parameters in order to minimize the difference $J_{\clS}-I_{\clS}$; the
solution can only be found through numerical methods, because of
the complexity of the integrand appearing in $J_{\clS}$. Further
details on this point are given in app.~\ref{app:zetaSub}.

We point out that the integral $J_{\clH}$ does not depend upon any free
parameters; thus, the number of negative-weight $\clH$ events is
a constant for fixed input (physical) parameters in eq.~(\ref{rMCatNLO}).
We remind the reader that these events will mainly populate the hard 
emission region.

We remark finally that, owing to the procedure described in 
sect.~\ref{sec:match}, the $2\to 2$ kinematic configurations resulting
from the integral in eq.~(\ref{Itwodef}) can all be obtained as the 
soft limit of the $2\to 3$ kinematic configurations identified by 
\mbox{$(z_1,z_2,\phi_3)$}. This suggests writing the integration
measure in eq.~(\ref{Itwodef}) with the two-body phase space explicitly
factorized, \mbox{$d\phi_3=d\phi_2 d\tilde{\phi}_3$}, and integrating
over $d\tilde{\phi}_3$ before passing the integrand function to 
{\small SPRING}. It should be clear that this intermediate integration 
step does not affect our capability of obtaining fully-exclusive final 
states; however, it does have an impact on the number of negative-weight
events produced by the $\MCatNLO$. In this paper, we do not exploit
this possibility, leaving the option open for future developments.

\subsection{Generalities on matched computations}\label{sec:generalities}
We now consider the case of an observable $O$ whose kinematically-%
allowed range is $O_s<O<O_h$. We assume that the region $O\simeq O_h$ 
receives only contributions due to  hard emissions,
and that large logarithms are present in any fixed-order
cross section for $O\simeq O_s$, where a resummation is therefore
necessary. In accordance with eq.~(\ref{rMCatNLO}), we write a differential
cross section obtained from the $\MCatNLO$ as follows:
\beq
\xsecO=\xsecOH-\xsecORSz+\xsecORS\,.
\label{MCatNLOres}
\eeq
The quantity $\xdsecORS$ is the contribution of all $\clS$ events, i.e.
of the MC evolutions $\IMC(O,\Ktwo)$. The quantities $\xdsecOH$ and
$\xdsecORSz$ are the contributions of the first and second terms
under the integral sign in eq.~(\ref{rMCatNLO}) respectively, and
thus both are associated with $\clH$ events. We expect them to be divergent
for $O\to O_s$, only their difference being finite. However, for $O\to O_h$,
we can assume that $\xdsecOH$ and $\xdsecORSz$ are separately finite.
This is because this region is by hypothesis dominated by hard emission,
which can only occur through real-emission diagrams; this prevents the 
emitted parton over whose variables we integrate in eq.~(\ref{rMCatNLO})
from being soft or collinear. The quantities introduced in 
eq.~(\ref{MCatNLOres}) have the following properties:
\beqn
&&\abs{\xsecOH-\xsecORSz}\ll \xsecORS\;\;\;\;\;\;
{\rm for}\;\;O\to O_s\,;
\label{HvsRSz}
\\
&&\xsecOH=\xsecONLO+{\cal O}(\as^2)\;\;\;\;\;\;
{\rm for}\;\;O\to O_h\,;
\label{HvsNLO}
\\
&&\xsecORS=\xsecORSz+{\cal O}(\as^2)\;\;\;\;\;\;
{\rm for}\;\;O\to O_h\,.
\label{RSvsRSz}
\eeqn
Eqs.~(\ref{HvsNLO}) and~(\ref{RSvsRSz}) imply that, for $O\to O_h$,
we recover the expected result $\xdsecO=\xdsecONLO$, which holds
up to ${\cal O}(\as^2)$ (NNLO) terms. Strictly speaking, eq.~(\ref{RSvsRSz})
is not necessary for this to happen; $\xdsecORS$ and $\xdsecORSz$ 
include the leading (and 
possibly next-to-leading) tower of logarithms \mbox{$\log O/O_h$},
or more precisely $\log [(O-O_s)/(O_h-O_s)]$, 
which become small for $O\to O_h$. However, eq.~(\ref{RSvsRSz}) implies
that $\xdsecO\simeq\xdsecONLO$ also for those values of $O$
which are not too close to $O_h$, but still outside the region where
resummation is necessary. The point here is the size of the unknown
${\cal O}(\as^2)$ coefficients. There is in fact no guarantee that
these coefficients are small; if they are big, the $\MCatNLO$ result
can differ substantially from the NLO one, although the accuracy of
the two computations is the formally of the same order
in perturbation theory.  This problem 
is always encountered when a fixed-order computation is matched with a
resummed computation. A possible solution (see for example
refs.~\cite{Cacciari:1998it,Cacciari:2001td}) would be to rewrite
eq.~(\ref{MCatNLOres}) as follows
\beq
\xsecO=\xsecOH+\Dfun(O)\left[\xsecORS-\xsecORSz\right]\,,
\label{MCatNLOdump}
\eeq
where $\Dfun(O)$ is a regular function, such that $\Dfun(O_h)\to 0$
for $O\to O_h$, and $\Dfun(O)\to 1$ for $O\to O_s$, up to terms
suppressed by powers of \mbox{$(O-O_s)/(O_h-O_s)$}. The perturbative and
logarithmic accuracies of eq.~(\ref{MCatNLOdump}) are the same as
those of eq.~(\ref{MCatNLOres}); however, the function $\Dfun$ can be
suitably chosen in order to suppress the ``MC'' contribution in the
region in which the NLO computation can be trusted. Notice that
the presence of $\Dfun(O)$ implies that the total rate predicted
by eq.~(\ref{MCatNLOdump}) differs from that predicted
at the NLO level, even when no cuts are applied.

The problem with eq.~(\ref{MCatNLOdump}) is that the function $\Dfun$
is specific to the observable considered; thus, eq.~(\ref{MCatNLOdump})
cannot be implemented directly in an $\MCatNLO$. However, one could 
try to devise a solution, similar to that of eq.~(\ref{MCatNLOdump}), that
works in the context of $\MCatNLO$'s. Clearly, in the hard-emission region
the function $\Dfun$ suppresses the contribution of $\clS$ events, and 
that part of $\clH$ events that corresponds to the ${\cal O}(\as)$ 
expansion of the MC result. In order to achieve this result without
any reference to a specific observable, a modification of the showering
mechanism in the MC is necessary. In particular, one could multiply
the Altarelli-Parisi kernel appearing in the Sudakov by a function
$Q(\xi,z)$ of the showering variables $\xi$ and $z$, that tends to
1 in the soft and/or collinear limits, and to zero in the hard emission
region. It should be noted that this function is completely analogous
to the function $Q(x)$ that we used in the toy MC, eq.~(\ref{Deltadef}).
Exactly as in the case of the toy model, in order to satisfy the requirement
that the ${\cal O}(\as)$ expansion of the $\MCatNLO$ result be identical to
the NLO result, we would need to perform the formal substitution
\beq
d\bSigma_{ab}\xMCB\;\;\longrightarrow\;\;
Q(\xi,z)d\bSigma_{ab}\xMCB\,,
\eeq
in eq.~(\ref{rMCatNLO}). Since this procedure would modify the integrand
in eqs.~(\ref{Ithreedef}) and~(\ref{Itwodef}), it would also have an
impact on the number of negative-weight events generated. In this paper, 
we shall not pursue this issue further, since we prefer to keep
the parton shower MC unmodified.

\subsection[Results: W$^+$W$^-$ and jet observables]{\boldmath Results:
W$^+$W$^-$ and jet observables}\label{sec:pheno}
In this section we present results obtained with our  W$^+$W$^-$
$\MCatNLO$,
implemented as described in sect.~\ref{sec:impl}. We do not seek to
give a phenomenological description that corresponds to a specific 
experimental configuration; rather, we wish to show the
differences between $\MCatNLO$, standard \HW\ MC, and NLO
results. For this reason, we also present distributions obtained
by integrating over the whole phase space, i.e., in most cases we 
do not apply any acceptance cuts. In the case in which cuts
are applied, we only consider the W$^+$ and W$^-$ variables, and
not the variables of their decay products. We also consider jet
distributions, the jets being reconstructed by means of the
$\kt$-clustering algorithm~\cite{Catani:1993hr}. We show
results for $pp$ collisions at $\sqrt{S}=14$~TeV.
We adopt the low-$\as$ set of MRST99 parton densities~\cite{Martin:1999ww},
since this set has a $\Lambda_{\sss QCD}$ value which is rather close
to that used as \HW\ default; this value of $\Lambda_{\sss QCD}$
($\Lambda_5^{\sss \overline{MS}}=164$~MeV) is used for all our
$\MCatNLO$, MC, and NLO runs. In the case of standard \HW\ MC 
runs, we give each event (we generate unweighted events) 
a weight equal to $\stot/N_{tot}$, with $N_{tot}$ the total number of
events generated. For fully exclusive distributions with no cuts
applied, this is equivalent to normalizing \HW\ results to the total
NLO rate $\stot$. All the $\MCatNLO$ and MC results (but not, of course,
the NLO ones) include the hadronization of the partons in the final state.

\begin{figure}[ht]
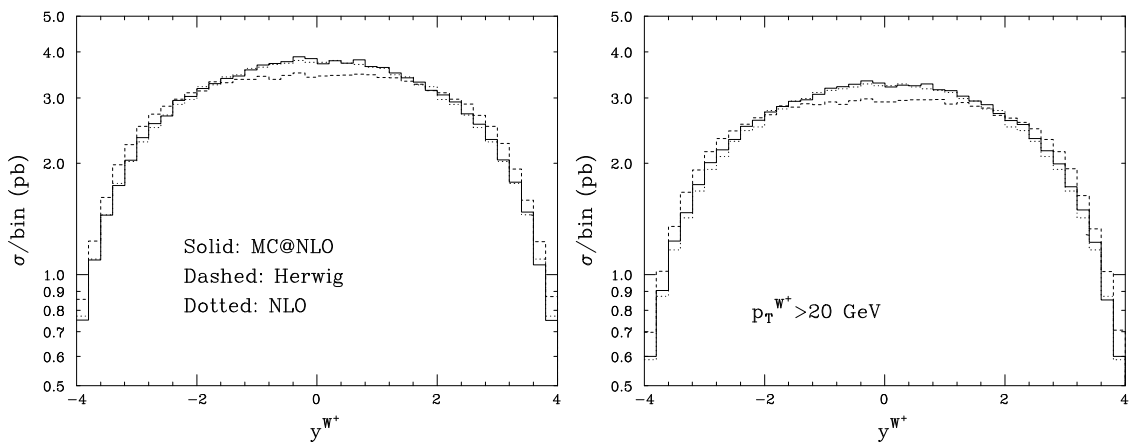

  \begin{center}
    \epsfig{figure=wwywp.eps,width=0.49\textwidth}
    \epsfig{figure=wwywp_cut.eps,width=0.49\textwidth}
\caption{\label{fig:ywp} 
  $\MCatNLO$ (solid), \HW\ (dashed) and NLO (dotted) results
  for the rapidity of the W$^+$. \HW\ results have been normalized
  as explained in the text.
}
  \end{center}
\end{figure}
We present in fig.~\ref{fig:ywp} the rapidity of the W$^+$, with (right
panel) or without (left panel) a cut on the transverse momentum $\ptwp$. 
The solid, dashed and dotted histograms show the $\MCatNLO$, 
MC and NLO results, respectively. This distribution
is fairly inclusive, and we expect it to be predicted well by NLO
QCD in a wide range. From the figure, we see that the NLO and $\MCatNLO$
results are extremely close to each other in the whole range considered,
whether or not there is a $\ptwp$ cut. The MC result is also
similar, but clearly broader than the other two.
\begin{figure}[ht]
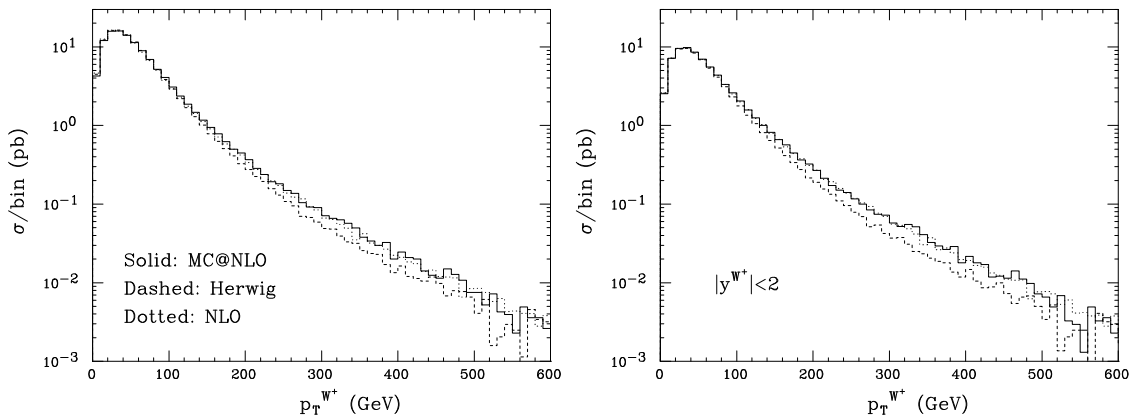

  \begin{center}
    \epsfig{figure=wwptwp.eps,width=0.49\textwidth}
    \epsfig{figure=wwptwp_cut.eps,width=0.49\textwidth}
\caption{\label{fig:ptwp} 
  As in fig.~\ref{fig:ywp}, for the transverse momentum of the W$^+$.
}
  \end{center}
\end{figure}
A similarly inclusive variable is presented in fig.~\ref{fig:ptwp}. The
transverse momentum of the W$^+$ is shown with (right panel) or without 
(left panel) a cut on the rapidity $\ywp$. The previous comment applies
in this case as well: NLO and $\MCatNLO$ basically coincide, whereas the
MC result is softer than the other two. The same pattern can for example
be found for the invariant mass of the W pair. We conclude that, for
this kind of observable, the ${\cal O}(\as^2)$ effects are very small,
and NLO and $\MCatNLO$ are almost equivalent. This also implies that
any possible reshuffling of the momenta, due to the hadronization phase
in $\MCatNLO$, has negligible impact on the colourless W's, as we expect.

\begin{figure}[t]
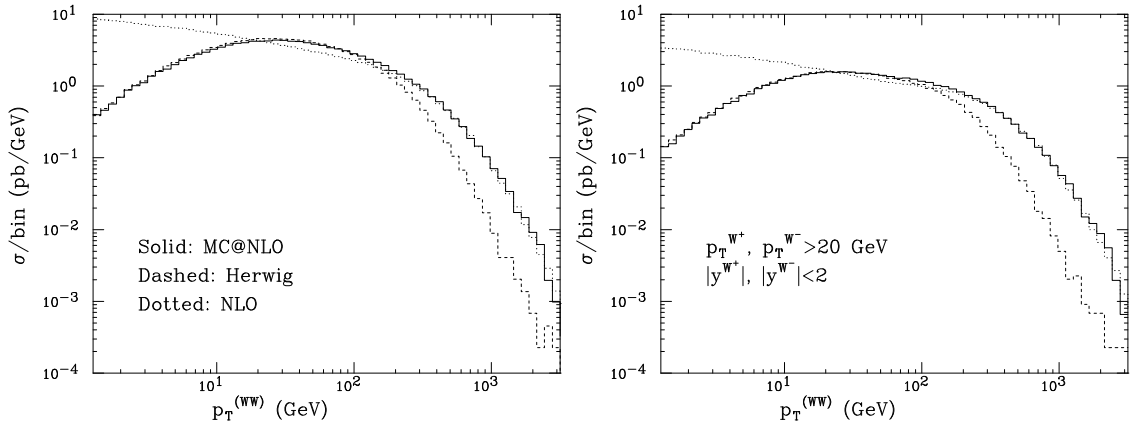

  \begin{center}
    \epsfig{figure=wwptpair.eps,width=0.49\textwidth}
    \epsfig{figure=wwptpair_cut.eps,width=0.49\textwidth}
\caption{\label{fig:ptpair} 
  As in fig.~\ref{fig:ywp}, for the transverse momentum of the
  W$^+$W$^-$ pair.
}
  \end{center}
\end{figure}

We now turn to the case of more exclusive quantities, such as
correlations between W$^+$ and W$^-$ variables. In fig.~\ref{fig:ptpair} 
we present $\ptww$, the modulus of the vector sum of the
transverse momenta of the two W's. NLO computations
cannot predict this observable in the region $\ptww\simeq 0$, because
of a logarithmic divergence for $\ptww\to 0$; on the other hand, NLO
is expected to give reliable predictions at large $\ptww$. The MC behaves
in the opposite way; thanks to the cascade emission of soft and collinear
partons, it can effectively resum the distribution around $\ptww=0$;
however, its results are not reliable in the large-$\ptww$ region, which 
is mainly populated by events in which a very hard parton recoils against 
the W pair.  Apart from the unreliability of the MC shower approximation to
the matrix element in the large-$\ptww$ region, this region cannot really
be filled efficiently by the shower;
this is easily understood by observing that, at fixed partonic c.m.\ energy, 
the large-$\ptww$ region coincides with the dead zone (the border of the 
dead zone in terms of $\ptww$ can be worked out by using the results of
app.~\ref{app:DZ}).

This complementary behaviour of NLO and MC approaches can be seen clearly 
in fig.~\ref{fig:ptpair}, independent of the cuts on the rapidities
and transverse momenta of the bosons. In the tail, the NLO cross section
is much larger than the MC, simply because hard emissions are correctly
treated only in the former. For $\ptww\to 0$, the difference between
the two histograms shows the effect of all-order resummation; clearly,
no meaningful comparison between NLO and data can be attempted in this
region. It is therefore reassuring that the $\MCatNLO$ result
interpolates the MC and NLO results smoothly. In the small-$\ptww$ region,
the shape of the $\MCatNLO$ curve is identical to that of the
MC result. This is evidence of the fact that MC and $\MCatNLO$
resum large logarithms at the same level of accuracy, as argued
in app.~\ref{app:logacc}.  When $\ptww$ grows large, the $\MCatNLO$
tends to the NLO result, as expected. Again, the hadronization has
no significant impact on the W$^+$W$^-$ system.

\begin{figure}[t]
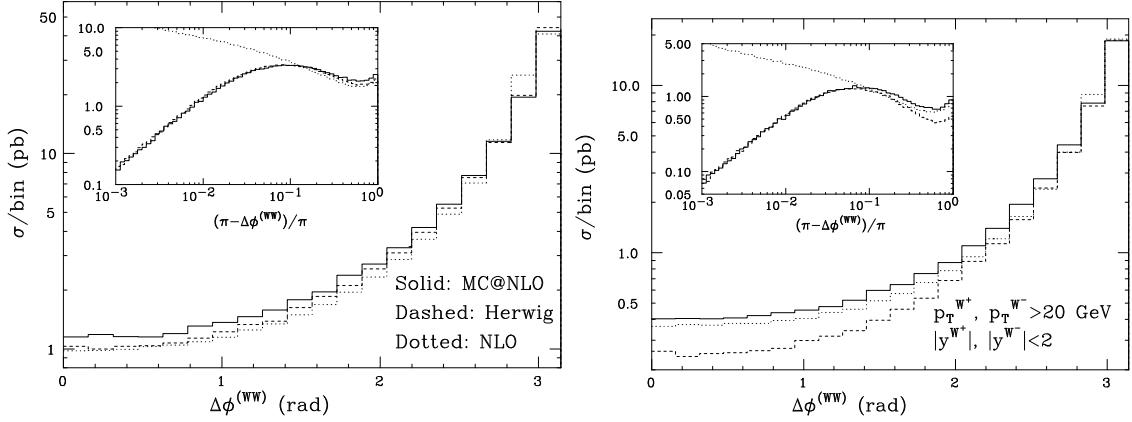

  \begin{center}
    \epsfig{figure=wwdphi.eps,width=0.49\textwidth}
    \epsfig{figure=wwdphi_cut.eps,width=0.49\textwidth}
\caption{\label{fig:dphi} 
  As in fig.~\ref{fig:ywp}, for the difference in the azimuthal angles
  of the W$^+$ and W$^-$. 
}
  \end{center}
\end{figure}

In fig.~\ref{fig:dphi} we present the distribution of the difference
between the azimuthal scattering angles (i.e., those in the plane transverse 
to the beam direction) of the W$^+$ and W$^-$. This distribution cannot be 
reliably predicted by fixed-order QCD computations in the region
$\dphiww\simeq\pi$; in fact, the NLO prediction diverges logarithmically
for $\dphiww\to\pi$. This can be seen in the insets of the plots,
where the variable $\dphiww$ has been plotted versus
\mbox{$(\pi-\dphiww)/\pi$} on a logarithmic scale, in order to visually
enhance the region $\dphiww\simeq\pi$. We can see that in this region the
$\MCatNLO$ and MC results have identical shapes, as in the case of 
the observable $\ptww$ near zero discussed above. The other end of the
spectrum, i.e. the tail $\dphiww\simeq 0$,
might be considered as analogous to the 
region of large $\ptww$, in that this region is dominated, in an NLO 
computation, by hard single-parton emission. However, there are important
differences from the point of view of MC treatment.
As discussed earlier, the region of high $\ptww$ is not filled
efficiently by the MC shower.
On the other hand, the $\dphiww\simeq 0$ tail for
low $\ptww$ is more easily populated by the MC.
This is confirmed in the left-hand panel of fig.~\ref{fig:dphi},
where we can see that the NLO and MC results are quite close to each other
at $\dphiww\simeq 0$ when no cuts are 
applied. Thus, the region $\dphiww\simeq 0$ receives contributions both 
from hard parton emissions (which dominate the configurations in which
the W's have large transverse momenta), and from multiple soft or 
collinear parton emissions (when the W's have small transverse momenta). 
NLO can treat the former but not the latter, and MC can treat the latter
but not the former. $\MCatNLO$, on the other hand, is expected to
handle both emissions correctly, while still avoiding any double counting.
When cuts on the transverse momenta of the W's are applied,
the contribution
from multiple soft or collinear emissions becomes less important,
as may be seen in the right-hand panel of fig.~\ref{fig:dphi}.
Overall, the $\MCatNLO$ results are close to the NLO predictions in
the $\dphiww\simeq 0$ tail; as we shall show later, the differences
are compatible with ${\cal O}(\as^2)$ effects.

\begin{figure}[t]
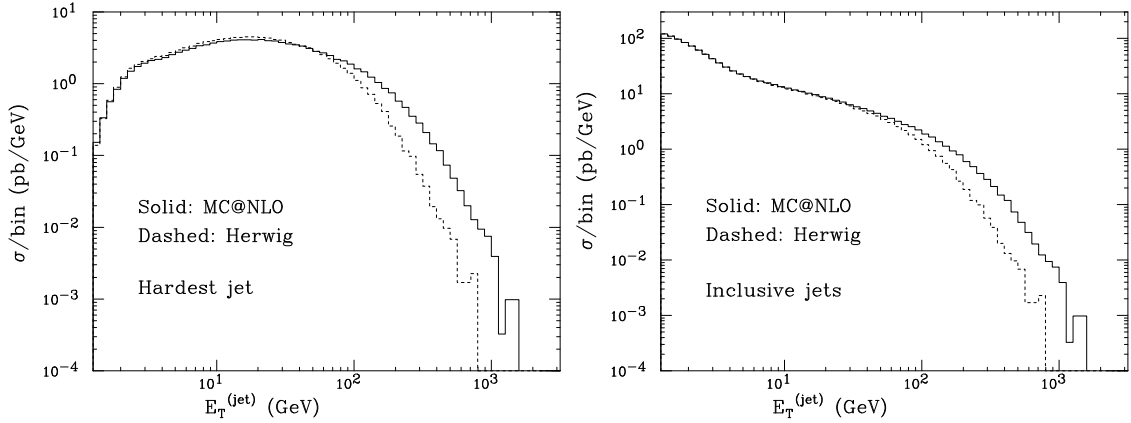

  \begin{center}
    \epsfig{figure=wwhrdjet.eps,width=0.49\textwidth}
    \epsfig{figure=wwincljet.eps,width=0.49\textwidth}
\caption{\label{fig:jets} 
  $\MCatNLO$ (solid) and \HW\ (dashed) results
  for the transverse energy of the hardest jet
  in each event (left panel), and of fully inclusive jets (right
  panel). \HW\ results have been normalized
  as explained in the text.
}
  \end{center}
\end{figure}

We now turn to the discussion of jet observables. In fig.~\ref{fig:jets}
we present the transverse energy distribution of the hardest jet
of each event (left), and of the inclusive jets (right). In this case,
we only compare $\MCatNLO$ and MC results; NLO jet results are trivial 
for this process, since there is only one ``jet'', which coincides with
a parton. In the case of the hardest jet, the same discussion as in the
case of $\ptww$ applies; the $\MCatNLO$ resums large logarithms at small
$\Et$ in the same way as the MC does, and can also treat the large-$\Et$
region, where the MC fails. In the case of inclusive jets, the spectrum
diverges for $\Et\to 0$, because of the increasing number of
jets with smaller and smaller $\Et$. In this region, the $\MCatNLO$
and MC results (rescaled by the K-factor) 
coincide. We stress that this is actually the same pattern that we
have already discussed in the toy model: see eqs.~(\ref{smallzxsec})
and~(\ref{smallzborn}). 

The NLO results presented so far have been obtained by setting 
the renormalization ($\mur$) and factorization ($\muf$) scales,
that we collectively refer to as NLO scales ($\muNLO$),
equal to a reference scale $\muz$, defined as follows
\beq
\muz^2=\frac{1}{2}\left[\left(\Etwp\right)^2+\left(\Etwm\right)^2\right],
\label{NLOscale}
\eeq
where $\Etwp$ and $\Etwm$ are the transverse masses of the W$^+$
and W$^-$ respectively. The same scale choice has been adopted
to compute the integrals given in eqs.~(\ref{Ithreedef}) 
and~(\ref{Itwodef}), that is, to obtain the sets of $\clH$ and
$\clS$ events eventually given to the MC as initial conditions.

In general, we refer to the scales used in the computations of
eqs.~(\ref{Ithreedef}) and~(\ref{Itwodef}) as $\MCatNLO$ scales
($\muMCatNLO$). We point out that the NLO and $\MCatNLO$ scales need 
{\em not} coincide. It actually seems more natural to choose different forms
for them.  The $\MCatNLO$ scale should in some way match the 
scale used in the parton shower; although any mismatch between the
two is formally of higher order than we are considering,
a choice of the $\MCatNLO$ scale motivated by
MC considerations should result in smaller coefficients for the
${\cal O}(\as^2)$ terms in the regions dominated by hard emissions.
Notice that, when the NLO and $\MCatNLO$ scales do not coincide, the
total rates predicted by NLO and $\MCatNLO$ computations are different.

\begin{figure}[t]
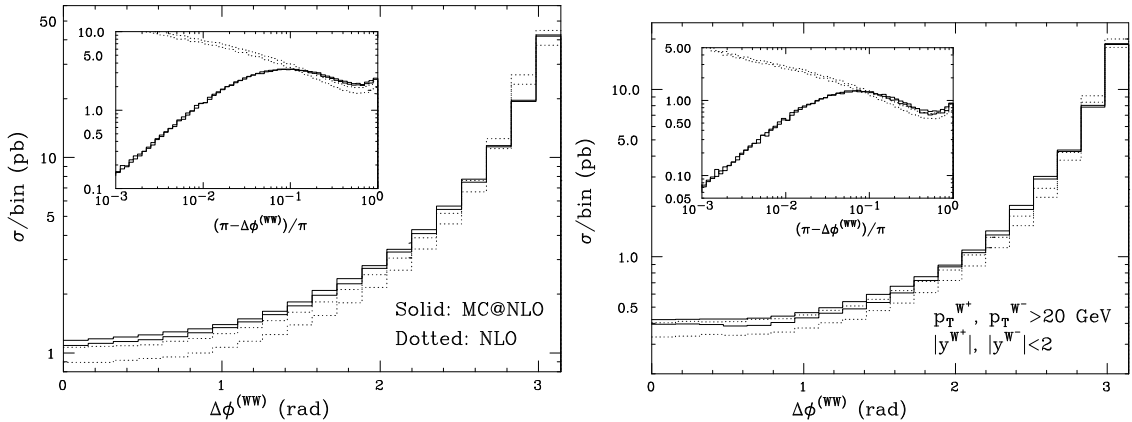

  \begin{center}
    \epsfig{figure=wwdphi_sc.eps,width=0.49\textwidth}
    \epsfig{figure=wwdphi_cut_sc.eps,width=0.49\textwidth}
\caption{\label{fig:dphi_sc} 
  Scale dependence of $\MCatNLO$ (solid band) and NLO (dotted band)
  results, obtained by varying the renormalization and factorization
  scales in the range $\muz/2<\mur,\muf<2\muz$. See the text for details.
}
  \end{center}
\end{figure}
We shall not pursue the issue of an appropriate $\MCatNLO$ scale choice
any further in the present paper.  We shall limit ourselves to considering 
the variation of NLO and $\MCatNLO$ predictions obtained by setting 
\mbox{$\mur=\muf=\muz/2$} and \mbox{$\mur=\muf=2\muz$}.
We stress that, while this procedure is a commonly accepted way to
obtain a rough estimate of the unknown higher-order terms in an NLO
computation, it cannot be used to guess the $\MCatNLO$ scale
that best matches the parton shower scale; for this purpose, functional 
forms other than that shown in eq.~(\ref{NLOscale}) would need to be 
considered. Still, the procedure can give an idea of the size
of the neglected higher-order terms in the $\MCatNLO$ formalism.
The results are presented in fig.~\ref{fig:dphi_sc} for the 
azimuthal correlation $\dphiww$; we have chosen this observable
since it is the one in which the scale effects are largest. The
solid and dotted bands show the variation in the $\MCatNLO$ and NLO predictions
respectively. We observe a sizable reduction of the scale dependence 
when going from NLO to $\MCatNLO$ results. Also, in the $\dphiww\simeq 0$
tail, the two bands are rather close to each other, and overlap in the
case that cuts are applied. This is an indication of
the fact that NLO and $\MCatNLO$ results indeed differ by ${\cal O}(\as^2)$
terms, and that the difference is less significant when hard cuts are
applied, as we should expect.

An issue related to the choice of scale for the parton showers is the
setting of the \HW\ parameter {\tt EMSCA}, which enforces an upper limit
on the transverse momenta of partons emitted in the showers.  The
coherence of gluon emission is ensured in \HW\ by using an angular
variable to perform the shower evolution.  In order to avoid
double-counting of hard processes, it is then necessary to veto
emissions with transverse momenta greater than the scale of the
primary hard process, which is specified by {\tt EMSCA}.  If such
emissions were allowed, they would constitute the true primary hard
process.  When the veto is applied, the harder emission does not occur  
but the evolution variable is reset to the value selected for the vetoed
emission.  In this way the constraint that the true hard scale is {\tt EMSCA}
is correctly taken into account together with QCD coherence in the shower.

In our case, when running the standard \HW\ MC for W$^+$W$^-$ production,
we set \mbox{{\tt EMSCA}=$\sqrt{s}$}.
In the case of the $\MCatNLO$, there is no reason to have the same
{\tt EMSCA} for events belonging to the classes $\clH$ and $\clS$. We have
already pointed out that the class $\clS$ must behave in the
same way as the ordinary MC, in order to reproduce (shape-wise) its 
results in those regions where a resummation is needed. Thus, for $\clS$
events, we set should also set \mbox{{\tt EMSCA}=$\sqrt{s}$}.
As far as $\clH$ events are
concerned, we notice that the initial condition for the MC in the toy
model can also be interpreted as a veto scale; in the case of toy
$\clH$ events, the initial condition is $\xm(x)$, which is imposed
to be a monotonic function, with $\xm(0)=1$ (see sect.~\ref{sec:Matching}
and sect.~\ref{sec:obs}). A possible choice that reproduces this
condition in the real case is \mbox{{\tt EMSCA}=$\sqrt{s}-2\pt$}, where
$\pt$ is the transverse momentum of the final-state light parton, which
is always present in $\clH$ events. Then, in the soft or collinear
limits, this scale tends to that adopted in the case of $\clS$ events.
However, such a choice implies that if a very hard emission
(i.e., \mbox{$\pt\simeq\sqrt{s}/2$}) occurs at the NLO, only
a very limited possibility is left for the MC to emit further,
which can lead to problems in the hadronization process.
Other choices are of course possible. In the toy model, it can be 
proven that different $\xm$'s only affect logarithms beyond the accuracy
of the MC. In the real case, we have verified that other reasonable
choices of {\tt EMSCA} do not induce visible differences in any of 
the observables studied.

Our $\MCatNLO$ results are obtained without soft matrix element
corrections~\cite{Seymour:1995df}, i.e. we set the \HW\ parameter
\mbox{{\tt SOFTME}=\tt{.FALSE.}}.
By doing so, we obtain the ${\cal O}(\as)$ \HW\ result reported 
in app.~\ref{app:DZ}, which is crucial in our implementation of
$\MCatNLO$ to avoid double counting. Furthermore, the purpose of
soft matrix element corrections is to ensure smooth matching
on the boundary between the regions dominated by parton showers and
by hard emissions, whereas in our formalism there is no such hard
boundary. In order to have meaningful comparisons with standard MC results,
we set {\tt SOFTME}={\tt .FALSE.} in that case as well. We verified 
that the choice of {\tt SOFTME} has in general only a small impact on
the MC results for the observables we studied; the only visible effects
are in the shapes of distributions in those regions in which a resummation
is needed: $\ptww\simeq 0$, $\dphiww\simeq\pi$, and 
$p_{\sss T}^{(jet)}\simeq 0$ for the hardest jet of the event. A non-negligible
effect ($\simeq 10$\%) is also seen at $\dphiww\simeq 0$, which is partly
related to the behaviour at $\dphiww\simeq\pi$, since the total rate
is independent of the value of {\tt SOFTME}.

We should mention finally that we performed many studies of the
stability of the $\MCatNLO$ results. We
studied the dependences of the physical
results upon the free parameters $\alpha$ and $\beta$ introduced
in eqs.~(\ref{Gfundef}) and~(\ref{txDZ}), and found them completely
negligible. This is a rather direct confirmation of the claim in
app.~\ref{app:DZ}, that the angular distribution of soft partons
is irrelevant in the determination of physical observables. We
performed test $\MCatNLO$ runs with increased values of the showering cutoffs
{\tt VQCUT} and {\tt VGCUT}, and found no significant variations with
respect to the results presented here; this supports the claim that MC 
and $\MCatNLO$ have the same dependences upon MC showering cutoffs.
Also, we find that the parton densities used for MC backward evolution
do not strongly influence the $\MCatNLO$ results for this production process.
The results obtained using low-$\as$ MRST99 in the MC evolution are
statistically compatible with those obtained using MRST-LO (low-$\as$ MRST99
being used in both cases for the NLO part).

\section{Other processes}\label{sec:other}
The extension of $\MCatNLO$ to other processes, at least those
with unambiguous colour flow, seems straightforward in principle. 
The key equation, (\ref{rMCatNLO}), is formally valid for any
kind of singularities. For final-state collinear singularities,
the choice of the Bjorken $x$'s and the flavour assignment are
trivial: one just has to give the MC the
lower-order process that preceded the collinear splitting.

In processes where there are several possible colour flows, the
procedure adopted in the \HW\ MC is first to select the kinematics
according to the full differential cross section, including
interference between colour flows, and then to assign a colour
flow according to the relative contributions to the cross section
neglecting interference (i.e. in the large-$N_c$ limit)~\cite{Odagiri:1998ep}.
This procedure can easily be extended to the $\MCatNLO$ case. We point
out that the colour flows can be determined independently for $\clS$
and $\clH$ events. The correct NLO results will still be recovered upon 
expansion in $\as$, since the cancellation of the MC terms~(\ref{QCDaddsub})
inserted in the modified subtraction formula, eq.~(\ref{rMCatNLO}), 
occurs on statistical basis, and not event by event. In order to
construct the MC-like cross sections~(\ref{MCexp}), the technique
presented in this paper can be adopted: namely, the pure MC result
is smoothly matched onto that predicted by perturbative QCD in the 
soft limit. In fact, independently of the choice of the showering 
variables, the former cross section is in general locally different from 
the latter, since it is only in the former that the radiation function 
is integrated over an azimuthal angle, in order to implement colour coherence 
in a MC-friendly manner. For processes involving emission from gluons,
there are also azimuthal correlations in the collinear limit, which
integrate to zero and are therefore not included in MC's but cannot
be neglected locally.

For some processes, such as QCD jet production, the total cross section
$\stot$ is not defined without a cut; for example one may require
the total amount of transverse energy to be bigger than a given quantity.
The formulation of the $\MCatNLO$ described in this paper should remain
valid in the presence of such a cut.

In processes involving incoming photons, pointlike and hadronic components
needs to be defined in a given subtraction scheme. Within the NLO
approximation, a change of scheme gives rise to terms that are of higher
order. However, for consistency the same scheme should be adopted for the
generation of the NLO cross section and the backward evolution of the
parton shower in the MC. We also point out that, in order for the
procedure described in this paper to be applicable, the photons should
have a continuous energy spectrum, such as that from an incoming electron
beam, in which the photon energies are distributed according to the
Weizs\"acker-Williams function.

\section{Conclusions and outlook}\label{sec:conc}

We have presented a practical proposal for constructing an event
generator which, we believe, displays the main advantages of the
NLO and MC approaches.  For brevity we call this $\MCatNLO$.
A key ingredient is the use of a modified
subtraction method for dealing with infrared and collinear
singularities.  The method has been illustrated using a toy model and
implemented in full for the process of hadroproduction of W boson
pairs.  The results there look encouraging (although of course
comparisons with data are for the future): all the observables
investigated show the expected behaviour in regions where
the NLO or MC results should be most reliable, with a smooth
transition between hard and soft/collinear regions.

The method generates a fraction of events with negative weights,
but in the cases we have studied the fraction is small and unweighted
(signed) event generation remains rather efficient.  One can 
reduce the number of negative weights for a given process by
tuning the parameters specific to that process (these parameters
have nothing to do with the global MC parameters, which are tuned
once and for all by comparison to data). We shall consider in future
upgrades of the implementation of our method the strategies outlined
in sect.~\ref{sec:impl} and in sect.~\ref{sec:generalities}, which
should reduce the number of $\clS$ and $\clH$ negative-weight events
respectively, using an intermediate integration step, and a modification
of the showering algorithm. One could also consider more exotic possibilities,
such as defining ``superevents'' by combining configurations that are
close together in three-body phase space.

The implementation presented here is based on the \HW\ MC generator,
but there appears no reason why other parton shower generators such as
{\small PYTHIA} should not be used.  The main points to be reconsidered
when changing the MC generator will be the dead zones for hard emission,
the matching of local counterterms in the soft region, and the relation
between the 2- and 3-body configurations connected by the backward
evolution in the showering initiated by initial-state partons
(corresponding to the event projection that we used in this paper).

\acknowledgments
We wish to thank CERN TH division for hospitality during
the preparation of this work. Many informative discussions with
S.~Catani, G.~Corcella, M.~Dobbs, J.-Ph.~Guillet, Z.~Kunszt,
M.~Mangano, P.~Nason, R.~Pittau, G.~Ridolfi, and M.~Seymour are gratefully
acknowledged.  We also thank S.~Catani, J. Collins, and M.~Dobbs for 
comments on the manuscript.

\appendix

\section{\boldmath W$^+$W$^-$ cross sections}\label{app:WWxsecs}
In this appendix, we give all the details relevant to cross sections
at ${\cal O}(\as^0)$ and ${\cal O}(\as)$. Although we study
explicitly the case of W$^+$W$^-$ production, some of the results
are based on fairly general ideas, which can be applied to any
other hard production process.

\subsection{Kinematics}\label{app:WWkin}
We start by defining some kinematic variables for W$^+$W$^-$ production.
The momenta in the $2\to 3$ partonic processes are assigned as follows:
\beq
a(p_1)+b(p_2)\;\longrightarrow\,c(k)+{\rm W^+W^-}(\kww)\,.
\label{partproc}
\eeq
Following ref.~\cite{Frixione:1993yp}, we introduce the quantities
\beq
x=\mwwt/s,\;\;\;\;\;\;y=\cos\theta,
\label{xydef}
\eeq
where $s=(p_1+p_2)^2$, $M_{\sss WW}$ is the invariant mass of the W
pair, and $\theta$ is the scattering angle of the outgoing light
parton in the c.m.\ frame of the colliding partons. In this frame,
$1-x$ is the energy fraction of the outgoing light parton (in units of
$\frac{1}{2}\sqrt{s}$).  These quantities unambiguously identify the
soft and initial-state collinear regions of the three-body phase
space. In particular
\beqn
&&x=1\;\;\;\Rightarrow\;\;\;{\rm soft}~(k^0=0)\,,
\label{softL}
\\
&&y=1\;\;\;\Rightarrow\;\;\;{\rm collinear}~(\vec{k}\parallel\vec{p}_1)\,,
\label{collLp}
\\
&&y=-1\;\;\;\Rightarrow\;\;\;{\rm collinear}~(\vec{k}\parallel\vec{p}_2)\,.
\label{collLm}
\eeqn
The ranges allowed for $x$ and $y$ are
\beq
\rho\le x\le 1\,,\;\;\;\;\;\;-1\le y\le 1\,,
\eeq
where $\rho=4\mwt/s$. Defining
\beq
\beta=\sqrt{1-\rho}\,,\;\;\;\;\;\;
\beta_x=\sqrt{1-\rho/x}\,,
\eeq
the two-body and three-body phase spaces are
\beqn
d\phi_2(s)&=&\frac{\beta}{16\pi}d\cos\thw\,,
\\
d\phi_2(xs)&=&\frac{\beta_x}{16\pi}d\cos\thw\,,
\\
d\phi_3(s)&=&\frac{s\beta_x}{512\pi^4}(1-x)dx\,dy\,d\cos\thw\,d\varphi\,,
\label{tbps}
\eeqn
where $\thw$ and $\varphi$ are the polar and azimuthal angles of the W$^+$
in the W$^+$W$^-$ c.m.\ frame. The normalization in eq.~(\ref{tbps}) is such 
that both $\thw$ and $\varphi$ range between 0 and $\pi$.

\subsection{Standard subtraction}\label{app:SSub}
Here we give the subtraction terms using the definition
of ref.~\cite{Frixione:1993yp}. The quantity ${\cal Q}_{ab}$ appearing
in eq.~(\ref{sigsv}) reads
\beq
{\cal Q}_{q\bar{q}}=\frac{\CF\as}{4\pi}\left(
\log\frac{s}{\mu^2}(6+16\log\tilde{\beta})
+32\log^2\tilde{\beta}-\frac{4}{3}\pi^2\right)\,,
\eeq
and ${\cal Q}_{ab}=0$ for $\{a,b\}\ne\{q,\bar{q}\}$.
We have introduced a free parameter $\tilde{\rho}$ such that
\beq
\rho\le\tilde{\rho} <1\,,\;\;\;\;\;\;
\tilde{\beta}=\sqrt{1-\tilde{\rho}}\,.
\eeq
The collinear terms of eq.~(\ref{sigcpm}) read
\beqn
d\hat{\sigma}_{ab}^{(c+)}(p_1,p_2)&=&\frac{\as}{2\pi}\Bigg\{
\Bigg[\log\frac{\omega s}{2\mu^2}\omxrhot
+2\lomxrhot\Bigg](1-x)P_{ca}^{(0)}(x)
\nonumber \\*&&\phantom{aaaa}
-P_{ca}^{(1)}(x)\Bigg\}\mborn_{cb}(xp_1,p_2) d\phi_2(xs)dx\,,
\label{Asigcp}
\eeqn
and analogously for $d\hat{\sigma}_{ab}^{(c-)}$. The real emission
contribution of eq.~(\ref{sigreal}) reads
\beqn
d\hat{\sigma}_{ab}^{(f)}(p_1,p_2)&=&
\frac{1}{2}\omxrhot\left[\omyom + \opyom\right]
\nonumber \\*&&\times
(1-x) (1-y^2)\mreal(p_1,p_2)d\phi_3(s)\,.
\label{Asigreal}
\eeqn
The distributions introduced in eqs.~(\ref{Asigcp}) and~(\ref{Asigreal})
define the subtraction terms as follows:
\beqn
\int_\rho^1 dx\,h(x)\omxrhot&=&
\int_\rho^1 dx\,\frac{h(x)-\stepf(x-\tilde{\rho})h(1)}{1-x},
\label{distonessub}
\\
\int_\rho^1 dx\,h(x)\lomxrhot&=&
\int_\rho^1 dx\,\frac{h(x)-\stepf(x-\tilde{\rho})h(1)}{1-x}
\log(1-x),
\label{disttwossub}
\\
\int_{-1}^1 dy\,g(y)\opmyom&=&
\int_{-1}^1 dy\,\frac{g(y)-\stepf(\mp y-1+\omega)g(\mp 1)}{1\pm y},
\label{distthreessub}
\eeqn
where $0<\omega\le 2$, and $h$ and $g$ are generic regular functions.
It is clear that a variation of the (arbitrary) parameters $\tilde{\rho}$
and $\omega$ amounts to a finite redefinition of the soft and
collinear counterterms respectively. The cross section is strictly
independent of the values of $\tilde{\rho}$ and $\omega$ in their
whole range (i.e., these parameters do not need to be small, as opposed
to the slicing method).

\subsection[$\zeta$--subtraction]{\boldmath $\zeta$--subtraction}
\label{app:zetaSub}
We give here an alternative form of the subtraction terms,
which has been used in our implementation of the $\MCatNLO$. 
Although the results of the $\MCatNLO$ do not 
depend on the particular form adopted for the subtraction terms, we
find that the form introduced in this section gives a superior numerical
stability, in a sense that, in the context of our $\MCatNLO$, allows
to reduce the number of negative-weight events. See sect.~\ref{sec:impl}
for more details. We define
\beq
\pfun(x,y)=(1-x)^2\,(1-y^2)\,,
\label{pfundef}
\eeq
where $x$ and $y$ are defined in eq.~(\ref{xydef}). The function $\pfun$
has a simple physical meaning; in fact
\beq
\ptww=\frac{\sqrt{s}}{2}\sqrt{\pfun(x,y)}\,,
\eeq
where
$\ptww$ is the transverse momentum of the W pair, which coincides
with that of the outgoing light parton at NLO. The $\zeta$--%
subtraction basically amounts to subtracting the counterterms when the 
condition
\beq
\pfun(x,y)\,<\,\zeta
\label{zetaregion}
\eeq
is met, $\zeta$ being a numerical parameter, $0<\zeta\le 1$.
More precisely, we define
\beqn
\int_\rho^1 dx\,h(x)\omxP&=&
\int_\rho^1 dx\,\frac{h(x)-\stepf(\zeta-\pfun(x,y))h(1)}{1-x},
\label{dPxdef}
\\
\int_\rho^1 dx\,h(x)\FomxP&=&
\int_\rho^1 dx\,\frac{h(x)-\stepf(\zeta-\pfun(x,y))h(1)}{1-x}
F(x),\phantom{aaaa}
\label{dPlxdef}
\\
\int_{-1}^1 dy\,g(y)\opmyP&=&
\int_{-1}^1 dy\,\frac{g(y)-\stepf(\zeta-\pfun(x,y))\stepf(\mp y)
g(\mp 1)}{1\pm y},
\label{dPydef}
\eeqn
where $y$ in eqs.~(\ref{dPxdef}), (\ref{dPlxdef}), and $x$ in
eq.~(\ref{dPydef}), are held fixed. We have
\beqn
&&{\cal Q}_{q\bar{q}}=\frac{\CF\as}{4\pi}\left(
\log\frac{s}{\mu^2}(6+16\log\beta)
+32\log^2\beta-\frac{4}{3}\pi^2
+\Fsoft\stepf(\by)\right)\,,
\\
&&d\hat{\sigma}_{ab}^{(c+)}(p_1,p_2)=\frac{\as}{2\pi}\Bigg\{
\Bigg[\log\frac{s}{2\mu^2}\omxP
+\left(\frac{\log(1-\Fcoll(x))}{1-x}\right)_{\pfun}
\nonumber \\*&&\phantom{a}
+2\lomxP\Bigg](1-x)P_{ca}^{(0)}(x)
-P_{ca}^{(1)}(x)\Bigg\}\mborn_{cb}(xp_1,p_2) d\phi_2(xs)dx\,,\phantom{aaa}
\\
&&d\hat{\sigma}_{ab}^{(f)}(p_1,p_2)=
\frac{1}{2}\omxP\left[\omyP + \opyP\right]
\nonumber \\*&&\phantom{d\hat{\sigma}_{ab}^{(f)}(p_1,p_2)=}
\times(1-x) (1-y^2)\mreal(p_1,p_2)d\phi_3(s)\,,
\eeqn
where
\beqn
\by=\left\{
\begin{array}{ll}
\sqrt{1-\zeta/\beta^4} &\phantom{aaaa} \sqrt{\zeta}<1-\rho\,,\\
0                     &\phantom{aaaa} \sqrt{\zeta}\ge 1-\rho\,,
\end{array}
\right.
\label{bydef}
\eeqn
and
\beqn
\Fsoft&=&4\log\frac{\zeta}{\beta^4}\log\frac{1+\by}{1-\by}
-2\log\frac{1+\by}{1-\by}\log(4(1-\by^2))
\nonumber \\*&&
+4{\rm Li}_2\left(\frac{1+\by}{2}\right)
-4{\rm Li}_2\left(\frac{1-\by}{2}\right)\,,
\\
\Fcoll(x)&=&\left\{
\begin{array}{ll}
\sqrt{1-\zeta/(1-x)^2} &\phantom{aaaa} x<1-\sqrt{\zeta}\,,\\
0                     &\phantom{aaaa} x\ge 1-\sqrt{\zeta}\,.
\end{array}
\right.
\eeqn
We point out that the subtraction scheme defined in this appendix
is rather general. In fact, the condition that determines the subtraction,
eq.~(\ref{zetaregion}), applies to the variables of the parton that
recoils against the system, regardless of the nature of the system itself.

We can now go back to the point raised in sect.~\ref{sec:impl}, concerning
the number of negative-weight events in our $\MCatNLO$. As mention there,
this number is a (very complicated) function of the free parameters
used to define the subtractions terms: $\tilde{\rho}$ and $\omega$
for standard subtraction, $\zeta$ for $\zeta$--subtraction. However, 
for a crude estimate of their best values, we can 
assume that the partonic cross sections do not depend
upon them, and thus that the parameter dependence of $J_{\clS}$ is
only due to the size of the phase-space regions in which the subtraction
of the counterterms is performed (this amounts to saying that the only
parameter dependence taken into account is that due to the $\stepf$
functions in eqs.~(\ref{distonessub})--(\ref{distthreessub}) for
the standard subtraction, and in eqs.~(\ref{dPxdef})--(\ref{dPydef})
for the $\zeta$--subtraction). In general, there is a cancellation
between the first two terms in eq.~(\ref{Itwodef}); to a certain
extent, the term \mbox{$-d\bSigma_{ab}^{(f)}/d\phi_3\cnt$}
can be tuned in order to cancel with 
\mbox{$d\bSigma_{ab}/d\phi_3\xMC$}. This explains why 
$\zeta$-subtraction is a better choice than standard subtraction
in the present case: the region in which the counterterms are
subtracted, eq.~(\ref{zetaregion}), is a better match to the 
dead zone, eq.~(\ref{DZdef}), compared with the corresponding
region for the standard subtraction.

In this paper we do not try to find an optimal value for the parameter
$\zeta$. The crude estimate mentioned above gives $\zeta=0.275$; the
fraction of negative-weight events that we get by using this value
is 12\%. This fraction appears to be remarkably stable against 
variations of $\zeta$ in the region $0.05\lesssim\zeta\lesssim 0.5$.

\subsection[Choice of Bjorken $x$'s]{\boldmath Choice of
Bjorken $x$'s\label{app:proj}}
Here we give the form of the change of variables in eq.~(\ref{xxtozz}). 
We stress that this is a convenient procedure for getting rid of the 
$\Kqtwo$ configurations that appear in eq.~(\ref{WWnlosubtint}), which 
also has the virtue of being useful for defining the MC-like terms of 
eq.~(\ref{QCDaddsub}), as we shall show in app.~\ref{app:DZ}.
These two problems could in principle be solved independently, and
by different means, as it may be necessary in other MC implementations; 
however, our master formula, eq.~(\ref{rMCatNLO}), would still hold. 
We denote by
\beq
\Kthree(x_1,x_2,\phi_3)
\label{thrbdconf}
\eeq
the three-body configuration that is obtained by using the three-body
phase-space variables $\phi_3$, and the momentum fractions $x_1$ and $x_2$;
the momenta in eq.~(\ref{thrbdconf}) are understood in the hadronic 
c.m.\ frame. Furthermore, we denote by
\beq
\Ktwo_s(x_1,x_2,\phi_3^{(s)})\equiv\Kthree(x_1,x_2,\phi_3^{(s)})
\label{ktwoLs}
\eeq
the two-body configuration that is obtained by setting to zero the 
energy of the final-state light parton; $\phi_3^{(s)}$ means ``soft limit''
of $\phi_3$. In the parametrization given in eq.~(\ref{tbps}), this
limit can be obtained by letting $x\to 1$ (see 
eq.~(\ref{softL})). It has to be pointed out that
any $\Ktwo$ configuration appearing in eq.~(\ref{WWnlosubtint}) can
be viewed as the soft limit of a $\Kthree$ configuration; thus, 
eq.~(\ref{ktwoLs}) is fully general. Analogously, we denote by
\beqn
\Kqtwo_{c+}(x_1,x_2,\phi_3^{(c+)})&\equiv&\Kthree(x_1,x_2,\phi_3^{(c+)})
\label{kqtwoLcp}
\\
\Kqtwo_{c-}(x_1,x_2,\phi_3^{(c-)})&\equiv&\Kthree(x_1,x_2,\phi_3^{(c-)})
\label{kqtwoLcm}
\eeqn
the three-body configurations that are obtained when the final-state 
light parton becomes collinear to the incoming parton coming from the left
($p_1\parallel k$) and from the right ($p_2\parallel k$) respectively; 
these limits correspond to $y\to 1$ and $y\to -1$ respectively 
(see eqs.~(\ref{collLp}) and~(\ref{collLm})).

We now look for changes of variables of the type given
in eq.~(\ref{xxtozz}). There, the possibility exists to have a 
different form for each of the terms contributing to eq.~(\ref{WWnlosubtint}).
However, since we are looking for a solution to the equation
$\Ktwo(x_i)=\Kqtwo(\tilde{x}_i)$, we shall actually have to find
a change of variables~(\ref{xxtozz}) for each kinematical configuration
$\Ktwo$ and $\Kqtwo$. Since only kinematics matters here, it follows
that the collinear counterterms to $d\Sigma_{ab}^{(f)}\ev$, and 
the terms $d\Sigma_{ab}^{(c\pm)}\ev$, must be treated exactly in the same 
way in eq.~(\ref{WWnlosubtint}), since they all have $\Kqtwo$ kinematics.
Similarly, the Born term $d\bSigma_{ab}^{(b)}$, the soft-virtual term 
$d\bSigma_{ab}^{(sv)}$, the soft counterterm to $d\Sigma_{ab}^{(f)}\ev$, and 
the counterterms $d\Sigma_{ab}^{(c\pm)}\cnt$, are treated in the same way,
since they have $\Ktwo$ kinematics. In summary, our task amounts to
finding the three changes of variables
\beq
x_1=x_1^{(\alpha)}(z_1,z_2,\phi_3)\,,\;\;\;\;
x_2=x_2^{(\alpha)}(z_1,z_2,\phi_3)\,,\;\;\;\;
\alpha=s, c+, c-\,,
\eeq
such that, at {\em fixed} $z_1$, $z_2$ and $\phi_3$,
\beqn
&&\Ktwo_s(x_1^{(s)}(z_1,z_2,\phi_3),x_2^{(s)}(z_1,z_2,\phi_3),\phi_3^{(s)})
\nonumber\\*&&\phantom{\Ktwo_s(x_1^{(s)}}
=\Kqtwo_{c+}(x_1^{(c+)}(z_1,z_2,\phi_3),x_2^{(c+)}(z_1,z_2,\phi_3),
\phi_3^{(c+)})
\nonumber\\*&&\phantom{\Ktwo_s(x_1^{(s)}}
=\Kqtwo_{c-}(x_1^{(c-)}(z_1,z_2,\phi_3),x_2^{(c-)}(z_1,z_2,\phi_3),
\phi_3^{(c-)})\,,
\label{ktwoeqkqtwo}
\eeqn
where the equality holds for the W$^+$ and W$^-$ variables only.
It is then apparent that eqs.~(\ref{WWnlosubtint}) and~(\ref{WWevproj})
are identical; in fact, if $\Ktwo=\Kqtwo$ for the W$^+$ and W$^-$
variables, then $O(\Ktwo)=O(\Kqtwo)$, since the outgoing light
parton cannot contribute to the kinematics of any infrared-safe
observables, being either soft or collinear to the beam line.

The requirement $\Ktwo=\Kqtwo$ amounts to two conditions, since
in both configurations the transverse momentum of the W pair is 
equal to zero, and only two components of the four-momentum
of the pair are left to be constrained. Notice that it is sufficient
to constrain the pair momentum, since the internal variables of the
pair are held fixed in eq.~(\ref{ktwoeqkqtwo}). Instead of imposing
eq.~(\ref{ktwoeqkqtwo}) directly, it turns out to be convenient
to follow the strategy developed in ref.~\cite{Mangano:jk}
(where the procedure described here is called
``event projection''), where $\Ktwo$ and $\Kqtwo$ are related
to $\Kthree$. This is done by imposing 
\beqn
O_1(\Kthree(z_1,z_2,\phi_3))&=&
O_1(\Ktwo_s(x_1^{(s)},x_2^{(s)},\phi_3^{(s)}))\,,
\label{Oonecond}
\\
O_2(\Kthree(z_1,z_2,\phi_3))&=&
O_2(\Ktwo_s(x_1^{(s)},x_2^{(s)},\phi_3^{(s)}))\,,
\label{Otwocond}
\eeqn
and analogous equations for $c+$ and $c-$; $O_1$ and $O_2$ are two
arbitrary observables, defined with the variables of the W$^+$W$^-$
pair; therefore, the conditions in eq.~(\ref{ktwoeqkqtwo}) are 
effectively implemented by relating $\Ktwo$ and $\Kqtwo$ to $\Kthree$.
It must be stressed that eqs.~(\ref{Oonecond}) and~(\ref{Otwocond}) are
imposed in the hadronic c.m.\ frame for those observables that are
not invariant under longitudinal boosts.

Notice that this procedure can be applied to any production
process; in the general case, the observables $O_1$ and $O_2$ are
defined with the variables of the {\em system} that recoils against 
the hard parton emitted by real diagrams at the NLO level.
In the present case, one obvious choice is
\beq
O_1=\mwwt\,,\;\;\;\;\;\;\;\;
O_2=\yww\,,
\label{Ochone}
\eeq
where $\mwwt$ and $\yww$ are the invariant mass squared and the rapidity
of the pair respectively. These two observables are easily expressed in
terms of the momentum fractions $x_{1,2}$, and of the phase space variables 
in the parametrization given in eq.~(\ref{tbps}):
\beqn
&&\mwwt(\Kthree(z_1,z_2,\phi_3))\equiv
\mwwt(z_1,z_2,x)=x\,z_1\,z_2\,S\,,
\label{mwwexpl}
\\
&&\yww(\Kthree(z_1,z_2,\phi_3))\equiv
\yww(z_1,z_2,x,y)=\log\xi + \frac{1}{2}\log\frac{z_1}{z_2}\,,
\label{ywwexpl}
\eeqn
where $S$ is the hadronic c.m.\ energy squared, and
\beq
\xi=\sqrt{\frac{2-(1-x)(1+y)}{2-(1-x)(1-y)}}\,\,.
\eeq
By imposing the conditions in eqs.~(\ref{Oonecond}) and~(\ref{Otwocond}), and
using the explicit forms for $\mwwt$ and $\yww$ given in eqs.~(\ref{mwwexpl})
and~(\ref{ywwexpl}), we get the following:
\beqn
&&
x_1^{(s)}=z_1\,\xi\,\sqrt{x}\,,\phantom{aaaaaa}
x_2^{(s)}=z_2\,\frac{\sqrt{x}}{\xi}\,,
\label{xtils}
\\&&
x_1^{(c+)}=x_1^{(s)}/x\,,\phantom{aaaaaa}
x_2^{(c+)}=x_2^{(s)}\,,
\label{xtilcp}
\\&&
x_1^{(c-)}=x_1^{(s)}\,,\phantom{aaaaaaaa}
x_2^{(c-)}=x_2^{(s)}/x\,.
\label{xtilcm}
\eeqn
It should be stressed that $x_1^{(\alpha)}$ and $x_2^{(\alpha)}$ as 
given in eqs.~(\ref{xtils}), (\ref{xtilcp}), and~(\ref{xtilcm}) do
not need to be less than 1 any longer; thus, the parton densities
contained in $\bSigma_{ab}$ have to be defined as equal to zero if
their arguments are larger than 1.

Although the choice made in eq.~(\ref{Ochone}) is arbitrary, we shall
show later that, in the context of our $\MCatNLO$, it is actually
dictated by the procedure used by the MC to generate the Bjorken $x$'s.
In this procedure, the MC typically constrains either the invariant
mass and the rapidity of the pair, or the invariant mass and the
longitudinal momentum of the pair. In the current version of \HW\,
the latter choice is adopted. We therefore also consider
\beq
O_1=\mwwt\,,\;\;\;\;\;\;\;\;
O_2=\kww\,,
\label{Ochtwo}
\eeq
where $\kww$ is the longitudinal momentum of the pair. With the
parametrization of eq.~(\ref{tbps}), we get
\beq
\kww(\Kthree(z_1,z_2,\phi_3))\equiv
\kww(z_1,z_2,x,y)=-\frac{\sqrt{S}}{4}\left[(1-x)y(z_1+z_2)
+(1+x)(z_2-z_1)\right].
\label{kwwexpl}
\eeq
By imposing again eqs.~(\ref{Oonecond}) and~(\ref{Otwocond}), we now obtain
\beqn
&&
x_1^{(s)}=\frac{1}{2}\left(\sqrt{B^2+4A}-B\right)\,,\phantom{aaaaaa}
x_2^{(s)}=\frac{1}{2}\left(\sqrt{B^2+4A}+B\right)\,,
\label{xtilskww}
\\&&
x_1^{(c+)}=x_1^{(s)}/x\,,\phantom{aaaaaaaaaaaaaaaaa}
x_2^{(c+)}=x_2^{(s)}\,,
\label{xtilcpkww}
\\&&
x_1^{(c-)}=x_1^{(s)}\,,\phantom{aaaaaaaaaaaaaaaaaaa}
x_2^{(c-)}=x_2^{(s)}/x\,,
\label{xtilcmkww}
\eeqn
where
\beqn
A&=&x\,z_1\,z_2\,,
\\
B&=&\frac{1}{2}\left[(1-x)y(z_1+z_2)
+(1+x)(z_2-z_1)\right]\,.
\eeqn
Note that eq.~(\ref{xtilcpkww}) is identical to eq.~(\ref{xtilcp}),
and eq.~(\ref{xtilcmkww}) is identical to eq.~(\ref{xtilcm}). This gives
a consistent picture when the results given above are used in the $\MCatNLO$,
eq.~(\ref{rMCatNLO}). In fact, event projection has been introduced
in the context of an NLO computation, and the constraints have been
imposed on final-state quantities only. However, in an $\MCatNLO$
the initial-state configuration of the hard process is relevant as well,
since it defines the initial conditions for the MC evolutions implicit in
$\IMC(O,\Kthree)$ and $\IMC(O,\Ktwo)$. Eqs.~(\ref{xtils}),
(\ref{xtilcp}), and~(\ref{xtilcm}) [and eqs.~(\ref{xtilskww}),
(\ref{xtilcpkww}), and~(\ref{xtilcmkww})] 
all enter the definition of $\clS$ events, and since in general
$x_i^{(s)}\ne x_i^{(c+)}\ne x_i^{(c-)}$, there is an ambiguity for the
prescription of the momenta of the initial state partons that initiate
the backward evolution. However, as explained in sect.~\ref{sec:impl},
the initial conditions for class $\clS$ events are the $2\to 2$ 
hard processes; the initial states of these processes coincide, by
definition, with those of the $2\to 2$ contributions to $\clS$
events, but do {\em not} coincide in general with those of the quasi-$2\to 2$ 
contributions. If we denote the momentum fractions of the initial-state
partons entering the $2\to 2$ processes as $x_1^\prime$ and $x_2^\prime$, 
owing to the fact that $1-x$ is the energy fraction of the outgoing light
parton (see app.~\ref{app:WWkin}), we have
\beqn
&&
x_1^{(s)^\prime}=x_1^{(s)}\,,\phantom{aaaaaaaaaaaaaaa}
x_2^{(s)^\prime}=x_2^{(s)}\,,
\label{xptils}
\\&&
x_1^{(c+)^\prime}\equiv x\,x_1^{(c+)}=x_1^{(s)}\,,\phantom{aaaaaa}
x_2^{(c+)^\prime}\equiv x_2^{(c+)}=x_2^{(s)}\,,
\label{xptilcp}
\\&&
x_1^{(c-)^\prime}\equiv x_1^{(c-)}=x_1^{(s)}\,,\phantom{aaaaaaa}\;
x_2^{(c-)^\prime}\equiv x\,x_2^{(c-)}=x_2^{(s)}\,.
\label{xptilcm}
\eeqn
Thus, $x_i^{(s)^\prime}=x_i^{(c+)^\prime}=x_i^{(c-)^\prime}$, and there
is no ambiguity in the momenta of the initial state partons. It should
be noted that an ambiguity here would in any case affect our results
at ${\cal O}(\as^2)$, since it would only be due to quasi-$2\to 2$
contributions; however, the absence of any ambiguity fits nicely
in a picture where MC and NLO give different descriptions of the
same physics.

In appendix~\ref{app:DZ}, we shall need to consider the soft and
collinear limits of $x_i^{(s)}$ and $x_i^{(c\pm)}$. By direct
computation, we obtain the following
\beqn
x\to 1&&\phantom{aaa}\Longrightarrow\phantom{aaa}\left\{
\begin{array}{ll}
\xos\to z_1\,,\phantom{aaaa}&\xts\to z_2\,,\\
\xocp\to z_1\,,&\xtcp\to z_2\,,\\
\xocm\to z_1\,,&\xtcm\to z_2\,,\\
\end{array}
\right.
\label{xxlims}
\\
y\to 1&&\phantom{aaa}\Longrightarrow\phantom{aaa}\left\{
\begin{array}{ll}
\xos\to xz_1\,,\phantom{aaaa}&\xts\to z_2\,,\\
\xocp\to z_1\,,&\xtcp\to z_2\,,\\
\end{array}
\right.
\label{xxlimcp}
\\
y\to -1&&\phantom{aaa}\Longrightarrow\phantom{aaa}\left\{
\begin{array}{ll}
\xos\to z_1\,,\phantom{aaaa}&\xts\to xz_2\,,\\
\xocm\to z_1\,,&\xtcm\to z_2\,,\\
\end{array}
\right.
\label{xxlimcm}
\eeqn
where we did not consider the $y\to 1$ limit of $x_i^{(c-)}$ and
the $y\to -1$ limit of $x_i^{(c+)}$, since these limits are not 
relevant in the computation of singular cross sections. We stress
that eqs.~(\ref{xxlims})--(\ref{xxlimcm}) hold regardless of
whether eq.~(\ref{Ochone}) or eq.~(\ref{Ochtwo}) has been used.
This is a manifestation of the factorization properties of QCD,
as will become clear in app.~\ref{app:DZ}.

\subsection{Dead zone and subtraction term\label{app:DZ}}
Here we define the quantity $d\bSigma_{ab}\xMC$, introduced 
in eq.~(\ref{MCexp}). We make use of some of the results of 
ref.~\cite{Corcella:1998rs}; although ref.~\cite{Corcella:1998rs} 
concerns only single vector boson production, the kinematics carry 
over to the present case if we simply replace the momentum of the
vector boson by that of the pair. According to the discussion of 
sect.~\ref{sec:generalization}, we have to consider the cross section
that we would obtain by keeping the ${\cal O}(\as)$ term in the
expansion of the result of the MC, which we denote by $d\sigma\xMC$. 
Roughly speaking, $d\sigma\xMC$ is equal to the Born cross section, 
times the ${\cal O}(\as)$ term in the expansion of the Sudakov
form factor. More precisely
\beqn
d\sigma\xMCB&=&\sum_{abc}d\xoMC d\xtMC \frac{\as}{2\pi}
d\sigma_{ab}^{(b)}(\xoMC P_1,\xtMC P_2)
\nonumber \\*&&\phantom{aa}\times
\Bigg(
\frac{d\xi_+}{\xi_+}\frac{dz_+}{z_+}
\stepf(z_+^2-\xi_+)P^{(0)}_{ac}(z_+)f_c^{\Hone}(\xoMC/z_+)f_b^{\Htwo}(\xtMC)
\nonumber \\*&&\phantom{aaa}
+\frac{d\xi_-}{\xi_-}\frac{dz_-}{z_-}
\stepf(z_-^2-\xi_-)P^{(0)}_{bc}(z_-)f_a^{\Hone}(\xoMC)f_c^{\Htwo}(\xtMC/z_-)
\Bigg)\,.\phantom{aa}
\label{sigmaMC}
\eeqn
Equation~(\ref{sigmaMC}) has been obtained directly from the QCD Sudakov 
form factor; in doing so, we have eliminated any cutoffs, precisely as in 
the case of the toy model. This implies that eq.~(\ref{sigmaMC}) has
soft and collinear divergences, which is what we want, since from it 
we derive $d\bSigma_{ab}\xMC$, namely a local counterterm to the
real emission matrix element.
The two terms on the r.h.s.\ account for the emissions from 
the incoming partons. If strongly-interacting partons were present in the
final state at the Born level, corresponding terms should be added.
The variables $\xi_\pm$ and $z_\pm$ are those used by the
\HW\ MC~\cite{Marchesini:1988cf} in the parton showering process.
The $\stepf$ functions define the region where an emission can occur
during the shower; therefore, from eq.~(\ref{sigmaMC}) we can
easily read the form of the dead zone, which is by definition the
region where {\em no emission} can occur. We have
\beq
{\rm Dead~zone\,:}\;\;\;\;
\xi_+>z_+^2\,,\;\;\;\;\;\;\xi_->z_-^2\,,
\label{DZxiz}
\eeq
where both conditions must be fulfilled. Finally, the $1/z_\pm$ factor
that appears in the arguments of the parton densities in eq.~(\ref{sigmaMC})
is due to the backward evolution in the shower from initial-state
partons. In the case of emission from final-state partons, the arguments
of the parton densities would simply read $\xoMC$ and $\xtMC$.

The values of $\xoMC$ and $\xtMC$ are determined by the MC (through
energy-momentum conservation) at the Born level, after the generation 
of the $\Ktwo$ configuration which is used as initial condition for
the shower. At the end of the shower, the momenta of the final-state
particles are boosted, the boost being determined by imposing a relation
between the variables of the W pair (in general, of the system produced
at the Born level) before and after the shower. It follows that, after the 
shower, $\xoMC$ and $\xtMC$ have still the values determined by the $\Ktwo$ 
configuration; however, after the shower $\xoMC$ and $\xtMC$ cannot be 
interpreted any longer as Bjorken $x$'s associated with the incoming partons.

This procedure is identical (when we consider the ${\cal O}(\as)$ result,
that is, the emission of a single parton) to the one we defined for
event projection in app.~\ref{app:proj}. In other words, the $\Kthree$
configuration implicit in the l.h.s.\ of eq.~(\ref{sigmaMC}) is related
to the $\Ktwo$ configuration that initiates the shower by means of
constraints such as those in eqs.~(\ref{Oonecond}) and~(\ref{Otwocond}).%
\footnote{This is the reason for introducing event projection in
app.~\ref{app:proj}, instead of just imposing eq.~(\ref{ktwoeqkqtwo}).
In a completely general case, one should map the three-body configurations
obtained from the MC onto the $\Kthree$ configurations of the NLO cross
section, thus defining a $d\bSigma_{ab}\xMC$ that acts as a local
counterterm in eq.~(\ref{rMCatNLO}).}
It follows that, if we label the $\Kthree$ configurations according to
the conventions of this paper, \mbox{$\Kthree\equiv\Kthree(z_1,z_2,\phi_3)$},
we have
\beq
\xoMC=x_1^{(s)}\,,\;\;\;\;\;\;\;\;
\xtMC=x_2^{(s)}\,.
\label{xMCeqxs}
\eeq
The explicit form of $x_i^{(s)}$ depends on the observables $O_i$ used
by the MC to implement eqs.~(\ref{Oonecond}) and~(\ref{Otwocond}). A priori,
these need {\em not} coincide with those given in eq.~(\ref{Ochone}) or
eq.~(\ref{Ochtwo}). In practice, \HW\ uses the variables in 
eq.~(\ref{Ochtwo}), and therefore the results in eq.~(\ref{xtilskww})
give the explicit form of $\xoMC$ and $\xtMC$. The relations between
$\xoMC$, $\xtMC$ and $z_1$, $z_2$ allow us to recast eq.~(\ref{sigmaMC})
in the following form
\beq
d\sigma\xMCB=\sum_{ab}dz_1 dz_2 d\bSigma_{ab}\xMCB\,,
\label{sigmaMCt}
\eeq
where
\beqn
d\bSigma_{ab}\xMCB&=&
\frac{1}{z_+}\frac{\partial(\xos,\xts)}{\partial(z_1,z_2)}
f_a^{\Hone}(\xos/z_+)f_b^{\Htwo}(\xts)d\sigma_{ab}^+\xMCB
\nonumber \\*&+&
\frac{1}{z_-}\frac{\partial(\xos,\xts)}{\partial(z_1,z_2)}
f_a^{\Hone}(\xos)f_b^{\Htwo}(\xts/z_-)d\sigma_{ab}^-\xMCB
\label{bSigmaMCdef}
\eeqn
is the quantity that the appears in eq.~(\ref{rMCatNLO}), and 
\beqn
d\sigma_{ab}^+\xMCB&=&\frac{\as}{2\pi}\,
\frac{d\xi_+}{\xi_+}dz_+
\stepf(z_+^2-\xi_+)P^{(0)}_{ca}(z_+)
d\sigma_{cb}^{(b)}(\xos P_1,\xts P_2)\,,
\label{SigmaMCplxiz}
\\
d\sigma_{ab}^-\xMCB&=&\frac{\as}{2\pi}\,
\frac{d\xi_-}{\xi_-}dz_-
\stepf(z_-^2-\xi_-)P^{(0)}_{cb}(z_-)
d\sigma_{ac}^{(b)}(\xos P_1,\xts P_2)\,,
\label{SigmaMCmnxiz}
\eeqn
where the sum over the parton label $c$ is now understood.

In order to prove that $d\bSigma_{ab}\xMC$ in eq.~(\ref{bSigmaMCdef})
can act as a local counterterm in eq.~(\ref{rMCatNLO}), we have to express
it in terms of the variables we use to describe the NLO cross section.
This can be easily done as explained in ref.~\cite{Corcella:1998rs}: we
relate the variables $\xi_\pm$ and $z_\pm$ used in the \HW\ parton 
showering to Lorentz invariant quantities, which we then express in
terms of the phase-space variables of eq.~(\ref{tbps}). Following 
ref.~\cite{Frixione:1993yp}, we define
\beqn
t_k=(p_1-k)^2=-\frac{s}{2}(1-x)(1-y)\,,
\label{tkdef}
\\
u_k=(p_2-k)^2=-\frac{s}{2}(1-x)(1+y)\,,
\label{ukdef}
\eeqn
where the momenta are assigned as in eq.~(\ref{partproc}), and $x$ and $y$ 
have been defined in eq.~(\ref{xydef}). We have~\cite{Corcella:1998rs}
\beqn
t_k &=& -\frac{1-z_+}{z_+^2}\xi_+\mwwt\,,
\label{tkxiz}
\\
\frac{t_k u_k}{s} &=& \frac{(1-z_+)^2}{2z_+^2} \xi_+(2-\xi_+)\mwwt\,,
\label{ukxiz}
\eeqn
for a shower initiated by parton $a$. A shower initiated by the other
incoming parton ($b$) is described in terms of $\xi_-$ and $z_-$, which 
are in turn related to $t_k$ and $u_k$ by the same functional forms given 
in eqs.~(\ref{tkxiz}) and~(\ref{ukxiz}), with the formal exchange 
\mbox{$t_k\leftrightarrow u_k$}.
We can now substitute eqs.~(\ref{xydef}), (\ref{tkdef}), and~(\ref{ukdef}) 
into eqs.~(\ref{tkxiz}) and~(\ref{ukxiz}), to find $\xi_\pm(x,y)$ and 
$z_\pm(x,y)$, whose explicit form we do not report here, since it is
not needed in the following.

We can now find the dead zone boundary in the $\langle x,y\rangle$ 
plane; from eq.~(\ref{DZxiz}) we obtain
\beq
{\rm Dead~zone\,:}\;\;\;\;
\rho\le x\le x_{\sss DZ}\,,\;\;\;\;
\abs{y} < Y_{\sss DZ}(x)\,,
\label{DZdef}
\eeq
where both conditions are met simultaneously, and
\beqn
x_{\sss DZ}&=&\frac{7+\sqrt{17}}{16}\,,
\\
Y_{\sss DZ}(x)&=&1-\frac{x}{1-x}\left(3-\sqrt{1+8x}\right)\,.
\eeqn
We now consider eqs.~(\ref{SigmaMCplxiz}) and~(\ref{SigmaMCmnxiz}).
Using eq.~(\ref{tbps}), we obtain
\beqn
\frac{d\sigma_{ab}^+(p_1,p_2)}{d\phi_3}\xMCBB &=&
\frac{16\pi\as}{s}\,\frac{P^{(0)}_{ca}(z_+)}{(1-x)\xi_+}\,
\frac{\partial(z_+,\xi_+)}{\partial(x,y)}\,
\mborn_{cb}(\xos P_1,\xts P_2)\,\stepf(y-Y_{\sss DZ}(x)),
\nonumber \\*&&
\label{HWxsecpl}
\\
\frac{d\sigma_{ab}^-(p_1,p_2)}{d\phi_3}\xMCBB &=&
\frac{16\pi\as}{s}\,\frac{P^{(0)}_{cb}(z_-)}{(1-x)\xi_-}\,
\frac{\partial(z_-,\xi_-)}{\partial(x,y)}\,
\mborn_{ac}(\xos P_1,\xts P_2)\,\stepf(-y-Y_{\sss DZ}(x)),
\nonumber \\*&&
\label{HWxsecmn}
\eeqn
where $p_i=z_iP_i$ (according to the definition given in 
eq.~(\ref{hatSigmadef})); therefore, $s=z_1 z_2 S$.

We can now consider the soft and collinear limits of $d\bSigma_{ab}\xMC$
given in eq.~(\ref{bSigmaMCdef}). We observe that, owing to the procedure
adopted by the MC to determine $\xoMC$ and $\xtMC$, the luminosities that 
appear in eq.~(\ref{bSigmaMCdef}) have non-trivial soft and collinear limits.
We make use of the fact, obtained by direct computation, that
\beq
\lim_{x\to 1}z_\pm=1\,,\;\;\;\;\;\;
\lim_{y\to 1}z_+=x\,,\;\;\;\;\;\;
\lim_{y\to -1}z_-=x\,.
\eeq
Therefore, using eqs.~(\ref{xxlims})--(\ref{xxlimcm}), we obtain
\beqn
&&
\lim_{x\to 1}f_a^{\Hone}(\xos/z_+)f_b^{\Htwo}(\xts)=
\lim_{x\to 1}f_a^{\Hone}(\xos)f_b^{\Htwo}(\xts/z_-)
\nonumber\\*
&=&
\lim_{y\to 1}f_a^{\Hone}(\xos/z_+)f_b^{\Htwo}(\xts)=
\lim_{y\to -1}f_a^{\Hone}(\xos)f_b^{\Htwo}(\xts/z_-)
\nonumber\\*
&=& f_a^{\Hone}(z_1)f_b^{\Htwo}(z_2)\,,
\label{limLumfour}
\eeqn
where the collinear limits have been considered only for the luminosities
attached to short-distance cross sections which are singular in those
limits. In the soft and collinear limits, $d\bSigma_{ab}\xMC$ will have
to cancel the singularities of $d\bSigma_{ab}^{(f)}$ and 
$d\bSigma_{ab}^{(c\pm)}$. Owing to the event projection procedure
described in app.~\ref{app:proj}, the luminosities attached to the
latter two terms behave as follows in the soft and collinear limits:
\beqn
&&
\lim_{x\to 1}f_a^{\Hone}(\xos)f_b^{\Htwo}(\xts)=
\lim_{y\to 1}f_a^{\Hone}(\xocp)f_b^{\Htwo}(\xtcp)
\nonumber\\*
&=&
\lim_{y\to -1}f_a^{\Hone}(\xocm)f_b^{\Htwo}(\xtcm)=
f_a^{\Hone}(z_1)f_b^{\Htwo}(z_2)\,,
\label{NlimLumfour}
\eeqn
where we used eqs.~(\ref{xxlims})--(\ref{xxlimcm}).
Eqs.~(\ref{limLumfour}) exactly match
eqs.~(\ref{NlimLumfour}). It is also 
straightforward to prove that
\beq
\frac{1}{z_\pm}\frac{\partial(\xos,\xts)}{\partial(z_1,z_2)}
\longrightarrow 1
\eeq
in the soft and collinear limits, and that the corresponding
Jacobians in the NLO cross section have the same behaviour.
This implies that the properties of $d\bSigma_{ab}\xMC$ as a local
counterterm are determined by the behaviour of its short-distance
parts, given in eqs.~(\ref{HWxsecpl}) and~(\ref{HWxsecmn}).

When considering eqs.~(\ref{HWxsecpl}) and~(\ref{HWxsecmn}), we notice
that these formulae are valid for any parton flavours $a$ and $b$.
However, in our case the soft and collinear limits are all non-trivial
only in the case $a=q$, $b=\bar{q}$, with $c=g$; we shall therefore
explicitly discuss only this case. We expect the forms of the leading
soft and collinear singularities of the cross sections given in
eqs.~(\ref{HWxsecpl}) and~(\ref{HWxsecmn}) to be dictated by QCD 
factorization theorem. In other words, we {\em expect} the following
equations to hold
\beqn
&&\frac{d\sigma_{q\bar{q}}^+(p_1,p_2)}{d\phi_3}\xMCBB 
+\frac{d\sigma_{q\bar{q}}^-(p_1,p_2)}{d\phi_3}\xMCBB 
\stackrel{x\to 1}{\longrightarrow}
8\pi\as\CF\frac{p_1\cdot p_2}{p_1\cdot k\,\,p_2\cdot k}
\mborn_{q\bar{q}}(p_1,p_2)
\nonumber
\\* &&
\phantom{\frac{d\sigma_{q\bar{q}}^+(p_1,p_2)}{d\phi_3}\xMCBB 
+\frac{d\sigma_{q\bar{q}}^-(p_1,p_2)}{d\phi_3}\xMCBB}
=\frac{64\pi\as}{s}\frac{\CF}{(1-x)^2(1-y^2)}\mborn_{q\bar{q}}(p_1,p_2)\,,
\phantom{aa}
\label{MCqqsoft}
\\
&&\frac{d\sigma_{q\bar{q}}^+(p_1,p_2)}{d\phi_3}\xMCBB 
\stackrel{y\to 1}{\longrightarrow}
\frac{4\pi\as}{p_1\cdot k}P_{qq}^{(0)}(x)
\mborn_{q\bar{q}}(xp_1,p_2)
\nonumber
\\* &&
\phantom{\frac{d\sigma_{q\bar{q}}^+(p_1,p_2)}{d\phi_3}\xMCBB}
=\frac{16\pi\as}{s}\frac{P_{qq}^{(0)}(x)}{(1-x)\,(1-y)}
\mborn_{q\bar{q}}(xp_1,p_2)\,,
\label{MCqqcollp}
\\
&&\frac{d\sigma_{q\bar{q}}^-(p_1,p_2)}{d\phi_3}\xMCBB 
\stackrel{y\to -1}{\longrightarrow}
\frac{4\pi\as}{p_2\cdot k}P_{qq}^{(0)}(x)
\mborn_{q\bar{q}}(p_1,xp_2)
\nonumber
\\* &&
\phantom{\frac{d\sigma_{q\bar{q}}^-(p_1,p_2)}{d\phi_3}\xMCBB}
=\frac{16\pi\as}{s}\frac{P_{qq}^{(0)}(x)}{(1-x)\,(1+y)}
\mborn_{q\bar{q}}(p_1,xp_2)\,.
\label{MCqqcollm}
\eeqn
Note that in the case of the soft limit we sum the short distance cross
sections, thanks to eq.~(\ref{limLumfour}).
Explicit computations of the collinear limits $y\to\pm 1$
in eqs.~(\ref{HWxsecpl}) and~(\ref{HWxsecmn}) give eqs.~(\ref{MCqqcollp})
and~(\ref{MCqqcollm}). On the other hand, in the soft limit $x\to 1$
we find the following:
\beqn
&&\frac{d\sigma_{q\bar{q}}^+(p_1,p_2)}{d\phi_3}\xMCBB 
+\frac{d\sigma_{q\bar{q}}^-(p_1,p_2)}{d\phi_3}\xMCBB 
\stackrel{x\to 1}{\longrightarrow}
\frac{64\pi\as}{s}\CF\mborn_{q\bar{q}}(p_1,p_2)
\nonumber \\*&&\phantom{aaaaaaaaaaaa}\times
\left(\frac{2\stepf(y+1/3)}{(1-x)^2(1-y)(3+y)}
+\frac{2\stepf(-y+1/3)}{(1-x)^2(1+y)(3-y)}\right)\,.
\phantom{aa}
\label{HWsoft}
\eeqn
It appears therefore that the choice of the showering variables in \HW\
violates the factorization properties of QCD cross sections. However, the
situation is not as bad as it may seem. In fact, the kernel in 
eq.~(\ref{HWsoft}) has the following property:
\beq
\lim_{\varepsilon\to 0}
\int_{-1+\varepsilon}^{1-\varepsilon}dy\,
\left(\frac{2\stepf(y+1/3)}{(1-y)(3+y)}
+\frac{2\stepf(-y+1/3)}{(1+y)(3-y)}
-\frac{1}{1-y^2}\right)=0\,.
\label{KerLim}
\eeq
Eq.~(\ref{KerLim}) implies that, although (in the absence of soft matrix
element corrections~\cite{Corcella:1998rs}) \HW\ generates an incorrect
angular distribution of soft gluons, the total amount of radiated
soft energy is identical to that computed in perturbative QCD. This
is reassuring, and explains why \HW\ correctly describes physical
observables, eq.~(\ref{HWsoft}) notwithstanding: any IR safe observable
is not sensitive to the details of the angular distribution of 
soft gluons. 

Nevertheless, eq.~(\ref{KerLim}) prevents us from using $d\bSigma\xMC$ in 
the construction of the $\MCatNLO$. In fact, $d\bSigma\xMC$ acts as a 
{\em local} counterterm in eq.~(\ref{rMCatNLO}), and its soft behaviour 
(eq.~(\ref{HWsoft})) prevents the necessary cancellation.

This problem can be solved in two different ways. The most 
straightforward procedure would be to define different 
showering variables, that would give short-distance cross sections
with the correct soft limit. However, this would imply a modification
of the existing MC code in one of its crucial parts, the showering
algorithm. As already stressed, in this paper we prefer not to 
modify any substantial part of \HW.
 
The second solution relies on the observation that the angular distribution
of soft gluons is irrelevant to $\MCatNLO$ physical results. We need
a local counterterm to cancel the divergences of the real matrix element,
but in the soft emission region the $\MCatNLO$ cross section is dominated
by the pure MC contribution. On the other hand, for hard MC emission, we
must have full control of the gluon emitted by the MC, since only in this
way will the $\MCatNLO$ result match the NLO result. It follows that we 
can choose a form for $d\Sigma\xMC$ identical to that we get by
using eqs.~(\ref{HWxsecpl}) and~(\ref{HWxsecmn}) 
for non-soft emission, and a form that can act as a local counterterm
for real matrix elements in the soft region. One possibility is
therefore the following
\beqn
\frac{d\sigma_{q\bar{q}}^+(p_1,p_2)}{d\phi_3}\xMCBB &=&
\frac{16\pi\as}{s}\Bigg(\frac{P^{(0)}_{qq}(z_+)}{(1-x)\xi_+}\,
\frac{\partial(z_+,\xi_+)}{\partial(x,y)}\,
\stepf(y-Y_{\sss DZ}(x))\Gfun(x)
\nonumber \\*&&
\phantom{a}
+\frac{P_{qq}^{(0)}(x)}{(1-x)\,(1-y)}(1-\Gfun(x))\Bigg)
\mborn_{cb}(\xos P_1,\xts P_2)\,,
\phantom{aa}
\label{HWCxsecpl}
\\
\frac{d\sigma_{q\bar{q}}^-(p_1,p_2)}{d\phi_3}\xMCBB &=&
\frac{16\pi\as}{s}\Bigg(\frac{P^{(0)}_{qq}(z_-)}{(1-x)\xi_-}\,
\frac{\partial(z_-,\xi_-)}{\partial(x,y)}\,
\stepf(-y-Y_{\sss DZ}(x))\Gfun(x)
\nonumber \\*&&
\phantom{a}
+\frac{P_{qq}^{(0)}(x)}{(1-x)\,(1+y)}(1-\Gfun(x))\Bigg)
\mborn_{ac}(\xos P_1,\xts P_2)\,,
\phantom{aa}
\label{HWCxsecmn}
\eeqn
where $\Gfun$ is a largely arbitrary function, such that
\beq
\lim_{x\to 1}\Gfun(x)=0\,,\;\;\;\;\;\;
\lim_{x\to {x_{\sss DZ}}}\Gfun(x)=1\,.
\eeq
A simple choice is 
\beqn
\Gfun(x)=\frac{(1-T(x))^{2\alpha}}
{T(x)^{2\alpha}+(1-T(x))^{2\alpha}}\,,\;\;
T(x)&=&\left\{
\begin{array}{ll}
(x-\tilde{x}_{\sss DZ})/(1-\tilde{x}_{\sss DZ})
&\phantom{aa} \tilde{x}_{\sss DZ}<x<1\,,\\
0 &\phantom{aa} x<\tilde{x}_{\sss DZ}\,,\\
\end{array}
\right.
\nonumber \\*&&
\label{Gfundef}
\eeqn
where
\beq
\tilde{x}_{\sss DZ}=1-(1-x_{\sss DZ})\beta\,.
\label{txDZ}
\eeq
The parameters in eqs.~(\ref{Gfundef}) and~(\ref{txDZ}) must be
chosen in the following ranges: $\alpha\ge 1$, $0<\beta\le 1$.
The dependence of $\MCatNLO$ results upon these parameters
will show the sensitivity of physical observables to the details
of soft gluon emission.

We point out that the procedure outlined above is not necessary when
the matrix element is not singular in the soft limit. This happens
for the partonic subprocesses $qg\to W^+W^-q$. In such a case,
we simply set $\Gfun\equiv 1$ in eqs.~(\ref{HWCxsecpl})
and~(\ref{HWCxsecmn}).

Finally, it is worth noting that the interesting features 
of eqs.~(\ref{HWsoft}) and~(\ref{KerLim}) are of a general nature, 
since the hard production process is factorized into $\mborn_{q\bar{q}}$ 
in eq.~(\ref{HWsoft}). Thus, the problem raised here is not specific 
to W$^+$W$^-$ production. We expect that the angular distributions of
soft gluons in \HW\ do not agree with those prescribed by the factorization
theorems of QCD. Therefore, the local counterterm $d\bSigma\xMC$ 
that we need in order to construct the $\MCatNLO$ will not coincide
in general with the ${\cal O}(\as)$ \HW\ result. As in the present 
case, we expect that physical observables will not display any dependence
upon the form chosen for $d\bSigma\xMC$ in the soft limit. In all
cases, we must verify that the total amount of soft gluon energy
given by the MC agrees with the QCD prediction, i.e., that the analogue
of eq.~(\ref{KerLim}) holds.

\section{\boldmath Technicalities of $\MCatNLO$}\label{app:MCatNLO}
In this appendix, we present formal proofs of some of the features
of $\MCatNLO$. We do so in the case of the toy model. However,
we believe that the corresponding properties hold in
the case of QCD as well, as we shall point out explicitly; the 
toy model just allows us to simplify the notation.

\subsection[Construction of $\IMC$]{\boldmath Construction of
$\IMC$\label{app:IMCdef}}
Our first aim is to give a precise definition of
$\IMC(O,\xm(x))$. We identify an MC event with the set of energies
of the $n$ photons emitted in the shower, $\{x_k\}_{k=1}^n$; the
dependence upon $\xm(x)$ is understood. We define an $\MCatNLO$ event as
the corresponding MC event plus the energy of the photon emitted at the
NLO level: $\{x,x_k\}_{k=1}^n$. In the case of soft emission, $x=0$,
the $\MCatNLO$ event simply coincides with the MC event. Any observable
$O$ will be in general a multi-valued function of $\MCatNLO$ events,
$O(\{x,x_k\}_{k=1}^n)=\{O_1,\ldots,O_p\}$, with $1\le p\le n+1$,
the $O_j$'s being values in the physical range of $O$. We then define
\beq
{\cal N}(\{x,x_k\}_{k=1}^n;\overline{O},\delta O)=
\sum_{j=1}^p \stepf(\overline{O}+\delta O/2-O_j)
\stepf(O_j-\overline{O}+\delta O/2),
\eeq
where $\overline{O}$ is any given value that the observable $O$ can assume 
in its physical range. ${\cal N}$ is then just the number of entries in
the bin \mbox{$(\overline{O}-\delta O/2,\overline{O}+\delta O/2)$}
resulting from the $\MCatNLO$ event $\{x,x_k\}_{k=1}^n$. Therefore
\mbox{$0\le {\cal N}\le p\le n+1$}. Starting from a given $\xm(x)$, 
we generate $N$ MC events, and we define
\beq
\IMC(\overline{O},\xm(x))=\lim_{N\to\infty}\lim_{\delta O\to 0}
\frac{1}{N}\frac{1}{\delta O}
\sum_{i=1}^N {\cal N}(\{x,x_k^{(i)}\}_{k=1}^{n_i};\overline{O},\delta O),
\label{IMCwithN}
\eeq
$n_i$ being the number of photons, with energies $x_k^{(i)}$,
emitted in event number $i$.  Thus
in general $\IMC(O,\xm(x))$ is the differential cross section
that one obtains by running the MC starting at $\xm(x)$, defining
$O$ through all photons emitted in the shower, {\em plus} that emitted
at the NLO level, normalized by the total number of MC events generated. 

If we assume that the MC is switched off, then from eq.~(\ref{IMCwithN}) 
we obtain
\beq
\IMC(O,\xm(x))=\delta(O-O(x)),
\label{IMCNLO}
\eeq
$O(x)$ being the single-valued function used at the NLO level to
describe the observable $O$. Eq.~(\ref{IMCNLO})
can be proven by using eq.~(\ref{IMCwithN}) 
in the computation of \mbox{$\int\!dO\IMC(O)f(O)$}, for any arbitrary
function $f(O)$. With the same technique, we can also find a closed
form for $\IMC(O,\xm(x))$ in the general case. We denote by 
\mbox{$P_n(x_1,\ldots,x_n)$} the probability that the toy MC emits
$n$ photons with energies\newline \mbox{$x_1>x_2>\ldots>x_n$}. This 
probability can be obtained from eq.~(\ref{Deltadef}), and reads as follows
\beqn
P_0&=&\Delta(x_0,\xm(x)),
\label{Probz}
\\
P_n(x_1,\ldots,x_n)&=&a^n\Delta(x_0,\xm(x))\stepf(\xm(x)-x_1)
\prod_{i=1}^n\frac{Q(x_i)}{x_i}\stepf(x_i-x_{i+1}),
\label{xProbn}
\eeqn
where in the case of $n$ emissions we define $x_{n+1}=x_0$. We also 
denote by $\Prob_n$ the probability of having $n$ emissions, regardless 
of the photon energies. Clearly, $\Prob_0\equiv P_0$. For $n>0$, we can 
compute this quantity by integrating eq.~(\ref{xProbn}), to obtain
\beq
\Prob_n\equiv\int_0^1\left(\prod_{i=1}^n dx_i\right) P_n(x_1,\ldots,x_n)
=\frac{a^n}{n!}\left[\int_{x_0}^{\xm(x)}dz\frac{Q(z)}{z}\right]^n 
\Delta(x_0,\xm(x)).
\label{Probn}
\eeq
With eqs.~(\ref{Probz}) and~(\ref{Probn}) we can check directly that
the toy MC conserves probability:
\beq
\Prob_0+\sum_{n=1}^\infty \Prob_n=1.
\label{ProbSum}
\eeq
We now introduce the functions \mbox{$O_{n;j}(x;x_1,\ldots,x_n)$}
that return the multi-valued observable $O$ (for single-value
observables, we simply have $j\equiv 1$). Observing that the sum
over all events in eq.~(\ref{IMCwithN}) can be split into sums
over events with a fixed number of emissions, we easily get
the following:
\beqn
&&\IMC(O,\xm(x))=\Prob_0\delta(O-O(x))
\nonumber \\*&&\phantom{aa}
+\sum_{n=1}^\infty\int_0^1\left(\prod_{i=1}^n dx_i\right) P_n(x_1,\ldots,x_n)
\sum_{j=1}^{p(x;x_1,\ldots,x_n)}\delta(O-O_{n;j}(x;x_1,\ldots,x_n)),
\phantom{aa}
\label{IMCwithPn}
\eeqn
where the integer $p(x;x_1,\ldots,x_n)$ in general depends on the kinematics
(this happens with jet observables, for example). Eq.~(\ref{IMCwithPn})
can be used, together with the explicit forms of the emission probabilities
$P_n$, to recover the results for $\IMC(y)$ and $\IMC(z)$ given
in eqs.~(\ref{eqIMCy1}), (\ref{eqIMCyxm}), (\ref{eqIMCzsoft}), and
(\ref{eqIMCz}); these equations have been obtained by direct computation,
which constitutes a cross check of eq.~(\ref{IMCwithPn}).

It should be noted that eq.~(\ref{IMCwithPn}) is obtained from
eq.~(\ref{IMCwithN}) without using the explicit form of the emission
probabilities $P_n$. Thus, both equations hold in the QCD case as well,
with obvious formal substitutions (the $\MCatNLO$ event $\{x,x_k\}_{k=1}^n$
gets replaced by an $\mathbf{n+1}$ kinematical configuration, and the 
information on parton flavours must be inserted).

\subsection{Total rate for exclusive observables\label{app:total}}
We define an observable to be exclusive if its values can be obtained
through a single-valued function, regardless of the number of the photons 
emitted in the shower. The hardest jet energy $y$, considered in
section~\ref{sec:obs}, is an observable of this kind, whereas
the fully inclusive photon energy $z$ is not. It is clear that
in the exclusive case the function ${\cal N}$ defined in 
app.~\ref{app:IMCdef} has the following properties:
\beqn
&&{\cal N}(\{x,x_k\}_{k=1}^n;\overline{O},\delta O)=0,1\phantom{aaaa}
\forall \overline{O};
\\
&&\exists !\overline{O}\;\;{\rm for}\;\;\delta O\to 0\;\;
{\rm such~that:}\phantom{aaaa}
{\cal N}(\{x,x_k\}_{k=1}^n;\overline{O},\delta O)=1.
\label{Nproptwo}
\eeqn
Eq.~(\ref{Nproptwo}) holds for any given $\MCatNLO$ event. Then
\beq
\int dO\IMC(O,\xm(x))=1.
\label{IMCnorm}
\eeq
We can also get this results directly from eq.~(\ref{IMCwithPn}),
using eq.~(\ref{ProbSum}), and the fact that the observable $O$ is 
a single-valued function (and thus \mbox{$p(x;x_1,\ldots,x_n)=1$}).
It is then apparent that eq.~(\ref{IMCnorm}) also holds in QCD.
We can use eq.~(\ref{IMCnorm}) in eq.~(\ref{IMCfive}); we get
\beqn
\int dO\left(\xsecO\right)_{\sss msub}
&=&\int_0^1 dx \Bigg[\left(\int dO\IMC(O,\xm(x))\right)
\frac{a[R(x)-BQ(x)]}{x}
\nonumber \\*
&&+\left(\int dO\IMC(O,1)\right)\left(B+aV+\frac{aB[Q(x)-1]}{x}\right)\Bigg]
\nonumber \\*
&=&\int _0^1 dx \Bigg[B + aV +a\frac{R(x)-B}{x}\Bigg]
\equiv\stot.
\label{exclrates}
\eeqn
In the particular case of $O\equiv y$, this result could be obtained from 
the results for $d\sigma/dy$ given in sec.~\ref{sec:obs}, by including the 
full $x_0$ dependence that has been neglected there.

Since eq.~(\ref{IMCnorm}) holds in QCD, we conclude that eq.~(\ref{exclrates})
also holds in QCD. This is in fact confirmed numerically by the results
for the exclusive observables (such as $\ptwp$, $\ptww$, $\dphiww$)
presented in sect.~\ref{sec:pheno} (in the case that no cuts are applied).

\subsection[Expansion to ${\cal O}(\as)$]{\boldmath Expansion to
${\cal O}(\as)$\label{app:expansion}}
We now consider the expansion to order $a$ and $\as$ of our formulae for
the toy model and QCD $\MCatNLO$, eqs.~(\ref{IMCfive}) and~(\ref{rMCatNLO})
respectively. We start with the toy model. Since we are not interested
in $\ordat$ terms, only zero- and one-photon emissions from the MC
are relevant. We have
\beq
\IMC(O,\xm(x))=\delta(O-O(x))+\orda.
\label{IMCnonsoftexp}
\eeq
In eq.~(\ref{IMCnonsoftexp}) we do not need
to know the $\orda$ term, since $\IMC(O,\xm(x))$ is already multiplied
by $a$ in eq.~(\ref{IMCfive}). On the other hand
\beqn
\IMC(O,1)&=&\Delta(x_0,1)\delta(O-O(0))
+a\int_{x_0}^1 dt\frac{Q(t)}{t}\Delta(x_0,1)\delta(O-O(t))+\ordat
\label{IMCsoftexp}
\\*
&=&\left(1-a\int_{x_0}^1 dt\frac{Q(t)}{t}\right)\delta(O-O(0))
+a\int_{x_0}^1 dt\frac{Q(t)}{t}\delta(O-O(t))+\ordat.\phantom{aa}
\nonumber \\*&&
\label{IMCsoftexpt}
\eeqn
The first term on the r.h.s.\ of eq.~(\ref{IMCsoftexp}) is the contribution
to the observable in the case that the MC does not emit any photons.
The corresponding observable must be computed in the soft limit, with
a coefficient which is the probability of no MC emission, that is,
the Sudakov form factor for the range $(x_0,1)$. The second 
term corresponds to the one-emission contribution. We stress that
eqs.~(\ref{IMCnonsoftexp}) and~(\ref{IMCsoftexp}) could have been
obtained directly from eq.~(\ref{IMCwithPn}). We can now substitute 
eqs.~(\ref{IMCnonsoftexp}) and (\ref{IMCsoftexpt}) into eq.~(\ref{IMCfive}). 
We get:
\beqn
\left(\xsecO\right)_{\sss msub}
&=&\int_0^1 dx \Bigg[\delta(O-O(x))\frac{a[R(x)-BQ(x)]}{x}
\nonumber \\*
&&+\delta(O-O(0))\left(B+aV-\frac{aB}{x}\right)
\nonumber \\*
&&+aB\delta(O-O(0))\left(\frac{Q(x)}{x}
-\int_{x_0}^1 dt\frac{Q(t)}{t}\right)
\nonumber \\*
&&+aB\int_{x_0}^1 dt\delta(O-O(t))\frac{Q(t)}{t}\Bigg]
+\ordat.
\eeqn
We thus have
\beqn
\left(\xsecO\right)_{\sss msub}
&=&\int_0^1 dx \Bigg[\delta(O-O(x))\frac{aR(x)}{x}
+\delta(O-O(0))\left(B+aV-\frac{aB}{x}\right)\Bigg]
\nonumber \\*&+&
aB\int_0^{x_0}dx\frac{Q(x)}{x}\Big[\delta(O-O(0))-\delta(O-O(x))\Big]
+\ordat.
\label{msuborda}
\eeqn
If we use eq.~(\ref{msuborda}) to compute 
\mbox{$\VEV{O}=\int dOO(d\sigma/dO)$}, we get the NLO result of 
eq.~(\ref{nlosubtint}), up to uncontrolled $\ordat$ terms, and up to
terms suppressed by powers of $x_0$, which are numerically very small,
and can be eliminated by letting $x_0\to 0$. In the case of QCD, the role 
of $x_0$ is played by some soft and collinear cutoffs, which cannot be
set to zero. Thus, 
in the real case we have power-suppressed corrections to the ${\cal O}(\as)$
result. Notice that MC's do not in any case control terms of this kind,
and the dependence upon the cutoff is in general power-like. Since,
as discussed in sect.~\ref{sec:msub}, the $\MCatNLO$ and the MC behave 
the same with respect to the cutoff, the result in eq.~(\ref{msuborda})
implies that the dependence upon the cutoff of the $\MCatNLO$ result beyond
${\cal O}(\as)$ is also power-like.

Had we applied the same procedure using naive subtraction,
eq.~(\ref{IMCnlonaive}), we would have obtained the following
(after letting $x_0\to 0$):
\beqn
\left(\xsecO\right)_{\sss naive}
&=&\int_0^1 dx \Bigg[\delta(O-O(x))\frac{aR(x)}{x}
+\delta(O-O(0))\left(B+aV-\frac{aB}{x}\right)\Bigg]
\nonumber \\*&&
+aB\int_0^1 dx\frac{Q(x)}{x}\Bigg[\delta(O-O(x))-
\delta(O-O(0))\Bigg]+\ordat.
\label{naiveorda}
\eeqn
The second integral on the r.h.s.\ of this equation vanishes upon integration
over the whole range of $O$, and the total rate (if meaningful) is then
identical to that obtained from eq.~(\ref{msuborda}). On the other hand,
if we compute $\VEV{O}$ the coefficient of $O(x)$ reads 
\mbox{$a(R(x)+BQ(x))/x$}, instead of the \mbox{$aR(x)/x$} that we get
from eq.~(\ref{msuborda}). This proves in general that the
naive subtraction method suffers from double counting, as already
shown in eq.~(\ref{xsecynaiveexp}) and~(\ref{xsecznaiveexp}).

We now turn to the case of the QCD $\MCatNLO$. The situation is
completely analogous to that of the toy model. The NLO result 
can be recovered only if the ${\cal O}(\as)$ terms of 
the expansion of $\IMC(O,\Ktwo)$, multiplied by the Born term,
cancel the contributions of the terms $d\bSigma\xMC$ in
eq.~(\ref{rMCatNLO}). The QCD case is however more involved,
owing to the fact that, in the case of emission from initial-state
partons, the Sudakov factor has a non-trivial dependence upon the
parton densities. The analogue of eq.~(\ref{IMCsoftexpt}) in the QCD
case reads as follows
\beqn
&&\IMC(O,\Ktwo)=\Bigg(1
-\frac{\as(\muMC)}{2\pi}\int_{q_c} \frac{d\xi_+}{\xi_+}\frac{dz_+}{z_+}
\stepf(z_+^2-\xi_+)P_{ac}^{(0)}(z_+)
\frac{f_c^{\Hone}(\xoMC/z_+,\muMC)}{f_a^{\Hone}(\xoMC,\muMC)}
\nonumber \\*&&\phantom{aaa}
-\frac{\as(\muMC)}{2\pi}\int_{q_c} \frac{d\xi_-}{\xi_-}\frac{dz_-}{z_-}
\stepf(z_-^2-\xi_-)P_{bc}^{(0)}(z_-)
\frac{f_c^{\Htwo}(\xtMC/z_-,\muMC)}{f_b^{\Htwo}(\xtMC,\muMC)}
\Bigg)\delta(O-O(\Ktwo))
\nonumber \\*&&\phantom{aaa}
+\cdots
\label{IMCsoftQCD}
\eeqn
where the dots indicate two more terms of ${\cal O}(\as)$, identical to those 
appearing in eq.~(\ref{IMCsoftQCD}) except for the formal substitution
$\delta(O-O(\Ktwo))\to \delta(O-O(\Kthree))$. In eq.~(\ref{IMCsoftQCD})
we understand that the initial-state partons that initiate the shower
have flavours $a$ and $b$. The lower bound on the integration range,
$q_c$, collectively denotes the cutoffs that prevent MC emissions
from being soft and/or collinear, and plays the same role as $x_0$
in eq.~(\ref{IMCsoftexpt}). The arguments of the parton densities 
have already been discussed in app.~\ref{app:DZ}. We also indicate
explicitly the dependence upon the hard scale $\muMC$, used by the MC 
in the shower; in principle, this scale depends on the kinematics 
of the emission. However, this is not important in what follows,
and the notation we use simplifies the discussion. The Born term
that gets multiplied by eq.~(\ref{IMCsoftQCD}) reads as follows:
\beq
\frac{d\bSigma_{ab}^{(b)}}{d\phi_2}=
\frac{\partial(x_1^{(s)},x_2^{(s)})}{\partial(z_1,z_2)}
f_a^{\Hone}(x_1^{(s)},\muMCatNLO) f_b^{\Htwo}(x_2^{(s)},\muMCatNLO)
\frac{d\sigma_{ab}^{(b)}}{d\phi_2}(x_1^{(s)}P_1,x_2^{(s)}P_2).
\label{bornQCD}
\eeq
The ${\cal O}(\as)$ terms in the product of eqs.~(\ref{IMCsoftQCD})
and~(\ref{bornQCD}) must cancel the contributions of $d\bSigma_{ab}\xMC$
in eq.~(\ref{rMCatNLO}). It is apparent that a necessary condition
for this to happen is \mbox{$x_i^{({\sss MC})}\equiv x_i^{(s)}$};
this is precisely eq.~(\ref{xMCeqxs}). However, the present
proof adds one piece of information which could not be inferred
from eq.~(\ref{xMCeqxs}): in fact, that equation merely states the
fact that the functional form of $x_i^{({\sss MC})}$ can be obtained
through the procedure of event projection; however, it does not
constrain the $x_i^{({\sss MC})}$ variables to be equal to the 
$x_i^{(s)}$ variables used in the event projection {\em at the NLO 
level}. This logical step is made only here. Once the condition
on the Bjorken $x$'s is fulfilled, we achieve the desired cancellation
only if the short-distance cross section resulting from 
eqs.~(\ref{IMCsoftQCD}) and~(\ref{bornQCD}) match those in
$d\bSigma_{ab}\xMC$. The former have the form reported in
eqs.~(\ref{HWxsecpl}) and~(\ref{HWxsecmn}), whereas we have
chosen the latter as in eqs.~(\ref{HWCxsecpl}) and~(\ref{HWCxsecmn});
therefore, we seemingly have uncancelled ${\cal O}(\as)$ terms.
However, as argued in app.~\ref{app:DZ}, the difference between
the two forms cannot induce any effects on infrared-safe observables.
This is in fact what we observe in practice (see sect.~\ref{sec:pheno}).
In other words, we do not have event-by-event cancellation,
but we do have cancellation on a statistical basis.
We remind the reader that the uncancelled ${\cal O}(\as)$ terms
would disappear event by event if the MC had the correct angular 
distribution for soft gluon emission. We point out that, owing
to the presence of the cutoff $q_c$ in eq.~(\ref{IMCsoftQCD}),
we still have a small uncancelled ${\cal O}(\as)$ contribution, which 
is the analogue of the last term on the r.h.s.\ of eq.~(\ref{msuborda}).
As in the case of the toy model, this contribution is suppressed
by powers of the cutoff. Notice that, because of the simultaneous
presence of soft and collinear singularities in QCD, these powers
can in general be multiplied by powers of the logarithm of the
cutoff, but they remain negligible.

We also mention the fact that the flavour structure of
eq.~(\ref{IMCsoftQCD}) is non-trivial. For the cancellation 
at ${\cal O}(\as)$ to occur, we need to ensure that the flavour
assignment in the initial conditions for $\clS$ events matches
the flavour structure of the Born term, eq.~(\ref{bornQCD}).
This is in fact guaranteed by the prescription described in
sect.~\ref{sec:impl}.

We remark finally that the mismatch between $\muMC$ and $\muMCatNLO$ is
formally of higher order. The two scales have fairly different
physical meaning, and one cannot in general assume that
$\muMC=\muMCatNLO$. However, since the evolution of the parton densities
in the MC is only leading-logarithmic, in the present paper we have made
that assumption.  A limited study of scale dependence has been presented
in sect.~\ref{sec:pheno}.

\subsection{Resummation of large logarithms\label{app:logacc}}
Here we give arguments to indicate that the $\MCatNLO$
resums large logarithms in the same way as the ordinary MC. As
previously, we start with the toy model, by reconsidering the 
results of section~\ref{sec:obs} for the variable $z$. We assume that 
eq.~(\ref{MCdz}) is replaced by the more general form
\beq
\dzMC=BS(z,a).
\label{MCdzS}
\eeq
The function $S$ is largely arbitrary, but it should at least have
the following properties:
\begin{itemize}
\item $S$ has a power expansion in $a$, starting with an $\orda$ term;
\item the leading behaviour at small $z$ is $a/z$.
\end{itemize}
The function in eq.~(\ref{MCdz}) has these properties.
Equation (\ref{MCdzS}) is, however, more general; we can for example
assume that $S$ correctly takes into account next-to-leading
logarithmic terms. Eqs.~(\ref{eqIMCzsoft}) and~(\ref{eqIMCz}) now become
\beqn
\IMC(z,1) &=& S(z,a),
\label{eqIMCzSsoft}
\\
\IMC(z,\xm(x)) &=& \delta(z-x)+
S(z,a)\stepf(\xm(x)-z).
\label{eqIMCzS}
\eeqn
We thus get
\beqn
\left(\xsecz\right)_{\sss msub}&=&S(z,a)\Bigg[
\stot+a\frac{R(z)-BQ(z)}{zS(z,a)}-
a\int_{\xm^{-1}(z)}^1 dx\frac{R(x)-BQ(x)}{x}\Bigg].\phantom{aaaa}
\label{dsigmadzfiveS}
\eeqn
Thus, since $S(z,a)$ diverges as $1/z$ for $z\to 0$, we have
\beq
\left(\xsecz\right)_{\sss msub}\approx\stot S(z,a)+\ldots
\label{smallzxsecS}
\eeq
where the neglected terms are either constant for $z\to 0$, or
vanish in this limit. Eq.~(\ref{smallzxsecS}) generalizes
eq.~(\ref{smallzxsec}). Formally, the $\orda$ terms in $\stot$ 
produce a new tower of NLL terms when multiplied by the LL tower in $S(z,a)$.
But since $\stot$ is factorized, this new tower of logarithms does not 
produce any change in the $z$ spectrum of the emitted photons
compared to the MC. This is in fact evident
if we normalize the $z$ distribution to the total exclusive
rate, which is $\stot$ in the case of $\MCatNLO$, and $B$ in the
case of MC: in both cases, we get $S(z,a)$.
Furthermore, the terms neglected in eq.~(\ref{smallzxsec}) do not 
lead to any logarithmic-enhanced behaviour. Thus, 
the leading and whatever subleading logarithmic behaviour in $z$ of soft 
photon emission is the same in MC and $\MCatNLO$ evolution.

Notice that eq.~(\ref{smallzxsecS}) implies that the ratio of the
number of soft photons to the number of showers is the same as in
the ordinary MC. This fact, and the fact that the energy
spectrum of the emitted photons is also the same, suggests that,
for an arbitrary observable $O$ that gets resummed
through ordinary MC methods, the resummation performed by the 
$\MCatNLO$  has the same logarithmic accuracy as the MC.

However, the preceding arguments are fairly closely linked to the
details of the toy model, which makes it difficult to extend them
to the QCD case. We therefore now give another argument, which
can easily be extended to QCD; this argument works only 
for a given class of observables, which however includes 
most of the common measurable quantities. Namely, we consider
the case of an exclusive observable $O$ such that, when 
$O\simeq O_s$, any fixed-order cross section contains large logarithms
\mbox{$\log(O-O_s)$}. Furthermore, we assume that, regardless of the
number of photons emitted, at NLO and/or by the MC, when $O\simeq O_s$
the energies of {\em all} photons emitted are constrained to be small;
in other words, any hard photon emission moves $O$ away from $O_s$.
This requirement is not very restrictive; still, there can be observables 
which do not meet it, and have large logarithms in the cross sections, 
but these logarithms will be multiplied by at least an extra factor of $a$ 
relative to the case we treat.

Our assumption implies that, when we
consider \mbox{$\IMC(O,\xm(x))$} for \mbox{$O\to O_s$}, the energy $x$
of the photon emitted at NLO is effectively a function of $O$:
\beq
x\equiv x(O)=\kappa(O-O_s)+{\cal O}\left((O-O_s)^2\right)\,,
\label{xvsO}
\eeq
where we used the fact that $x(O_s)=0$. 
We can use eq.~(\ref{xvsO}) to compute the NLO integrated cross section.
We have
\beqn
&&\overline\sigma_{\sss NLO}(O)\equiv\int_{O_s}^O d\OO\xsecOONLO
\nonumber \\*&&\;\;
=\int_0^1 dx \left[\int_{O_s}^O d\OO\delta(\OO-O(x))\frac{aR(x)}{x}
+\int_{O_s}^O d\OO\delta(\OO-O(0))\left(B+aV-\frac{aB}{x}\right)\right],
\nonumber \\*&&
\label{largeLNLO}
\eeqn
where we have used eq.~(\ref{nlosubtint}). With eq.~(\ref{xvsO}) we obtain
\beq
\overline\sigma_{\sss NLO}(O)=B+aV+
\int_0^1 dx\left[\frac{aR(x)}{x}\stepf(\kappa(O-O_s)-x)-\frac{aB}{x}\right].
\label{vsigO}
\eeq
Consistently with eq.~(\ref{limreal}), we assume that
\beq
R(x)=B+\sum_{n=1}^\infty R_n x^n\,,
\label{Rxexpand}
\eeq
and thus eq.~(\ref{vsigO}) gives
\beq
\overline\sigma_{\sss NLO}(O)=B+aB\log\left(\kappa(O-O_s)\right)+aV
+a\sum_{n=1}^\infty \frac{R_n}{n} \left(\kappa(O-O_s)\right)^n\,.
\label{RSxsecNLO}
\eeq
The first two terms on the right-hand side are the first two
terms in the LL tower that is resummed by MC means. The third and
fourth terms are NLL, the latter being also power suppressed. We point
out that since the only large logarithm in the cross section has to be 
\mbox{$\log(O-O_s)$}, the coefficient $\kappa$ must be of order unity.

We now consider the case of the $\MCatNLO$. We denote the normalized
pure MC result as follows:
\beq
\overline\sigma_{\sss MC}(O)\equiv\int_{O_s}^O d\OO\IMC(\OO,1).
\eeq
We also notice that, owing to the fact that $\xm(0)=1$, and assuming 
that $\xm(x)$ has a power expansion in $x$ around $x=0$, we have
(for $O\simeq O_s$)
\beq
\int_{O_s}^O d\OO\IMC(\OO,\xm(x))=
\int_{O_s}^O d\OO\left(\IMC(\OO,1)+{\rm NLL}\right)
\stepf(\kappa(O-O_s)-x)\,.
\label{sigOOH}
\eeq
We can now repeat the same computation that led us to eq.~(\ref{RSxsecNLO}),
for the $\MCatNLO$, eq.~(\ref{IMCfive}). In the case of $\clS$ events,
analogously to what happens for the contribution of the 0-body events
in eq.~(\ref{largeLNLO}), the restricted range in $O$ does not constrain
the integration range in $x$. Thus
\beqn
\overline\sigma_{\clS}(O)&\equiv &\int_{O_s}^O d\OO\xsecOOclS
\nonumber \\*&=&
\overline\sigma_{\sss MC}(O)\left[B+aV
+aB\int_0^1\,dx\frac{Q(x)-1}{x}\right]\;;
\label{RSxsecclS}
\eeqn
we therefore recover the pure MC result, up to $\orda$ terms
in the normalization. On the other hand, the restricted range in $O$ 
{\em does} constrain the integration range in $x$ in the case of $\clH$
events, as shown in eq.~(\ref{sigOOH}). We obtain
\beqn
\overline\sigma_{\clH}(O)&\equiv &\int_{O_s}^O d\OO\xsecOOclH
\nonumber \\*&=&
a\left(\overline\sigma_{\sss MC}(O)+{\rm NLL}\right)
\sum_{n=1}^\infty \frac{R_n-BQ_n}{n} \left(\kappa(O-O_s)\right)^n\,,
\label{RSxsecclH}
\eeqn
where $Q_n$ are the coefficients of the expansion of $Q(x)$ around
$x=0$. Therefore, eq.~(\ref{RSxsecclH}) does not contribute to the 
large-logarithm resummation for $O\to O_s$. Taking into account 
eq.~(\ref{RSxsecclS}), this allows us to conclude that the $\MCatNLO$ 
resums large logarithms in the same way as the ordinary MC does.

It is clear that the argument given above works in the same way in
the QCD case. One has to pay attention to the fact that, in QCD,
soft and collinear singularities exist. In general, the restricted
range in $O$ constrains the emissions to be either soft {\em or} 
collinear, not necessarily soft {\em and} collinear. This implies
that in QCD each term of the power series that appears in 
eq.~(\ref{RSxsecclH}) can be multiplied by a power of \mbox{$\log(O-O_s)$}.
However, this does not change our conclusion, that $\clH$ events
do not contribute large logarithms to the physical cross section.

\section{\boldmath $\MCatNLO$ based upon slicing method}
In this appendix, we comment on the approach of 
refs.~\cite{Potter:2001an,Dobbs:2001gb}, where a formalism is
presented to construct an $\MCatNLO$ using the slicing method at 
the NLO level. This formalism is called $\Phi$-space Veto in
ref.~\cite{Dobbs:2001gb}.  It has the main objective and virtue
of ensuring that only positive event weights are generated.
However, we argue that this is at the expense of being
able to recover the NLO results upon expansion in $\as$
(this is because the resulting $\MCatNLO$ does {\em not} have an expansion
in $\as$), and a smooth matching between different kinematic regions.

In the course of the discussion, we
use the toy model, which is sufficient to highlight the main
features of the method. We also propose a formula for an $\MCatNLO$
based upon the slicing method, which does not have any of the problems
that affect the $\Phi$-space Veto method, and fulfills the definition
of $\MCatNLO$ given in sect.~\ref{sec:obj} (but as a consequence is
not guaranteed to generate positive event weights).

The $\Phi$-space Veto relies on the result for the NLO cross section
obtained with the slicing method, eq.~(\ref{nloslicing}). The first
step consists in fixing the parameter $\delta$, by imposing the 
condition that eq.~(\ref{nloslicing}) does not contain any contribution
proportional to $O(0)$. In other words, one seeks $\delta_0$ such 
that\footnote{This idea was first introduced, in a slightly different form,
in ref.~\cite{Baer:1991ca}.}
\beq
B+a\left(B\log\delta_0 +V\right)=0\,.
\eeq
Thus
\beq
\log\delta_0=-\frac{B+aV}{aB}\;\;\;\;\Longleftrightarrow\;\;\;\;
\delta_0=\exp\left(-\frac{B+aV}{aB}\right)\,.
\label{deltazero}
\eeq
With such a choice, eq.~(\ref{nloslicing}) becomes
\beq
\VEV{O}_{\sss slice}=a\int_{\delta_0}^1 dx\, \frac{O(x)R(x)}{x} 
+ {\cal O}(\delta_0).
\label{slicePhi}
\eeq
Given the fact that the parameter $\delta_0$ is not free any longer,
being related to the Born and virtual contributions as shown in
eq.~(\ref{deltazero}), the uncontrolled ${\cal O}(\delta_0)$ terms in
eq.~(\ref{slicePhi}) may be non-negligible. An improvement of 
eq.~(\ref{slicePhi}) is proposed in ref.~\cite{Potter:2001an},
which uses a subtraction method in a restricted part of the phase
space (this procedure is called ``hybrid subtraction''): however, 
this improvement is of no importance in what follows, and will be 
neglected here (also, it is not used in current implementations of 
refs.~\cite{Potter:2001an,Dobbs:2001gb}). As anticipated, the reason for 
fixing the slicing parameter $\delta$ can be seen in eq.~(\ref{slicePhi});
this equation does {\em not} contain any negative contributions,
and this fact is used in refs.~\cite{Potter:2001an,Dobbs:2001gb}
to define an $\MCatNLO$ with no negative-weight events. However,
the resulting $\delta_0$ is not analytic around $a=0$, which seems
to us to cause problems, as will be discussed below.

The next step in the $\Phi$-space Veto method is the introduction of 
a free parameter $\delta_{\sss PS}$, which fulfills the condition 
\mbox{$\delta_0<\delta_{\sss PS}<1$}. In ref.~\cite{Dobbs:2001gb},
it is suggested that one set $\delta_{\sss PS}$ equal to the boundary
of the dead zone, in the case that \HW\ is used to construct 
the $\MCatNLO$. In general, it seems difficult to prove that 
this condition is consistent with the requirement 
\mbox{$\delta_0<\delta_{\sss PS}$}, especially since the boundary of
the dead zone has a different (process dependent) shape from the
region removed by slicing, and may depend on different variables.

In the language of the present paper, the introduction
of $\delta_{\sss PS}$ implies the definition of a modified
slicing formula:
\beq
\VEV{O}_{\sss mslice}=a\int_0^1 dx\left[
O(x)\frac{R(x)}{x}\stepf(x-\delta_{\sss PS})
+O(0)\frac{R(x)}{x}\stepf(x-\delta_0)\stepf(\delta_{\sss PS}-x)\right].
\label{mslicePhi}
\eeq
From this equation, we can construct a toy $\MCatNLO$ as explained in
sect.~\ref{sec:Matching}, paying attention to the fact that, in
the $\Phi$-space Veto method, the maximum energy available to the
photon in the first branching is required to be $\delta_{\sss PS}$
(instead of 1) for class $\clS$ events. Therefore
\beqn
\left(\xsecO\right)_{\sss mslice}&=&a\int_0^1 dx\Bigg[
\IMC(O,\xm(x))\frac{R(x)}{x} \stepf(x-\delta_{\sss PS})
\nonumber \\*&&\phantom{a\int_0^1 dx}
+\IMC(O,\delta_{\sss PS})\frac{R(x)}{x} 
\stepf(x-\delta_0)\stepf(\delta_{\sss PS}-x)\Bigg].
\label{PhiVMC}
\eeqn
Eq.~(\ref{PhiVMC}) can now be used to compute the predictions of
the $\Phi$-space Veto method, in the same way that eq.~(\ref{IMCfive})
was used for our $\MCatNLO$.  We consider here the exclusive observable
$y$, introduced in sect.~\ref{sec:obs}.
We can substitute eq.~(\ref{eqIMCyxm}) directly into eq.~(\ref{PhiVMC});
on the other hand, we need to use
\beq
\IMC(y,\delta_{\sss PS})= a\frac{Q(y)}{y}\Delta(y,\delta_{\sss PS})
\stepf(\delta_{\sss PS}-y)
\eeq
instead of eq.~(\ref{eqIMCy1}). We find the following:
\beq
\left(\dszero\right)_{\sss mslice}
= a^2\frac{Q(y)}{y}\Delta(y,\delta_{\sss PS})\stepf(\delta_{\sss PS}-y)
\int_{\delta_0}^{\delta_{\sss PS}}dx\,\frac{R(x)}{x}\,.
\label{mdszeromslice}
\eeq
We also find:

\noindent
$\diamond$ for $y<x_e:$ 
\beq
\left(\dsone\right)_{\sss mslice}=\Bigg[a\frac{R(y)}{y}\Delta(y,\xm(y))
+a^2\frac{Q(y)}{y}\int_{\delta_{\sss PS}}^y 
\frac{dx}{x}R(x)\Delta(y,\xm(x))\Bigg]\stepf(y-\delta_{\sss PS})\;;
\label{mslicecaseone}
\eeq

\noindent
$\diamond$ for $y>x_e:$ 
\beq
\left(\dsone\right)_{\sss mslice}=a\frac{R(y)}{y}\stepf(y-\delta_{\sss PS})
+a^2\frac{Q(y)}{y}\stepf(\xm^{-1}(y)-\delta_{\sss PS})
\int_{\delta_{\sss PS}}^{\xm^{-1}(y)}\frac{dx}{x}R(x)\Delta(y,\xm(x))\;.
\label{mslicecasetwo}
\eeq
At $y\to 0$, we expect the result of the ordinary MC, eq.~(\ref{MCdy}),
to be reproduced, up to terms of higher logarithmic order, and up to
multiplicative factors of higher order in $a$, i.e., we expect an
equation similar to~(\ref{smallyxsec}) to hold. Eq.~(\ref{mdszeromslice}) 
appears to fail to achieve this result, since it starts at $\ordat$.
However, $\delta_0$ also depends upon $a$; thus, in order to reach
firm conclusions, the integral in eq.~(\ref{mdszeromslice}) needs to
be evaluated explicitly. To do that, we assume again that $R(x)$ has
a power series expansion in $x$, eq.~(\ref{Rxexpand}).
Eq.~(\ref{mdszeromslice}) then becomes
\beq
\left(\dszero\right)_{\sss mslice}
=a^2\frac{Q(y)}{y}\Delta(y,\delta_{\sss PS})\stepf(\delta_{\sss PS}-y)
\left[B\log\frac{\delta_{\sss PS}}{\delta_0}
+\sum_{n=1}^\infty \frac{R_n}{n}\left(\delta_{\sss PS}^n-\delta_0^n\right)\right]\,.
\eeq
Using eq.~(\ref{deltazero}) we now find
\beqn
\left(\dszero\right)_{\sss mslice}
&=&a\frac{Q(y)}{y}\Delta(y,\delta_{\sss PS})\stepf(\delta_{\sss PS}-y)
\Bigg[B+aV+aB\log\delta_{\sss PS}
\nonumber \\*&&
+a\sum_{n=1}^\infty \frac{R_n}{n}\left(\delta_{\sss PS}^n-
\exp\left(-n\frac{B+aV}{aB}\right)\right)\Bigg]\,.
\label{dzmslicetwo}
\eeqn
This equation has the same functional form in $y$ as eq.~(\ref{MCdy}),
up to uncontrolled NLL terms (due to the fact that the second argument of the
Sudakov form factor is $\delta_{\sss PS}$ instead of 1), and {\em formally}
starts at $\orda$. However, it has the rather unpleasant feature that it
cannot be expanded as a power series in $a$, since it is not analytic at $a=0$.
We therefore have the paradoxical situation that a quantity which is
based on perturbation theory cannot be expanded in powers of 
the coupling constant. This is in fact due to the form of $\delta_0$,
eq.~(\ref{deltazero}). Still, if we assume that $a^2\delta_0^n$ is of
$\ordat$, then eq.~(\ref{dzmslicetwo}) has the correct Sudakov
behaviour for $y\to 0$ (this is also formally consistent with the
necessity of neglecting powers of $\delta_0$ in the slicing procedure;
however, we believe that this is not justified here, since $\delta_0$ is not 
a free parameter any longer). However, with the same assumption we conclude
that the $\Phi$-space Veto method has double counting problems, since for
$y<\delta_{\sss PS}$ the result of eq.~(\ref{dzmslicetwo}) does not coincide 
with the pure NLO result. On the other hand, for $y>\delta_{\sss PS}$ 
the $\MCatNLO$ and NLO results agree up to $\ordat$ terms (see
eqs.~(\ref{mslicecaseone}) and~(\ref{mslicecasetwo})). Therefore, the
$\MCatNLO$ prediction has a discontinuity at $y=\delta_{\sss PS}$.

We now define an $\MCatNLO$ based upon the slicing method that does not
have the problems outlined above. We can easily accomplish this task by
noting that the slicing formulae can be obtained, at the NLO level,
by making suitable approximations in the subtraction formulae. We start from
eq.~(\ref{nloslicing}), which we re-write as follows
\beq
\VEV{O}_{\sss slice}=\int_\delta^1 dx \left[O(x)\frac{aR(x)}{x}
+O(0)\left(B+aV+aB\log\delta\right)\right].
\label{nloslicetwo}
\eeq
From now on, we do not write explicitly the terms of order ${\cal O}(\delta)$
that are always neglected in slicing formulae. Notice that 
eqs.~(\ref{nloslicing}) and~(\ref{nloslicetwo}) in fact coincide up to 
${\cal O}(\delta)$ terms; the two equations would be exactly identical had we 
multiplied $O(0)$ by \mbox{$1/(1-\delta)$}. Using eq.~(\ref{nloslicetwo})
we now define an $\MCatNLO$, following the procedure described in
sect.~\ref{sec:Matching}. We obtain
\beqn
\left(\xsecO\right)_{\sss mslice}
&=&\int_\delta^1 dx \Bigg[\IMC(O,\xm(x))\frac{a[R(x)-BQ(x)]}{x}
\nonumber \\*
&&+\IMC(O,1)\left(B+aV+aB\log\delta+aB\frac{Q(x)}{x}\right)\Bigg]\;.
\label{MCatNLOslice}
\eeqn
We stress that this formula has the same features as a ``standard''
slicing formula, such as eq.~(\ref{nloslicetwo}): the parameter
$\delta$ is free (and small), the integration
over $x$ does {\em not} extend down to zero, and the counterterm
of the subtraction method is replaced by a term proportional to
$\log\delta$. There is, however, one important difference. In
eq.~(\ref{nloslicetwo}), the weights attached to $O(x)$ and $O(0)$
are divergent when $\delta\to 0$ (and can now be negative).
This does not happen in 
eq.~(\ref{MCatNLOslice}) for the weights attached to $\IMC(O,\xm(x))$
and $\IMC(O,1)$. In the latter term, however, this property holds
only if the integration over the phase space is performed before the 
limit $\delta\to 0$ is carried out. This is inconvenient, because the 
unweighting procedure is performed in practice {\em before} the integration.
In the spirit of the slicing method, we should integrate the term
\mbox{$aBQ(x)/x$} in the weight attached to $\IMC(O,1)$ analytically,
which may or may not be possible, depending upon the form of $Q(x)$.
In order to simplify the discussion, we assume
\beq
Q(x)=\stepf(x_{\sss dead}-x)\,.
\eeq
Using this form in eq.~(\ref{MCatNLOslice}) we obtain
\beqn
\left(\xsecO\right)_{\sss mslice}
&=&\int_\delta^1 dx \,\IMC(O,\xm(x))\frac{a[R(x)-B\stepf(x_{\sss dead}-x)]}{x}
\nonumber \\*
&&+\IMC(O,1)\left(B+aV+aB\log x_{\sss dead}\right).
\label{MCatNLOslicesimp}
\eeqn
In general, the weights attached to $\IMC(O,\xm(x))$ and $\IMC(O,1)$
in this equation are not positive definite. However, it is clear that
unweighting (in the sense of generating events with weights of equal
{\em magnitude}) can easily be achieved, since all the cancellations between
large numbers occur under the integral sign.
In this sense, the $\MCatNLO$ based upon the slicing method is similar 
to the $\MCatNLO$ based upon the subtraction method, eq.~(\ref{IMCfive}). 
However, while in eq.~(\ref{IMCfive}) the cancellation is achieved by
numerical methods, the generalization of eq.~(\ref{MCatNLOslicesimp})
can be obtained from eq.~(\ref{MCatNLOslice}) only through an
analytical integration. This is similar to what happens at the NLO
level; in the subtraction method, the residue of the soft/collinear
singularity is integrated numerically, whereas in the slicing method
an analytical integration is performed.

In the general case in which $Q(x)$ does not give an easy analytical
integral, we can use a mixture of analytical and numerical methods.
Namely, we use the identity
\beq
Q(x)=\stepf(x_{th}-x)+Q(x)-\stepf(x_{th}-x)
\eeq
in the last term of eq.~(\ref{MCatNLOslice}), where $x_{th}$ is
an arbitrary constant. We readily obtain
\beqn
\left(\xsecO\right)_{\sss mslice}
&=&\int_\delta^1 dx \Bigg[\IMC(O,\xm(x))\frac{a[R(x)-B Q(x)]}{x}
\nonumber \\*
&&+\IMC(O,1)\left(B+aV+aB\log x_{th}
+aB\frac{Q(x)-\stepf(x_{th}-x)}{x}\right)\Bigg]\!.\phantom{aaaaaa}
\label{MCatNLOslicesimpt}
\eeqn
We can now prove that eq.~(\ref{MCatNLOslice}) has no 
double counting, and resums soft logarithms in the same way as
standard MC does. We follow the procedures introduced in 
app.~\ref{app:expansion} and \ref{app:logacc}. Using
eqs.~(\ref{IMCnonsoftexp}) and~(\ref{IMCsoftexpt}), we get
\beqn
\left(\xsecO\right)_{\sss mslice}
&=&\int_\delta^1 dx \left[\delta(O-O(x))\frac{aR(x)}{x}
+\delta(O-O(0))\left(B+aV+aB\log\delta\right)\right]
\nonumber \\*&+&
aB\int_0^1 dx\frac{Q(x)}{x}\Big[\delta(O-O(0))-\delta(O-O(x))\Big]
\!\Big[\stepf(x-\delta)-\stepf(x-x_0)\Big]
\nonumber \\*&+&\ordat,
\label{msliceorda}
\eeqn
which has to be compared to eq.~(\ref{msuborda}). As in that case, we 
recover the NLO result by considering \mbox{$\int\!dO\,O(d\sigma/dO)$},
up to $\ordat$ and power-suppressed terms. Notice that we also have
terms of ${\cal O}(\delta)$ in eq.~(\ref{msliceorda}), consistently
with the approximation upon which the slicing method is based.
The distribution of the photon energies can be obtained as explained
in app.~\ref{app:logacc}. We get
\beqn
\left(\xsecz\right)_{\sss mslice}&=&S(z,a)\Bigg[
\stot+a\frac{R(z)-BQ(z)}{zS(z,a)}+
aB\int_{\max(\delta,\xm^{-1}(z))}^1 dx\frac{Q(x)}{x}\Bigg],\phantom{aaaa}
\label{dsigmadzsliceS}
\eeqn
which is the analogue of eq.~(\ref{dsigmadzfiveS}). It is apparent that
the same conclusions as in app.~\ref{app:logacc} apply here for $z\to 0$.

Finally, notice that, by using eq.~(\ref{IMCnorm}), we obtain, in
the case of an exclusive variable
\beq
\int dO\left(\xsecO\right)_{\sss mslice}
=\int_\delta^1 dx \left[B+aV+aB\log\delta
+\frac{aR(x)}{x}\right]
\equiv\stot,
\label{exclratesslice}
\eeq
which is the result that we also find in app.~\ref{app:total}.
We conclude that eq.~(\ref{MCatNLOslice}) defines an $\MCatNLO$
that has the same formal properties as that defined in eq.~(\ref{IMCfive}).

\end{document}